\documentclass[a4paper,12pt]{article}
\usepackage{aas_macros}
\pdfoutput=1 

\usepackage{jcappub} 

\usepackage[utf8x]{inputenc}
\usepackage[T2A]{fontenc}
\usepackage[english]{babel}
\usepackage{amsmath}
\usepackage{framed}
\usepackage{amssymb}
\usepackage{bm}
\usepackage{epsfig}
\usepackage[dvipsnames,usenames]{xcolor} 
\usepackage[colorlinks=true,citecolor=blue]{hyperref}
\usepackage{todonotes}
\usepackage[numbers,sort&compress]{natbib}

\graphicspath{{./plots/}}

\newcommand{\be}{\begin{equation}}
\newcommand{\ee}{\end{equation}}
\newcommand{\bea}{\begin{eqnarray}}
\newcommand{\eea}{\end{eqnarray}}

\title{From dwarf galaxies to galaxy clusters: Self-Interacting Dark Matter over 7 orders of magnitude in halo mass}

\author[1,2]{Kyrylo~Bondarenko,}
\author[3]{Anastasia~Sokolenko,}
\author[4]{Alexey~Boyarsky,}
\author[5]{Andrew~Robertson,}
\author[4]{David~Harvey,}
\author[2]{Yves Revaz}

\emailAdd{kyrylo.bondarenko@cern.ch}
\emailAdd{anastasia.sokolenko@oeaw.ac.at}
\emailAdd{boyarsky@lorentz.leidenuniv.nl}
\emailAdd{andrew.robertson@durham.ac.uk}
\emailAdd{harvey@lorentz.leidenuniv.nl}
\emailAdd{yves.revaz@epfl.ch}

\affiliation[1]{Theoretical Physics Department, CERN, 1 Esplanade des Particules, Geneva 23, CH-1211, Switzerland}
\affiliation[2]{L'Ecole polytechnique f\'ed\'erale de Lausanne, Route Cantonale, 1015 Lausanne, Switzerland}
\affiliation[3]{Institute of High Energy Physics, Austrian Academy of Sciences, Nikolsdorfergasse 18, 1050 Vienna, Austria}
\affiliation[4]{Intituut-Lorentz, Leiden University, Niels Bohrweg 2, 2333
CA Leiden, The Netherlands}
\affiliation[5]{Centre for Extragalactic Astronomy, Durham University, South Road, Durham DH1 3LE, UK}

\abstract{ In this paper we study the density profiles of self-interacting dark matter (SIDM) haloes  spanning the full observable mass range, from dwarf galaxies to galaxy clusters. Using realistic simulations that model the baryonic physics relevant for galaxy formation, we compare the density profiles of haloes simulated with either SIDM or cold and collisionless dark matter (CDM) to those inferred from observations of stellar velocity dispersion, gas rotation curves, weak and strong gravitational lensing, and/or  X-ray maps. We make our comparison in terms of the maximal surface density of haloes, circumventing the need for semi-analytic or parametric models for dark matter density profiles. We find that the maximal surface density as a function of halo mass is well reproduced by CDM simulations that include baryons, while for SIDM with a velocity-independent cross-section of 1 cm$^2/$g, the simulated galaxy clusters have mean maximal surface densities that are below those of observed systems by an amount greater than the standard deviation of the observed maximal surface density at fixed mass. For less massive systems both CDM and SIDM agree with the observation equally well.
}

\begin{document}
\maketitle

\section{Introduction}

Understanding the nature of dark matter (DM) has become one of the most pressing questions in modern science. Despite its ubiquity, evidence for its existence is exclusively based on its gravitational interactions (see e.g.~\cite{Bertone:2010zza}). As such we know very little about its particle properties. Should there be any confirmed detection of non-gravitational interactions, a window to new physics beyond the standard model would be opened, defining the direction of astro- and particle physics in the coming decades. As such, it is vital that we test DM and constrain its properties in the most model independent ways possible. 

The strength of DM's interactions with standard model particles is extremely well constrained by terrestrial detectors and colliders (see e.g.~\cite{Aprile:2017iyp,Cui:2017nnn,Bondarenko:2019vrb} and references therein). Despite this, the bounds on interactions with {\it itself} remain distinctly loose, with the limits $\sim 20$ orders of magnitude higher, allowing for interaction cross-sections similar to those for strong interactions between nucleons (see \cite{Tulin:2017ara} for a review). Self-interacting dark matter (SIDM) has been cited as a way to solve potential discrepancies between theory and observations on small scales, including the diversity of rotation curves \citep{Oman:2015xda,Koda2011}, the number of DM substructures in the Milky Way and the density profile of DM dominated dwarf galaxies (see e.g.~\cite{Tulin:2017ara} and references therein). However, whether these inconsistencies exist (when observational uncertainties and baryonic physics are taken into account) \cite{harvey:2018dwarf,baryonCores,Wang:2012sv,Guo:2015jia,Fattahi:2016nld} and whether they can be solved with SIDM remain disputable \citep{diversitySIDM,2019arXiv191109116S}. It has been argued, for example, that Draco, a DM dominated dwarf galaxy, actually harnesses a central density cusp~\citep{dracoCusp}, potentially placing a strong constraint on the cross section of dark matter at these velocity scales.

Important constraints on the self-interaction cross section divided by the DM particle mass, $\sigma / m$, come from studies of merging and relaxed galaxy clusters \cite{Markevitch:2003at,Randall:2007ph,walker_private,Kahlhoefer:2013dca,Harvey:2015hha,Robertson:2016xjh,Wittman:2017gxn,Harvey2017:wobble,Harvey2018:BAHAMAS}, where the dynamics and distribution of dark matter can be inferred through gravitational lensing. The large number of studies from galaxy clusters has led to robust constraints at a velocity of $\sim 1000$~km/s, however it would be quite natural to have a velocity-{\it dependent} SIDM cross-section. In this scenario, constraints derived from galaxy clusters would have  little bearing on dwarf galaxy scales since this would allow a degree of freedom whereby the cross-section could be much higher in these environments (see~\cite{Valli:2017ktb,Sameie:2019zfo,Kahlhoefer:2019oyt} for a discussion of the possible effects of SIDM in dwarf galaxies). 

\begin{figure}[t!]
  \centering
  \includegraphics[width=0.47\textwidth]{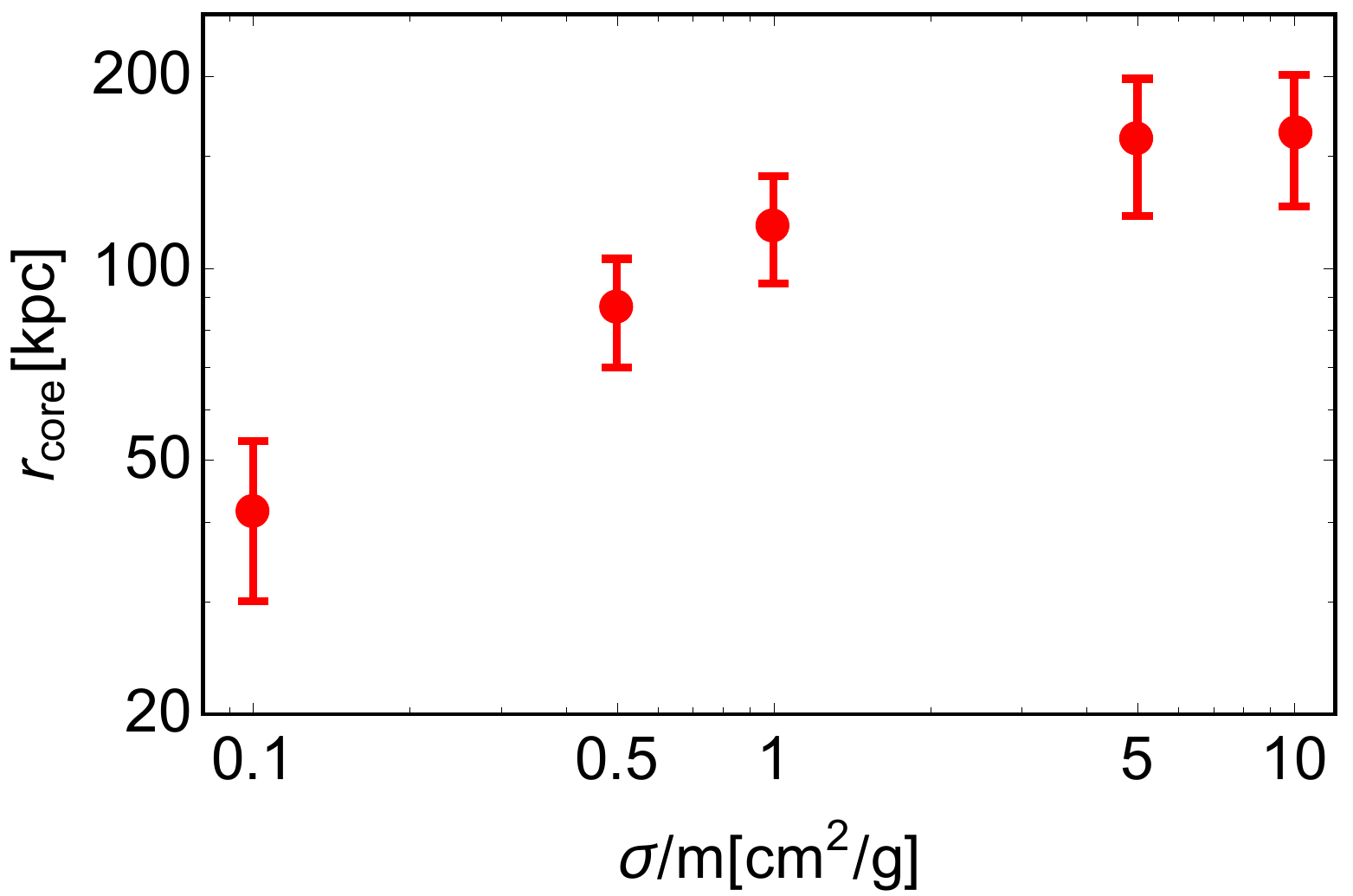}~\includegraphics[width=0.49\textwidth]{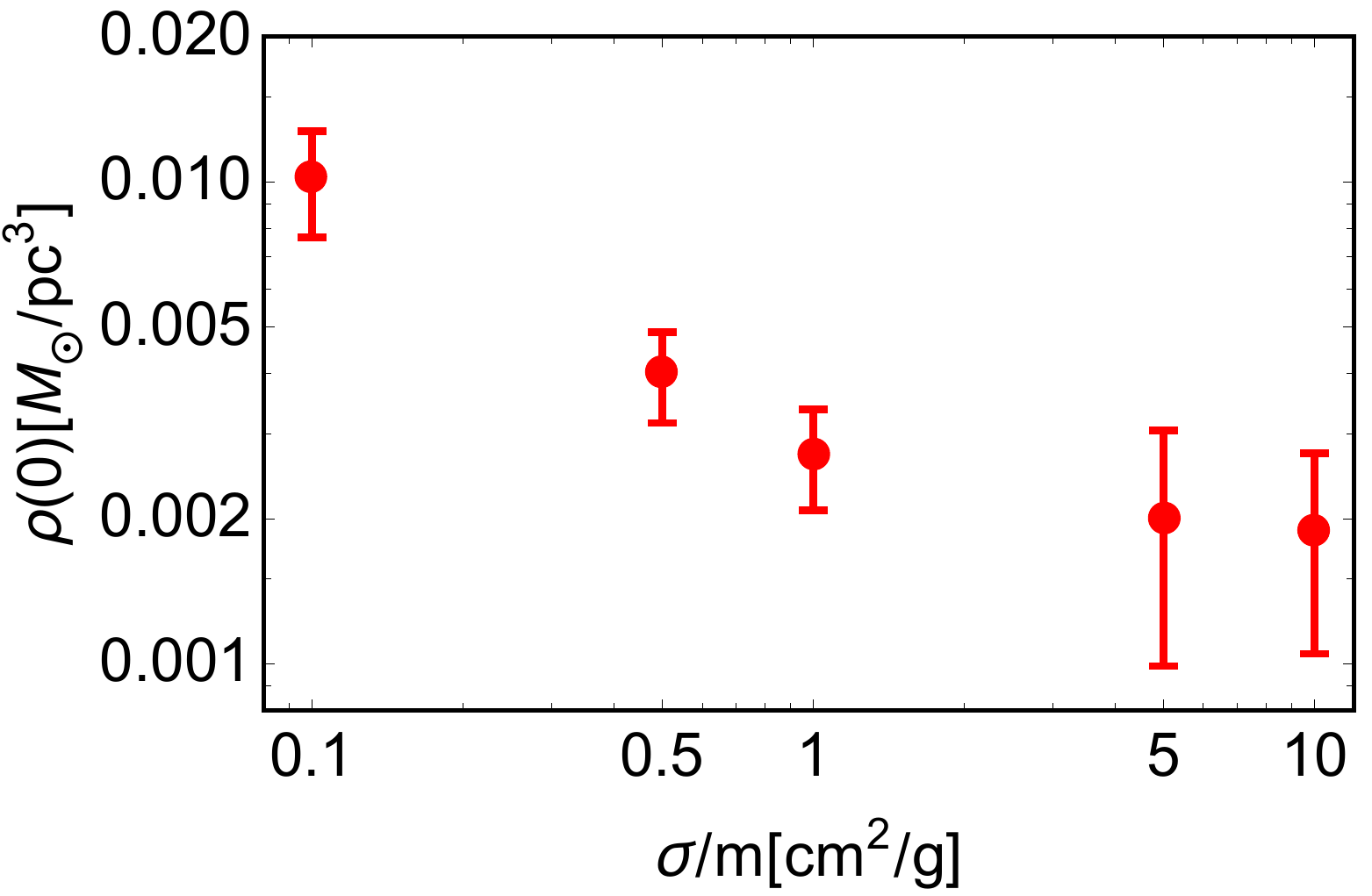}
  \caption{\textit{Left panel:} The dependence of $r_\text{core}$ on the cross section as predicted by dark matter only simulations of clusters of galaxies~\citep{Brinckmann:2017uve}, where we define $r_{\rm core}$ as the radius at which $\rho_{\text{CDM}}(r_{\rm core}) = \rho_{\text{SIDM}}(0)$, see~\cite{Sokolenko:2018noz} for more details. We see that below $\sim1$cm$^2$/g the core radius is a linear function of the cross section, however above this it saturates and becomes indistinguishable. The error bars represent  the standard deviation of the distribution. \textit{Right panel:} The density in the center for the same SIDM simulations.
  This figure is taken from~\cite{Sokolenko:2018noz}.
  }
  \label{fig:rMofsigma}
\end{figure}

Therefore, in the paper we will use the inner properties of DM halos of all sizes to constraint SIDM. 
There is, however, a theoretical difficulty that makes such an analysis non-trivial. \textcolor{black}{First, the dependence of the radii of constant-density central cores (that are expected to form in SIDM halos) on the cross section  appears to be non-linear and saturates at large cross sections (see Fig.~\ref{fig:rMofsigma} that we take from~\cite{Sokolenko:2018noz}, see that paper for discussion).
Also, some of the assumptions that are often used to make semi-analytic predictions for the properties of SIDM haloes (see e.g.~\cite{Tulin:2017ara,Kaplinghat:2015aga}) are not always satisfied in the simulations~\cite{Sokolenko:2018noz}. 
At the same time, despite some ingredients of the 
isothermal Jeans model being in contradiction with SIDM simulations, the resulting SIDM profiles often provide a reasonable match to the profiles of simulated halos
~\cite{Nishikawa:2019lsc,Elbert:2014bma,Ren:2018jpt,Essig:2018pzq, Robertson:2020pxj}. To be on the safe side,} in this work we will avoid any analytic model for the density profiles of SIDM halos and will compare SIDM simulations with observational results directly.

$N$-body simulations of SIDM have advanced dramatically in the last decade and are now readily available, see e.g.~\cite{Harvey2018:BAHAMAS,Robertson:2018anx,Brinckmann:2017uve}.
To address this ambitious task we will therefore adopt simulations of halos that span the entire mass range of the observations.

To summarise,  in this paper we are going to treat SIDM consistently over several orders of magnitudes in halo mass (and therefore over a range of relative particle velocities) by

 \begin{enumerate}
     \item adopting an ensemble of objects at each characteristic velocity (mass scale) that will allow us to sample the diversity in halo properties, allowing us (to some extent) to marginalize over specific features of individual haloes, as done in \cite{Boyarsky:2009rb};
     \item using several state-of-the-art simulations suites (with the same implementation of SIDM)  to produce theoretical predictions, in order to avoid (potentially unjustified) simplifications  of analytical models.
\end{enumerate}

This paper is structured as follows. In Section~\ref{sec:SD} we introduce the main measurable quantity we are going to use to compare the inner properties of different halos, the \emph{surface density}.
In Section \ref{sec:sims} we describe the simulations, and in Section \ref{sec:data} we outline the observational data used. In Section \ref{sec:compare} we present the maximal surface density as a function of the halo mass, both for simulated haloes and observed ones, and compare between the two. Finally, in Section \ref{sec:conc}, we discuss the implications of our results.

\section{Combining halos ranging six orders of magnitude in halo mass}
\label{sec:SD}

The goal to robustly constrain SIDM, as described above,
raises a number of challenges.
\begin{enumerate}
\item We need to find a way to compare different haloes in a homogeneous way over many orders of magnitude in their masses and sizes. 
\item We have to use a measurable quantity ("an observable") that is derived from the observational data from the same region where this quantity is calculated
rather than rely on extrapolations as is often done with parametric fits. 
\item The observable needs to be sensitive to self-interactions and have a monotonic relation with the cross section.
\end{enumerate}

To this end we choose to adopt the mass (3D) surface density, $S$ defined as 
\begin{equation}
    S(r) = \frac{M(r)}{\frac{4}{3} \pi r^2} \equiv \langle\rho(<r)\rangle r,
    \label{eqn:surfaceDensity}
\end{equation}
(where $M(r)$ and $\langle\rho(<r)\rangle$ are the mass and the average density within some three dimensional radius, $r$)
as our primary observable.

The mass surface density is a good choice as it has been shown to change relatively slowly and obey 
a simple scaling law as function of the virial halo mass ($M_{200}$) over many objects of very different masses~\cite{Boyarsky:2009rb,Boyarsky:2009af,Bondarenko:2017rfu,Sokolenko:2018noz}.  We emphasize, however, that the DM surface density in those works was calculated inside certain characteristic radii -- the radius of a central core or inside the scale radius ($r_s$) of NFW profiles. Below we choose a slightly different approach. 

Notably, if the DM density profile has a core, as is expected  for SIDM~\cite{Rocha:2012jg,Vogelsberger:2012sa}, the surface density as a function of radius, $S(r)$ will have a maximum at a certain radius $r_{S_{\text{DMmax}}}$ and will decrease towards the centre inside of $r_{S_{\text{DMmax}}}$. In CDM, the DM density is predicted to scale as $\rho(r)\propto 1/r$ near the center (this is the case for the NFW profile, for example). For such a profile the surface density increases with decreasing radius and then plateaus to a constant value in the center (in the $\rho(r)\propto 1/r$ regime). Even in CDM, some halos have inner DM density profiles that are shallow than $1/r$ and therefore the surface density will have a maximum. However, this maximum typically corresponds to a much larger value of the surface density and is located much closer to the centre than in the case of SIDM.

This presents us with an opportunity to define a model-independent quantity, the maximum of the surface density $S_{\max}$, that does not require any parametric fits. Finally, as we have seen above (see Fig.~\ref{fig:rMofsigma} and corresponding discussion), in SIDM the average core size for a group of halos with similar masses is expected to change  monotonically as a function of cross section (at least for small enough values of $\sigma/m$\footnote{For large self-interaction cross sections ($\sigma/m\gtrsim 10\text{ cm}^2/\text{g}$) simulations show that haloes undergo gravothermal collapse~\cite{Koda2011}, making the DM density profile cuspier than an NFW, with $\rho_{\text{DM}}\propto r^{-2}$.}). Therefore  $S_{\max}$ will have a similar monotonic behaviour in the same range of cross section values, potentially allowing us to derive bounds on the latter.

Although surface density is defined in terms of observational data in relatively clear and largely model independent way, there are of course observational challenges. For example the method to probe $S(r)$ varies depending on which astrophysical object is observed. As such, how we derive  $S_{\max}$ will depend on the type of object and the observational data involved. We will discuss this in details for each class of objects (dwarf galaxies, spiral and elliptical galaxies and galaxy clusters).

Furthermore, some objects will exhibit no clear maximum and the observed surface density will continue to rise until the smallest radius available. In these cases we can use the maximum among the measured values allowing us to put a lower bound on $S_{\max}$.

\section{SIDM in simulations}\label{sec:sims}

\begin{table}[!t]
    \centering
    \begin{tabular}{|l|c|c|c|}
        \hline
     Simulation (DM types)  & Number of haloes & $M_{200}$, $M_{\odot}$ & $r_{\text{trust}}$, kpc \\
         \hline
         GEAR (All) & 60 & $3.8\cdot 10^{8} - 3.6\cdot 10^{9}$ & 0.2 \\
         APOSTLE Low res. (All) & 142 & $1.0\cdot 10^{10} - 2.4\cdot 10^{12}$ & 1 \\
         APOSTLE High res. (No SIDM1b) & 193 & $3.0\cdot 10^{9} - 1.6\cdot 10^{12}$ & 0.4 \\
         APOSTLE Subhalos (No SIDM1b) & 51 & $3.0\cdot 10^{9} - 6\cdot 10^{10}$ & 0.4 \\
         EAGLE 50 Mpc (All) & 1750 & $1.6\cdot 10^{11} - 1.6\cdot 10^{14}$ & 2 \\
         EAGLE 100 Mpc (CDM, CDMb) & 3000 & $2.2\cdot 10^{11} - 3.8\cdot 10^{14}$ & 2 \\
         BAHAMAS (All) & 3499 & $4.4\cdot 10^{13} - 3.1\cdot 10^{15}$ & 30 \\
         C-EAGLE (CDMb, SIDM1b) & 31 & $1.1\cdot 10^{14} - 1.7\cdot 10^{15}$ & 2 \\
         \hline
    \end{tabular}
    \caption{Properties of simulations, see Appendix~\ref{sec:simulations} for details. Column description: (1) simulation name and DM types in simulations; there are 4 different possible types: CDM, CDM with baryons (CDMb), SIDM1, SIDM1 with baryons (SIDM1b), (2) total number of halo in all types of DM simulations, (3) range of virial masses of halos, (4) trust radius in the simulation (approximately 3 times the Plummer-equivalent gravitational softening length).}
    \label{tab:simulations}
\end{table}

We use ensembles of simulated objects, across a wide range of scales (clusters of galaxies, elliptical and spiral galaxies, dwarf galaxies) for both CDM and SIDM (with a cross section of $1\text{ cm}^2/\text{g}$, SIDM1) in both DM-only simulations and in simulations including baryons (CDMb, SIDM1b), see Table~\ref{tab:simulations} with the main properties of our simulations.
We choose to use a ``trust'' radius of approximately 3 times the Plummer-equivalent gravitational softening length, meaning that we do not use surface density values at radii less than this. 
A detailed description of the simulations and halo selection is given in Appendix~\ref{sec:simulations}.
We study the dependence of the surface density on radius for CDM and SIDM as well as the effects of baryons on this dependence. Also, we analyze the difference in the behaviour of the DM and the total mass surface densities.
This data will be used for the final comparison between the simulations and observations that we will perform in Section~\ref{sec:compare}.

\subsection{Surface density in DM-only simulations}

\begin{figure}[h!]
  \centering
  \includegraphics[width=0.49\textwidth]{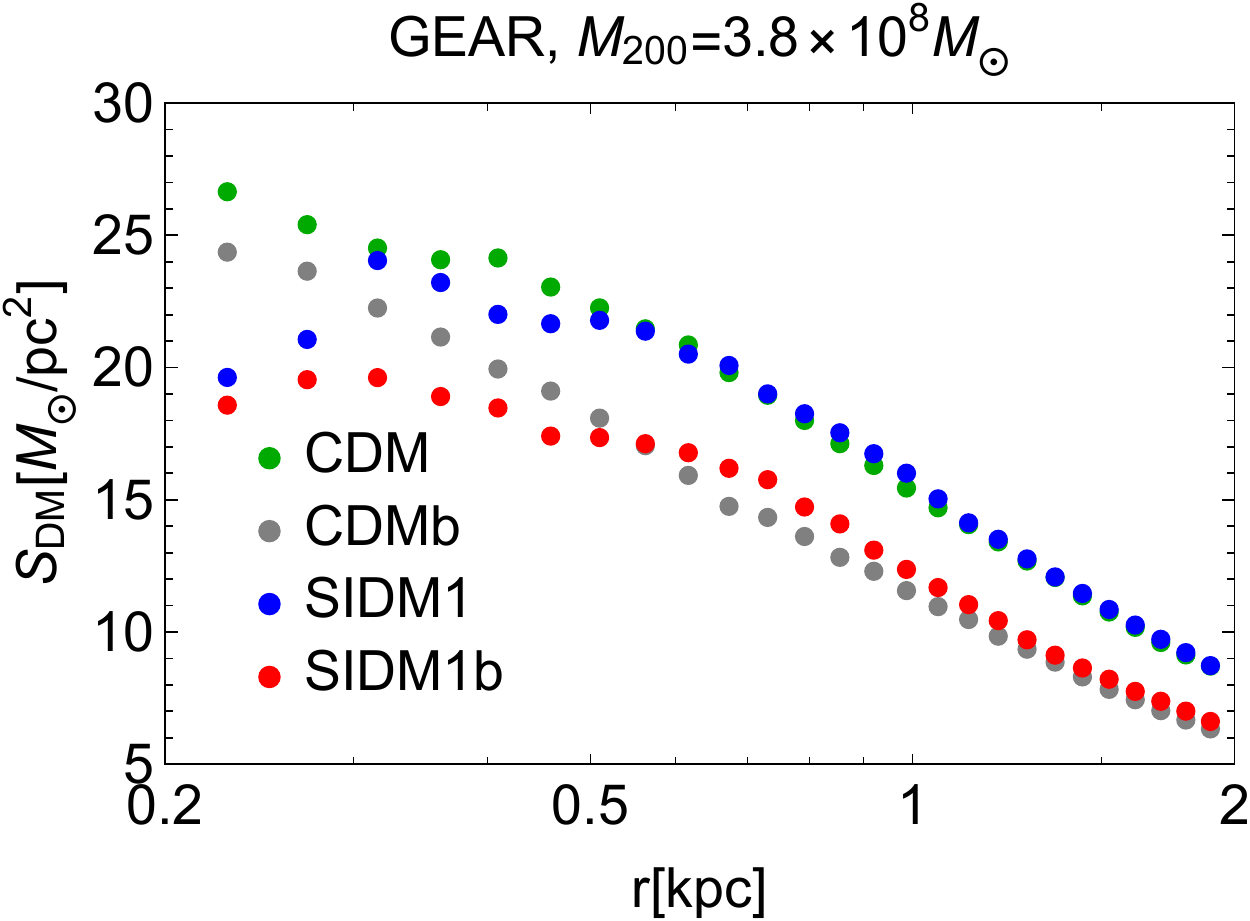}
  ~\includegraphics[width=0.49\textwidth]{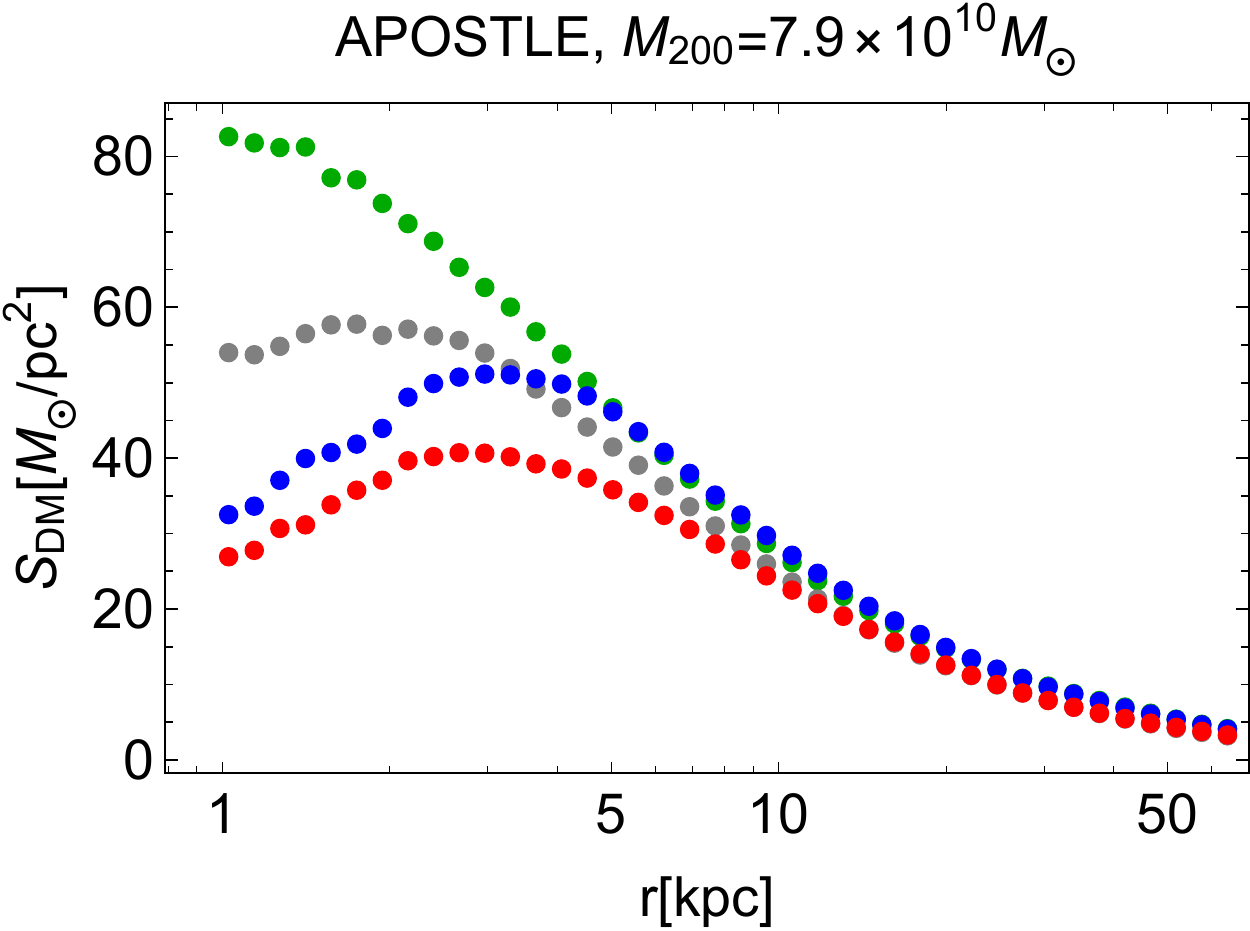}\\
  \includegraphics[width=0.49\textwidth]{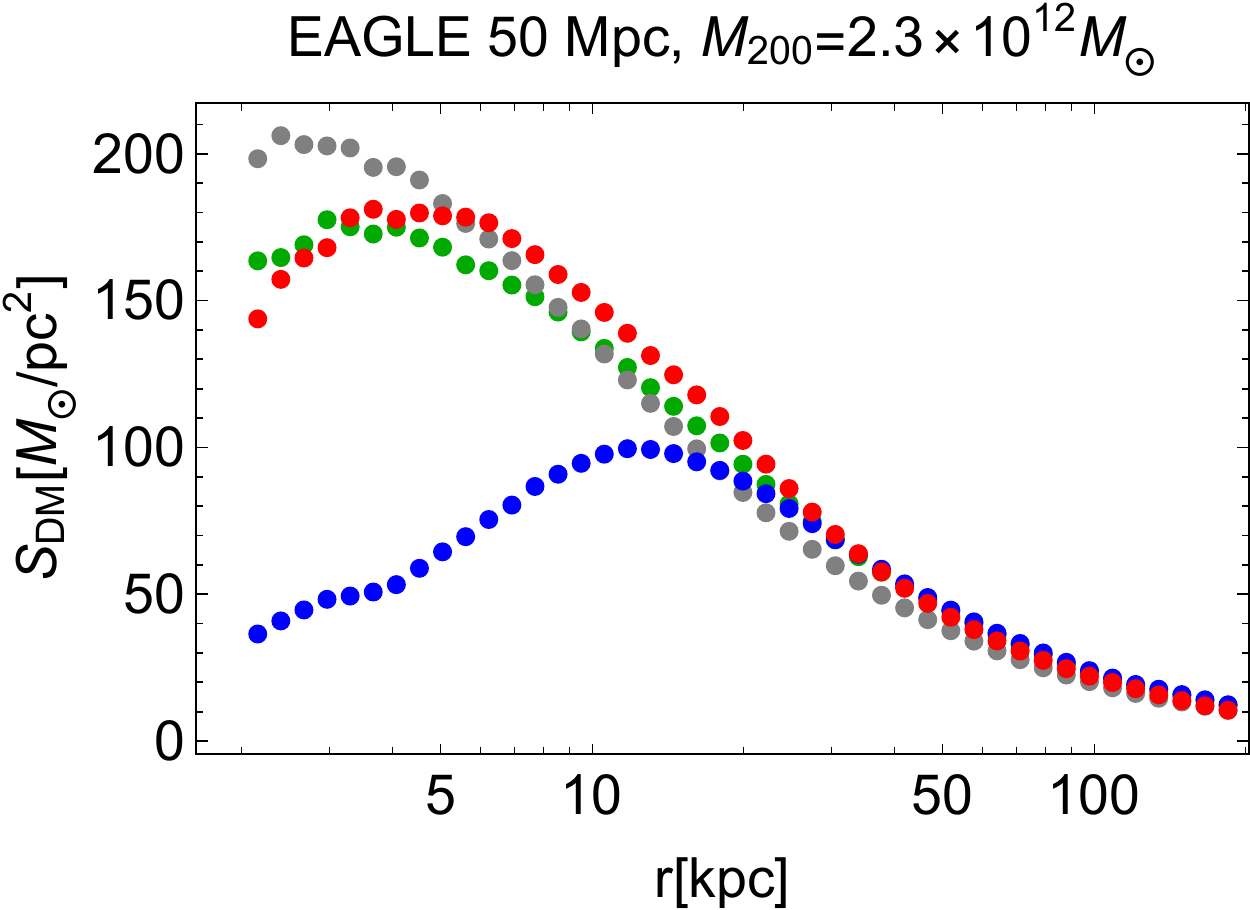}
  ~\includegraphics[width=0.49\textwidth]{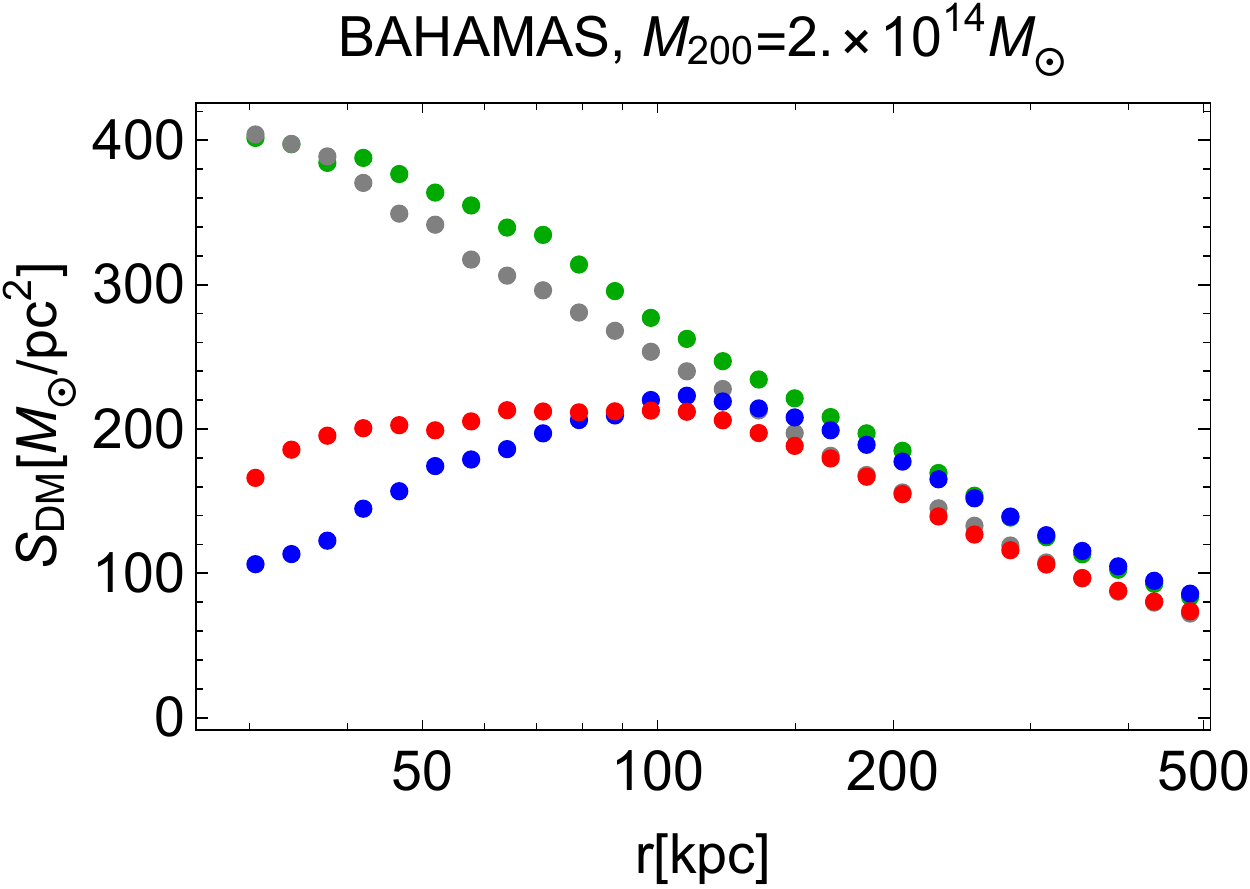}
  \caption{Dark matter surface density profiles, $S(r) = \langle\rho(<r)\rangle r$,  for halos from different simulations with various cosmology. We plot CDM profiles in green, CDMb in gray, SIDM1 profiles in blue, and SIDM1b in red.}
  \label{fig:SD_DM}
\end{figure}

To clearly illustrate the differences between the two DM models we study, we start by comparing DM-only (DMO) simulations. 

Examples of the behaviour of the DM surface density as a function of radius for different DMO simulated haloes are shown in Fig.~\ref{fig:SD_DM}. The difference between the CDM and SIDM1 models becomes apparent at small radii: as expected, for the SIDM1 the surface density first reaches a maximum and then 
goes to zero near the center, while for CDM it grows. 
For most of the CDM halos, there is no maximum of the surface density outside of the trust radius. 
Therefore,  in such cases we use for comparison a lower bound on the maximum of $S(r)$ -- its value at the trust radius
- the smallest radius where we trust the enclosed mass profiles from the simulations (see Table~\ref{tab:simulations}).
Of course, to really constrain SIDM we will need to compare the maximum of $S(r)$ in the radial range that can be robustly measured from observations. We will discuss this question in Section~\ref{sec:compare}.

\begin{figure}[h!]
  \centering
  \includegraphics[width=0.48\textwidth]{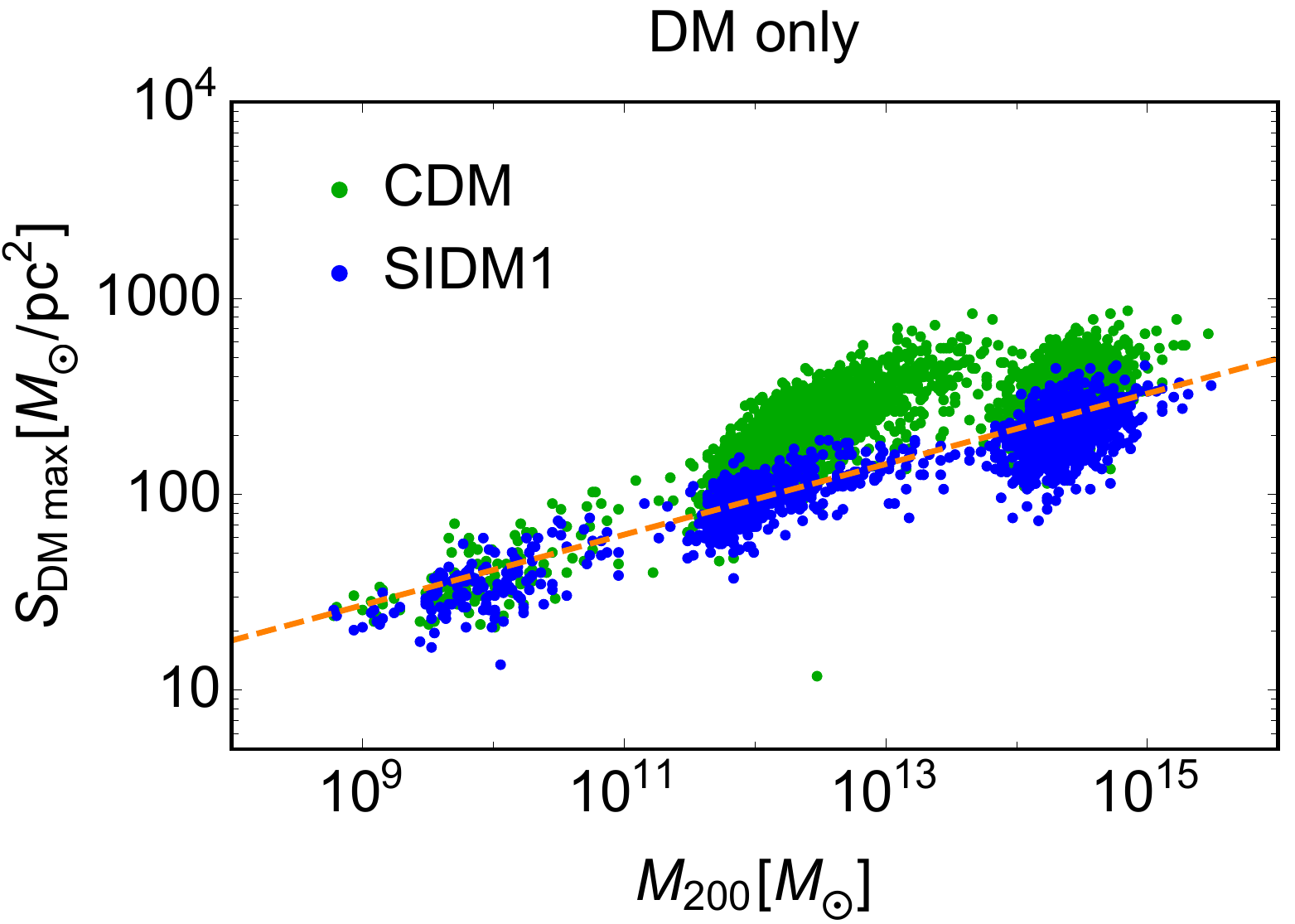}~\includegraphics[width=0.48\textwidth]{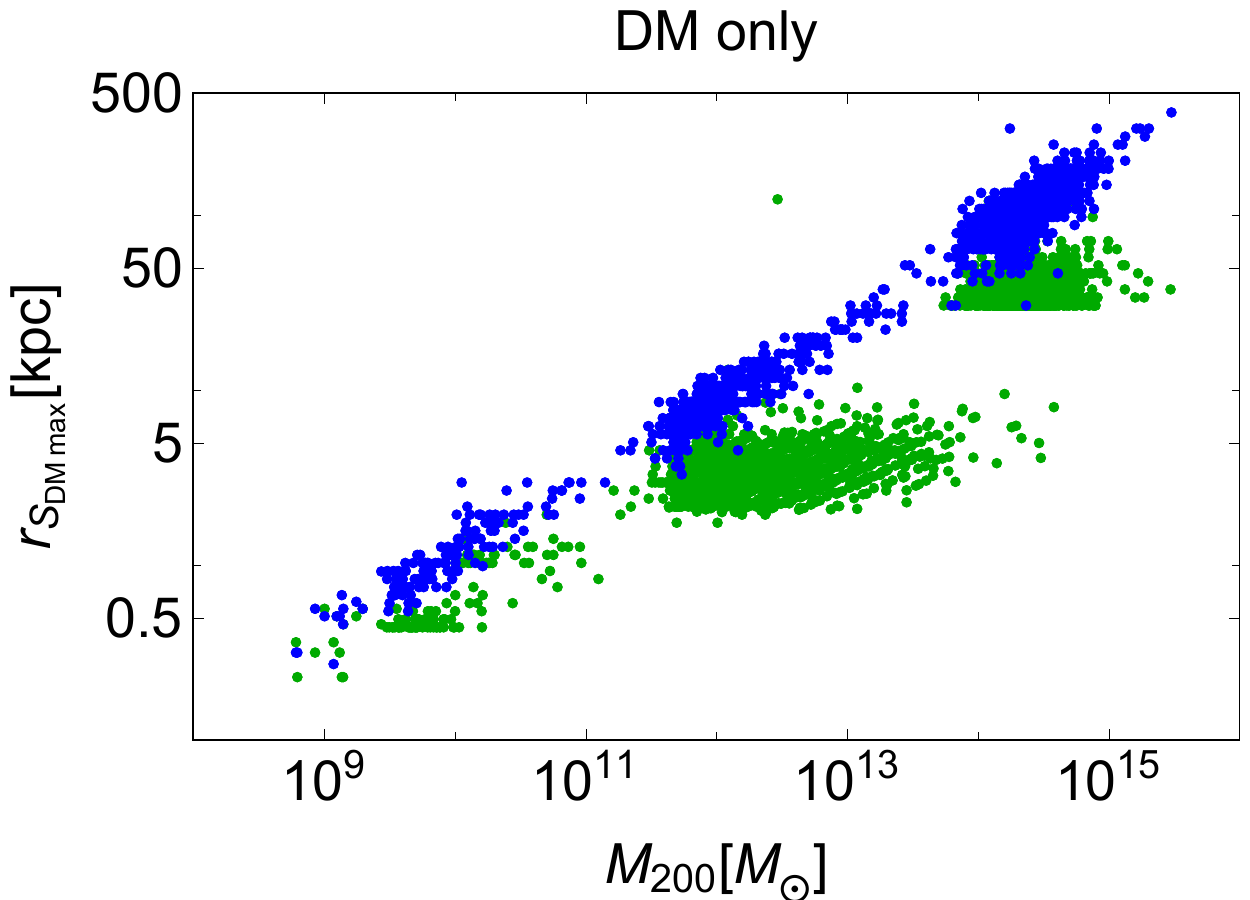}
  \caption{\textit{Left panel}: The ensemble maximum surface density as derived from the density profiles as a function of halo mass, $M_{\rm 200}$, for all DM-only simulations. We show CDM in green and SIDM in blue. Where CDM does not exhibit a maximal surface density we show a lower limit, coming from the surface density at the trust radius. \textcolor{black}{The orange dashed line shows a scaling relation for SIDM halos obtained in~\cite{Lin:2015fza}.} \textit{Right panel:} The same for the radius of the maximum surface density. }
  \label{fig:MaxSD_DMonly}
\end{figure}

The maximal surface density for CDM and SIDM1 simulations, as a function of the virial mass of the objects, is shown in Fig.~\ref{fig:MaxSD_DMonly}. We see that the maximal surface density is systematically higher in the CDM case and appears at lower radii, in agreement with Fig.~\ref{fig:SD_DM}. We see that the difference in the maximal surface density between the two models is not larger than the scatter between different halos
for objects with  $M_{200}< 10^{11}M_{\odot}$. The difference between the models is more visible for more massive objects. Among simulated objects that we use here, this difference is the most profound for the  objects with masses around $10^{13}M_{\odot}$, although the reason this difference does not continue to grow with halo mass above this is because for more massive halos we are limited by the 30 kpc trust radius of BAHAMAS.

\subsection{Effects of baryons on DM density profiles}

\begin{figure}[h!]
  \centering
  \includegraphics[width=0.48\textwidth]{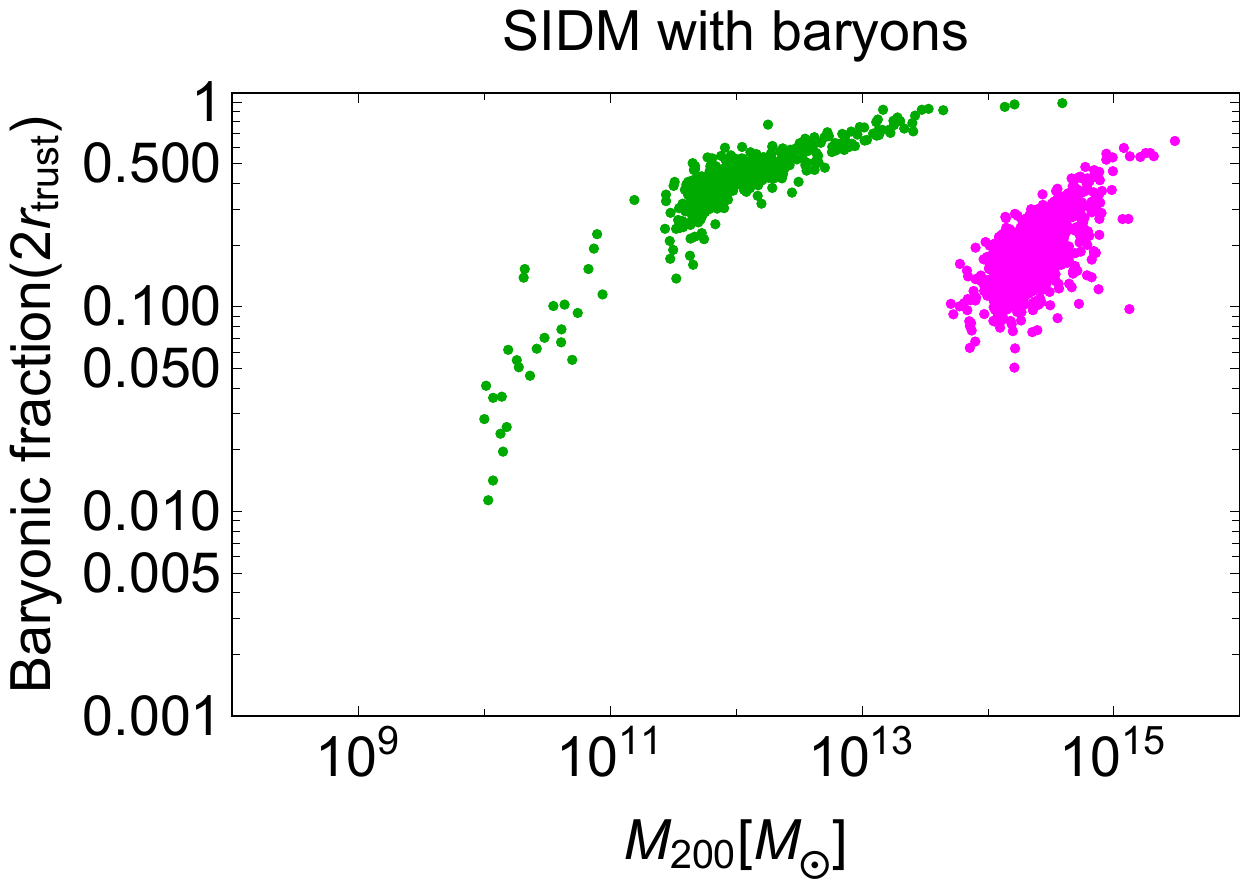}~\includegraphics[width=0.48\textwidth]{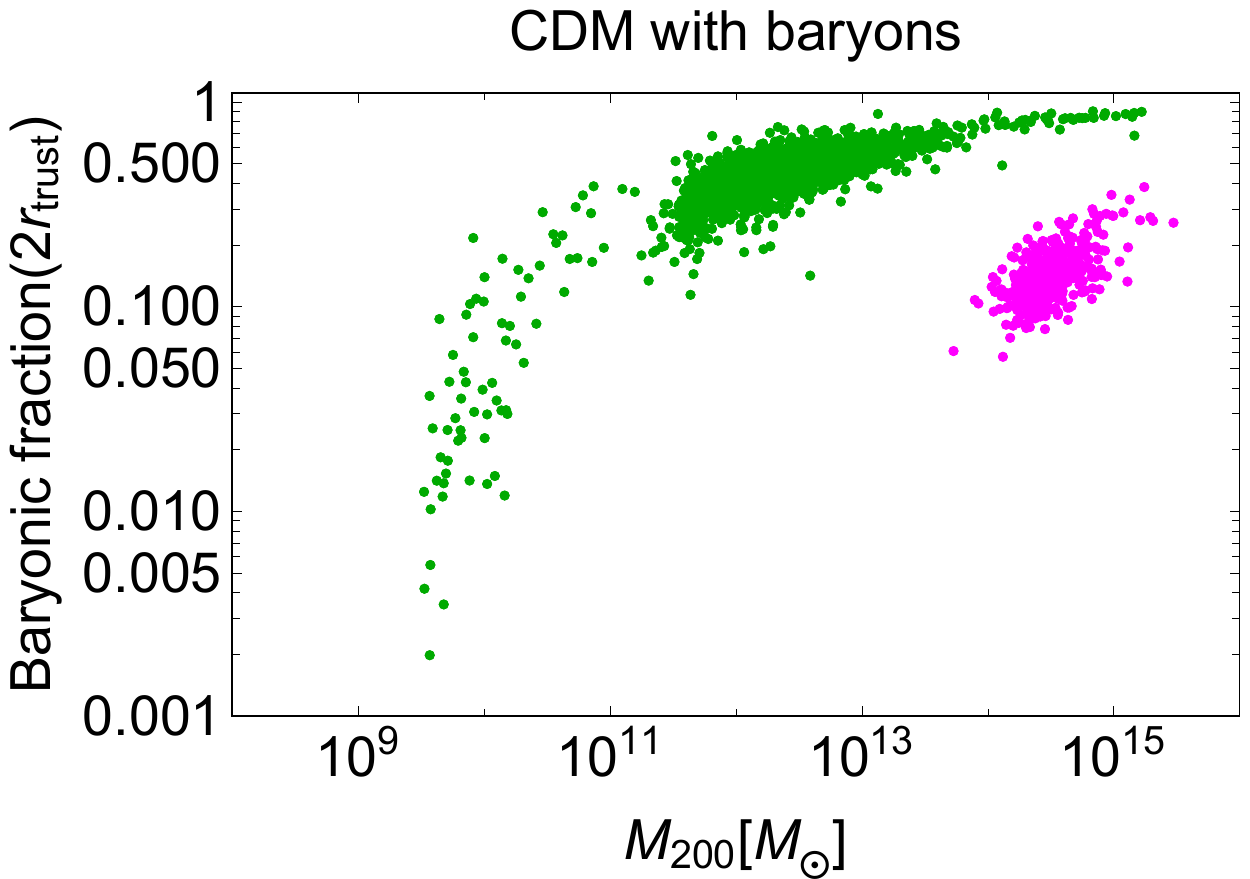}
  \caption{The fraction of mass within two trust radii that is baryonic, versus virial mass. For galaxy clusters we present both high-resolution simulations with trust radii of  2kpc (green) and the
  BAHAMAS simulations with trust radii of 30 kpc (magenta), see Table~\ref{tab:simulations}.}
  \label{fig:baryon_fraction} 
\end{figure}

Until now we have discussed  DM-only simulations.
Of course, the real Universe contains baryons, so to directly compare the predictions of simulations with the real observational data we use more realistic simulations that include baryons.
In this Section we want to study how the inclusion of baryons changes the predictions for the two DM models and the differences between them.
In Fig.~\ref{fig:baryon_fraction} we show  how the fraction of baryons in the inner part of the halo (inside 2 trust radii) changes with $M_{200}$ in the simulations that we use. We see that the fraction of baryons is small for dwarf galaxies, while for larger galaxies and clusters, baryons dominate the total mass in the central regions.
This means that: (i) the predictions of 
our simulations for both models depend strongly on the realistic modeling of baryonic effects; (ii) observationaly, it 
may be difficult to separate 
the DM mass from the total mass; (iii) 
the difference between DM models may be masked by baryonic effects in the inner parts of the halos.
Therefore, we may have to search for the maximum of the surface density in a range of larger radii, where baryons play a less important role.\footnote{See e.g.  pink points in Fig.~\ref{fig:baryon_fraction}, calculated for BAHAMAS simulations of the galaxy clusters, where the trust radius 
as large as 30 kpc, as compared to 2 kpc
for C-EAGLE simulations, represented by the green points with the largest $M_{200}$ in Fig.~\ref{fig:baryon_fraction}.}
Even if the difference between CDM and SIDM are less pronounced when averaged inside of larger radii, this may be more efficient for distinguishing between the two models, provided all theoretical and observational uncertainties are understood.

\begin{figure}[h!]
  \centering
  \includegraphics[width=0.48\textwidth]{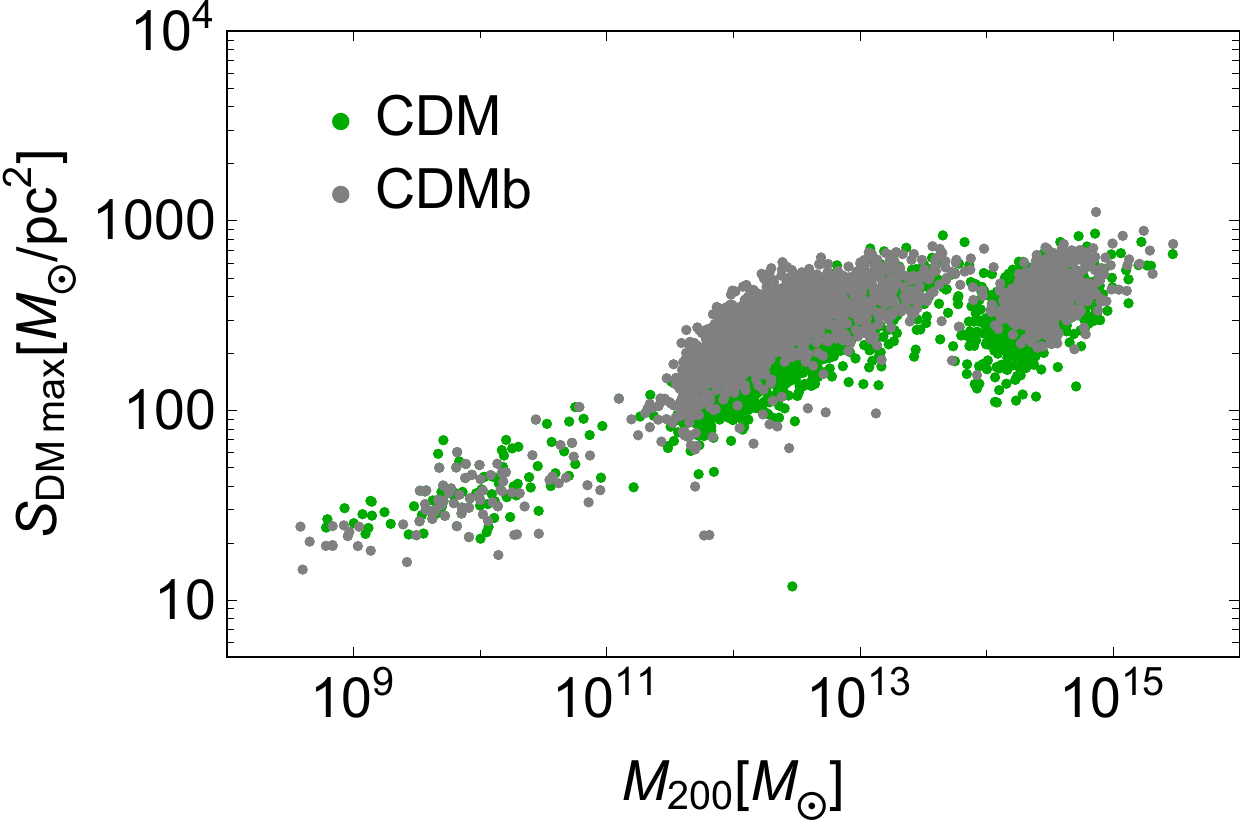}~\includegraphics[width=0.48\textwidth]{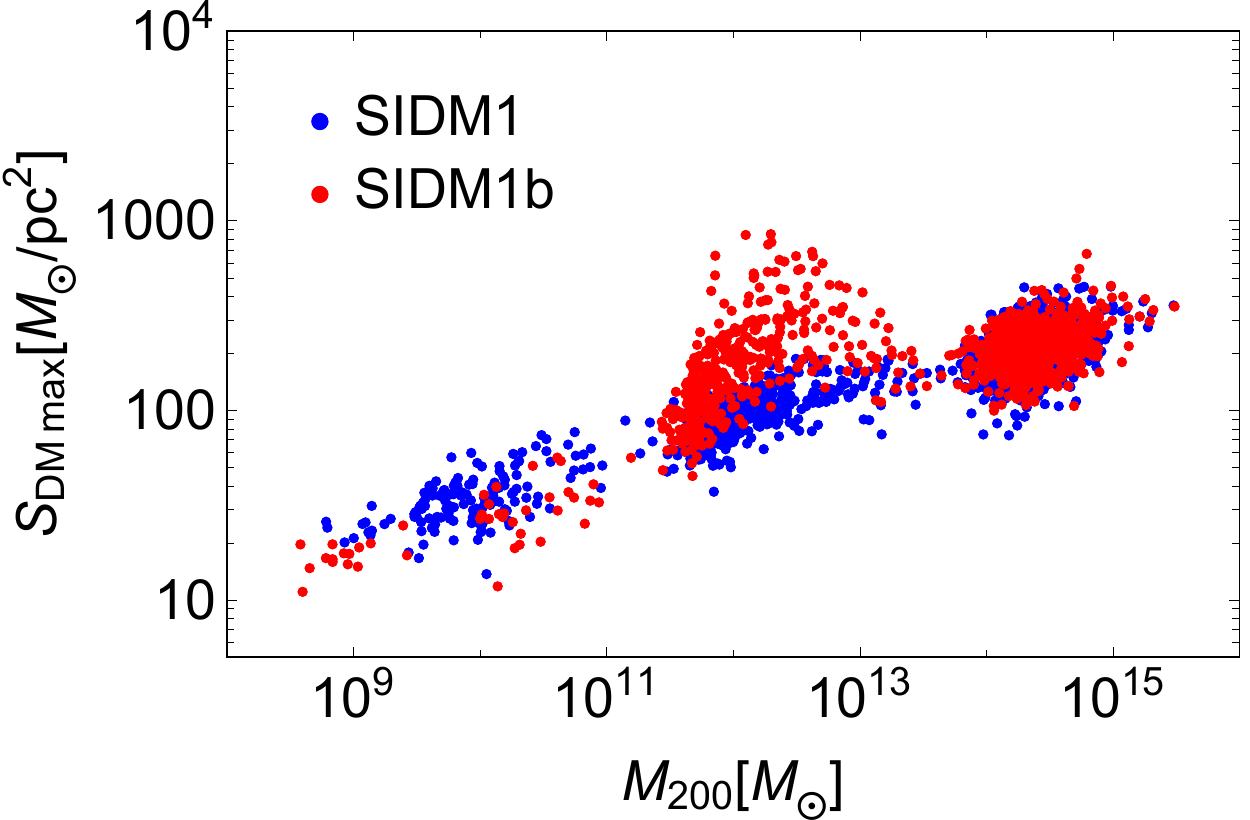}
  \caption{The impact of baryonic feedback on the maximal value of dark matter surface density obtained in the simulations:
  CDM on the left panel (CDM dark matter only in grey, the version with baryons (CDMb)
  in green); SIDM on the right panel (SIDM1 dark matter only in blue, the version with baryons (SIDM1b) in red. For objects where  $S_{\text{DM}}(r)$ does not have maximum outside trust radius we show the lower bound on the maximum -- the value at the trust radius.}
  \label{fig:SD_max_c} 
\end{figure}

Baryons can change the behaviour of the maximal dark matter surface density quite significantly. In Fig.~\ref{fig:SD_DM} we show the examples of the surface density $S(r)$ for the same halos, simulated both for SIDM and CDM with and without baryons. Baryonic effects can significantly contract the DM distribution making it steeper. This effect is stronger for SIDM, such that baryons make the differences between the two models smaller. In some cases baryonic effects could create small cores in CDM halos (see e.g.~\cite{Read:2018fxs} and references therein). 

\begin{figure}[h!]
  \centering
  \includegraphics[width=0.49\textwidth]{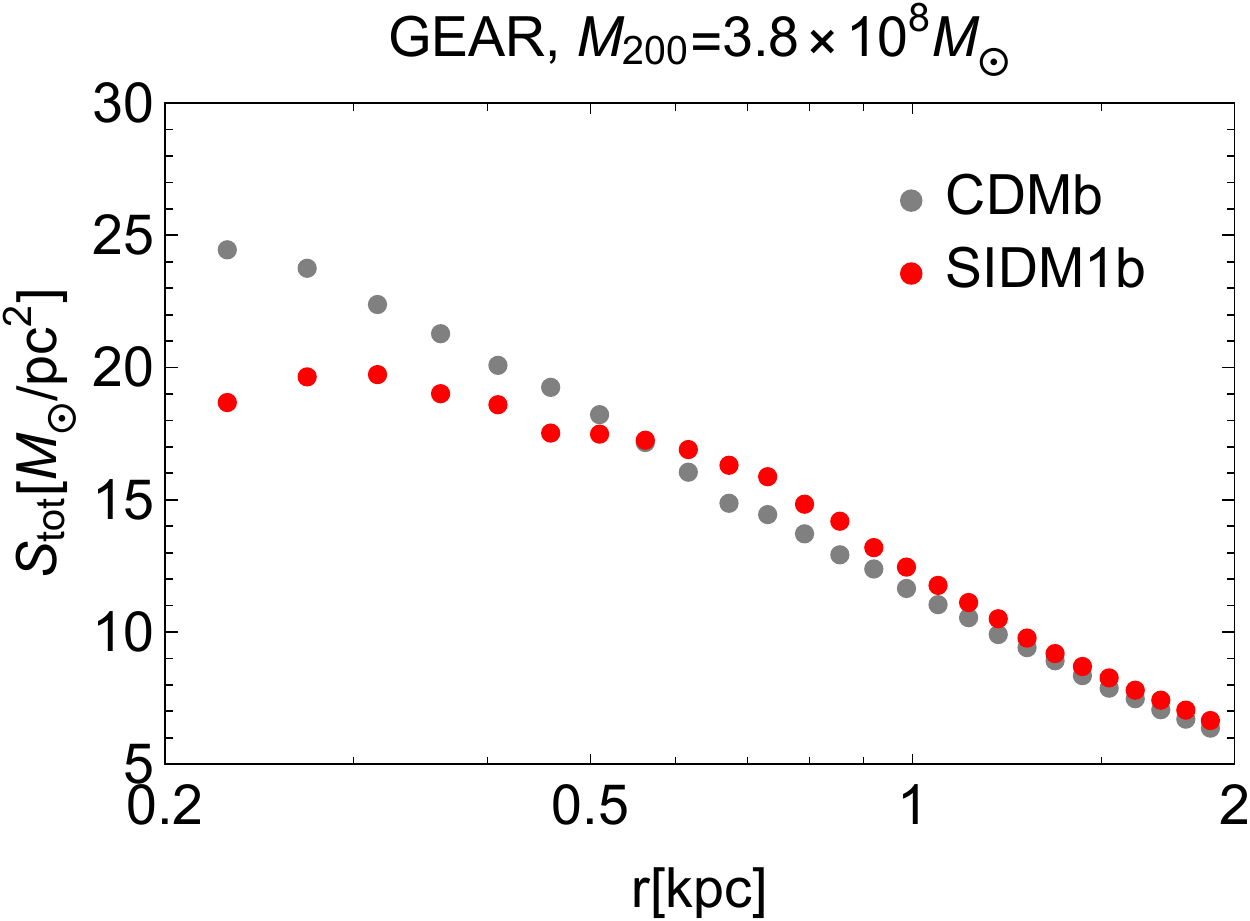}
  ~\includegraphics[width=0.49\textwidth]{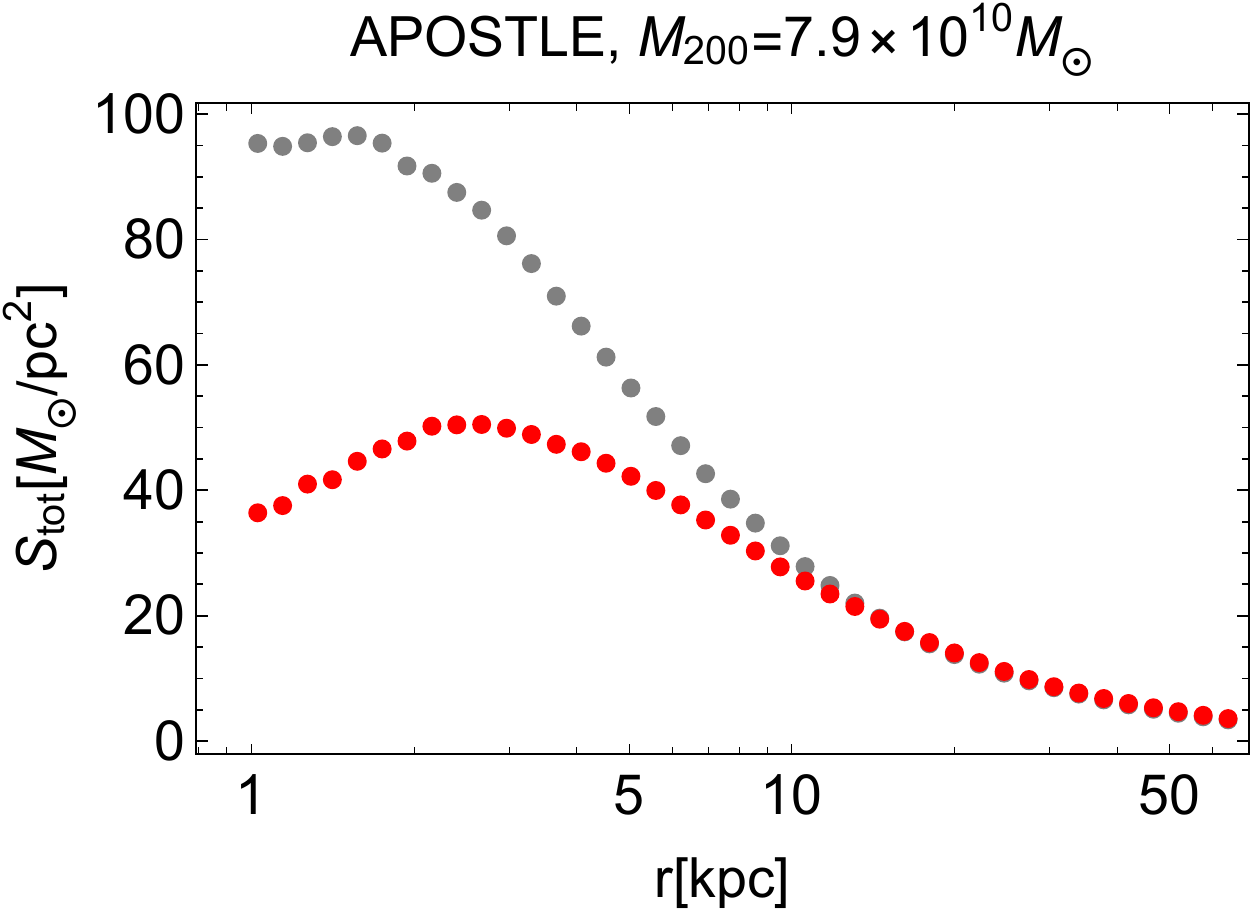}\\
  \includegraphics[width=0.49\textwidth]{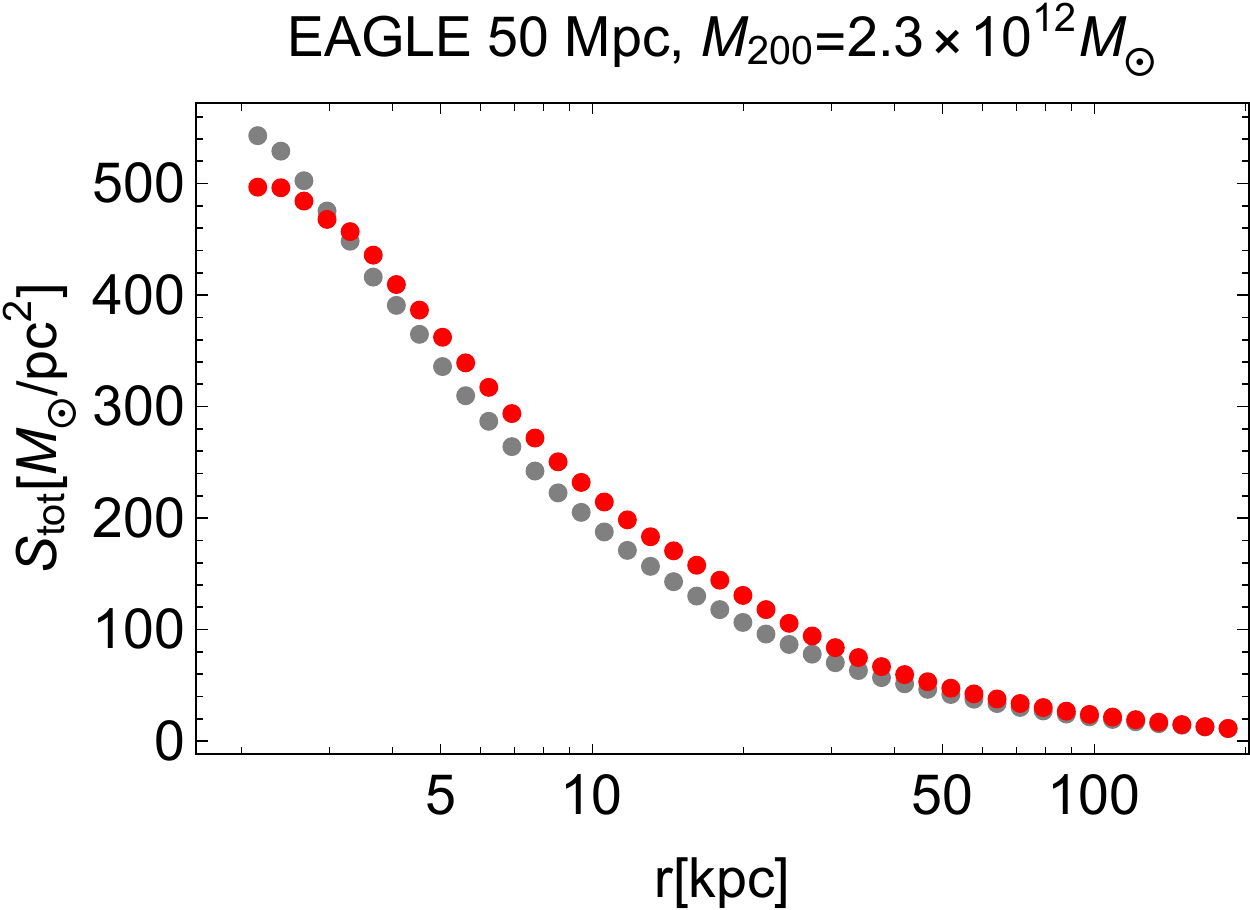}
  ~\includegraphics[width=0.49\textwidth]{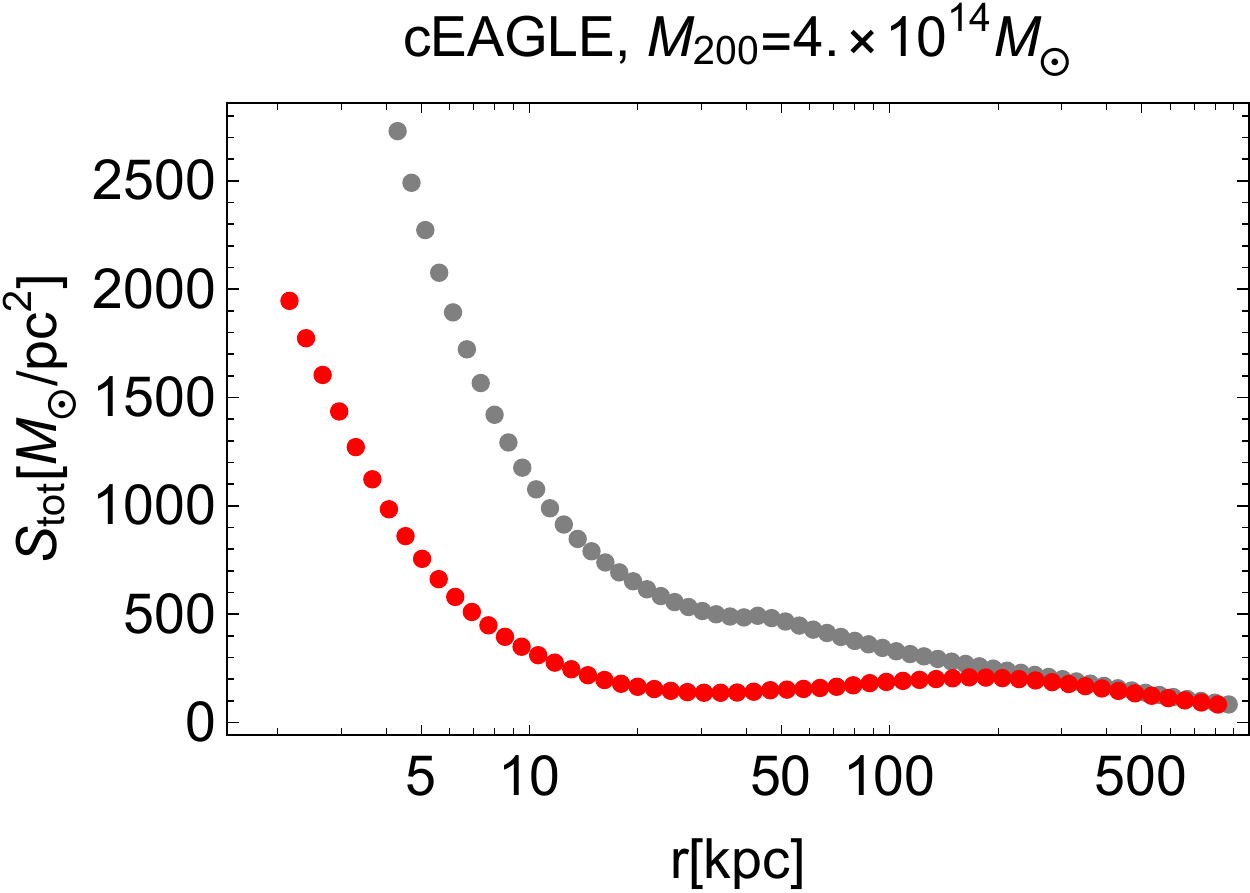}
  \caption{Total mass surface density profiles for example halos over a wide range of halo masses. We plot CDMb in gray and SIDM1b in red.}
  \label{fig:SD_tot}
\end{figure}

The comparison of the maximal surface density values for the whole ensemble of halos simulated for CDM and SIDM with and without baryons is shown in Fig.~\ref{fig:SD_max_c}. 
We see that baryonic effects on the DM surface density are the strongest for the mass range around
$10^{12}-10^{13}M_{\odot}$ and are stronger in SIDM than in CDM (see e.g.~\cite{Kaplinghat:2019dhn} for  previous discussion).
Dwarf galaxies (DM haloes with mass $10^{11}M_{\odot}$ and lower ) are DM-dominated even in the central parts (see Fig.~\ref{fig:baryon_fraction}), so the influence of baryons is smaller than for more massive galaxies.
The same is true for galaxy clusters (halos with mass larger than $10^{14}M_{\odot}$), but for a different reason. As we can see in Fig.~\ref{fig:SD_tot} (right panel), the maximum of the DM surface density for galaxy clusters with SIDM1 occurs at large distances (outside 30 kpc), while the objects are baryon dominated only at smaller radii.

\textcolor{black}{The fact that the maximal surface density changes quite dramatically when including baryons suggests that a robust comparison with the observations requires the simulated halos to have realistic baryon distributions. This requirement does not only apply to the total (DM + baryon) surface density, but extends to the case of considering the DM surface density as well, because adiabatic contraction \cite{2004ApJ...616...16G} can cause the DM density to increase as gas cools towards the centre of the halo. In the inner regions of galaxy or cluster scale halos (where SIDM can alter the density profile), the dominant baryonic component is the stars. This means that it is important that the simulated halos contain realistic stellar distributions.}

\textcolor{black}{The EAGLE simulations were calibrated to have a stellar mass function and galaxy stellar mass--size relation in agreement with observations \cite{Schaye:2014tpa}, suggesting that the resulting effects on the DM halo should be similar to those in observed systems. BAHAMAS also has a realistic stellar mass function \cite{McCarthy:2016mry}, and we have verified that the radii enclosing half of the stellar mass (in projection) of BAHAMAS brightest cluster galaxies are in rough agreement with the effective radii measured from observations of brightest cluster galaxies \cite{2015MNRAS.453.4444Z}.}

\begin{figure}[h!]
  \centering
  \includegraphics[width=0.49\textwidth]{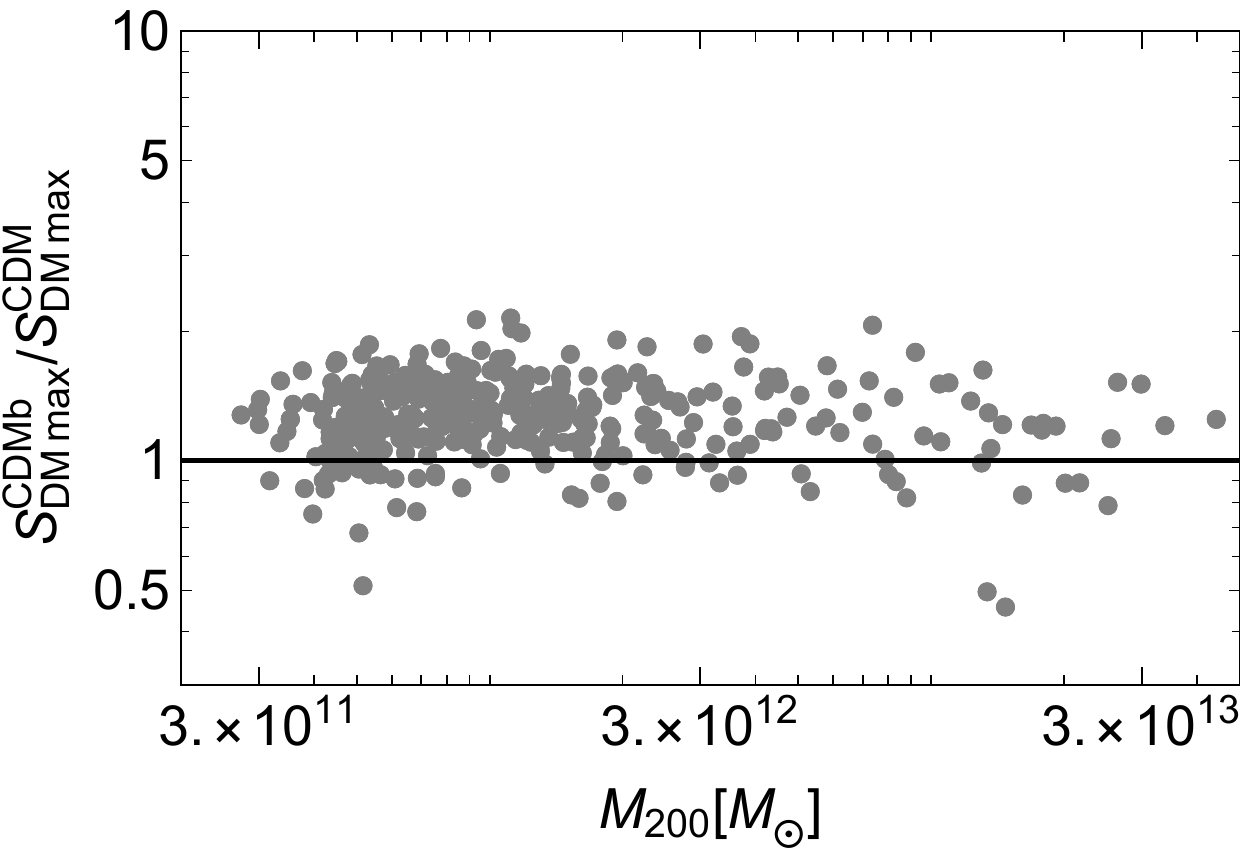}~
  \includegraphics[width=0.49\textwidth]{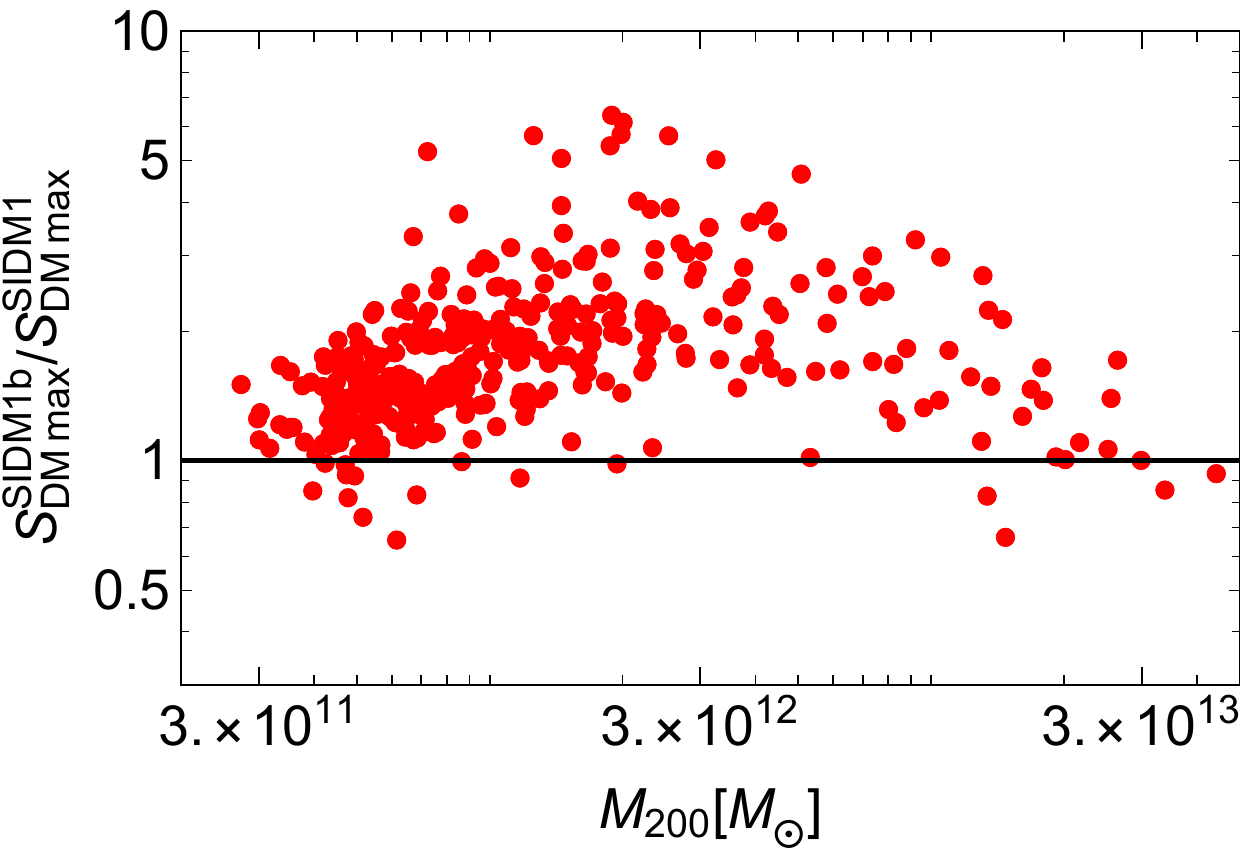}
  \caption{The ratio between maximal surface density of galaxies in simulations with baryons and pure DM simulations versus halo mass. Left panel for CDM, right panel from SIDM1. Black line shows ratio equal to one.}
  \label{fig:contraction}
\end{figure}

\begin{figure}[h!]
  \centering
  \includegraphics[width=0.48\textwidth]{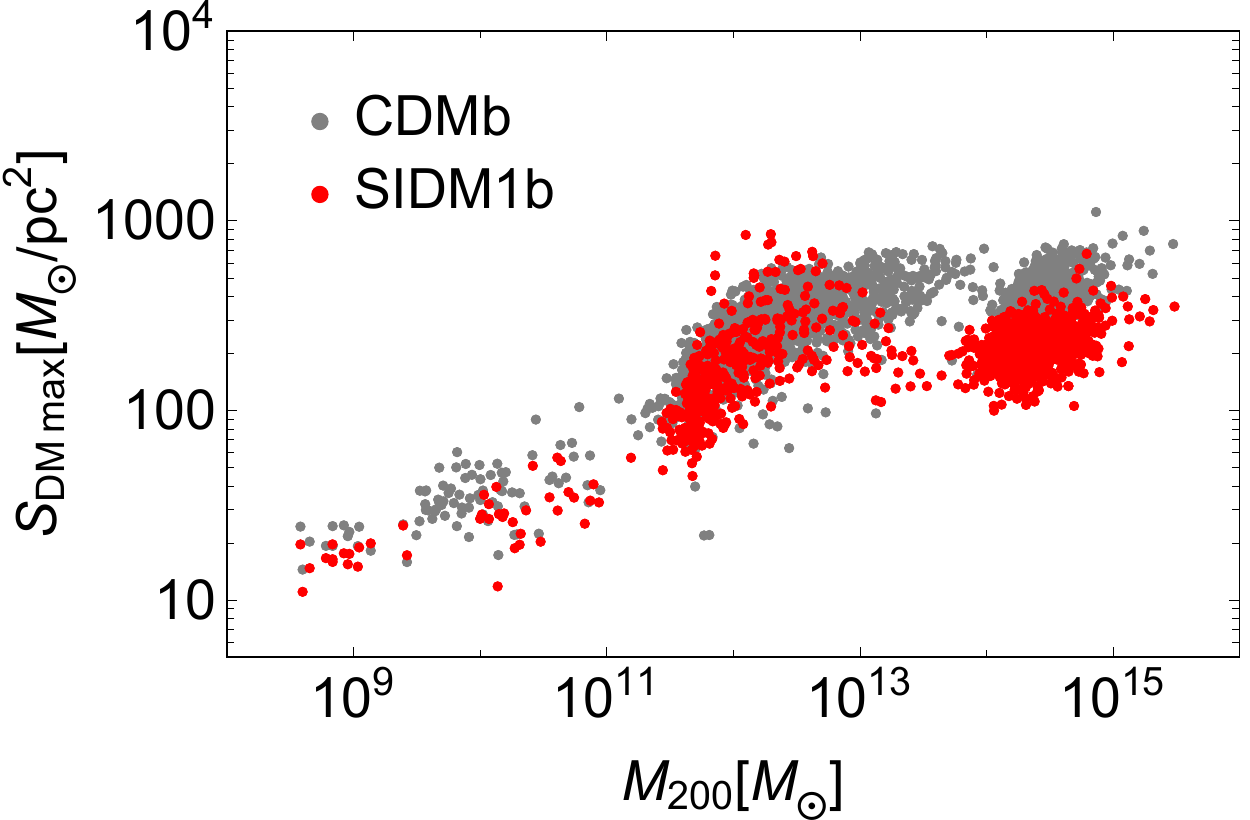}~\includegraphics[width=0.48\textwidth]{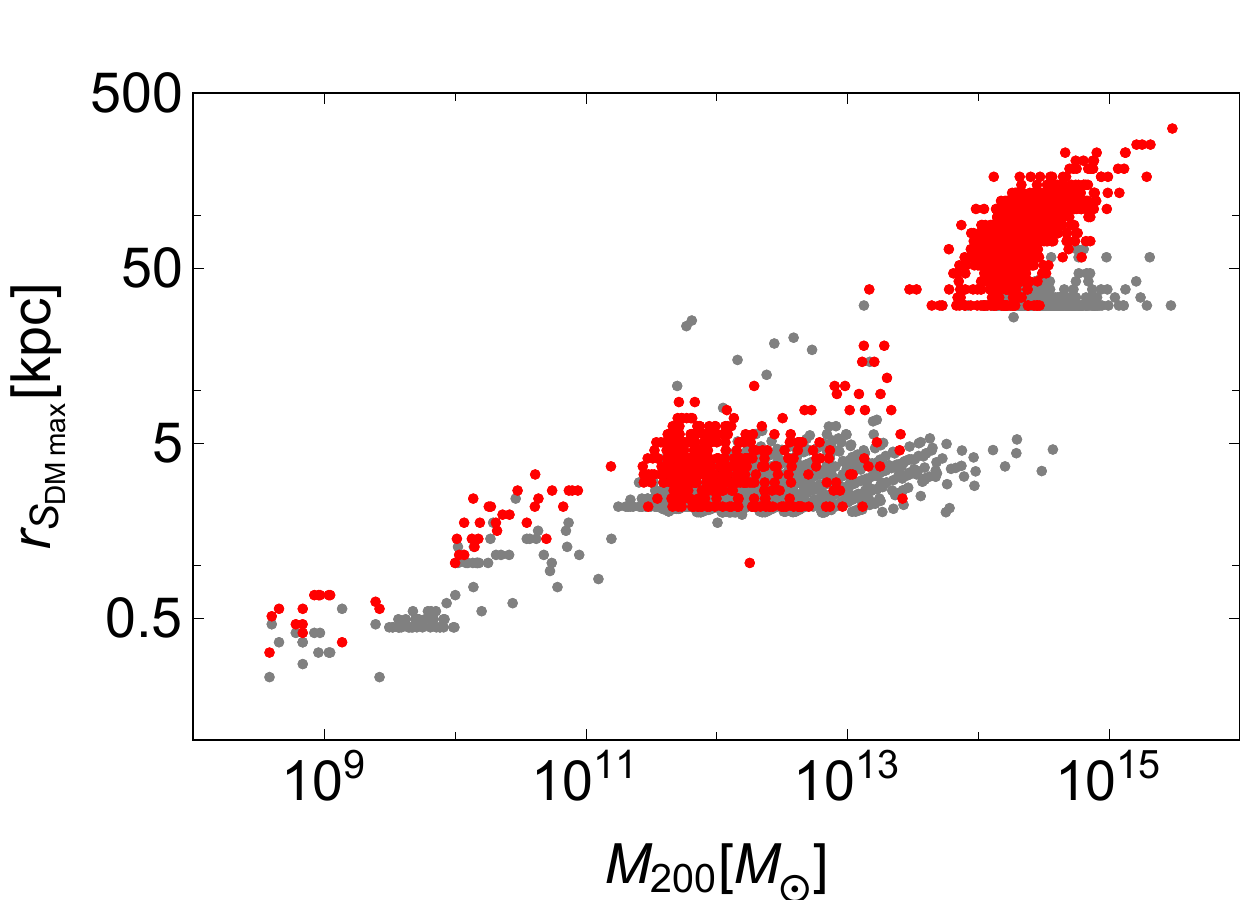}
  \caption{\textit{Left panel:} Maximum surface density for DM mass versus the halo mass $M_{200}$ for simulations with baryons: gray points for CDMb, red points for SIDM1b. \textit{Right panel:} The radius where the maximum of DM surface density is achieved in our simulation, the same color scheme.}
  \label{fig:SDmax_baryons}
\end{figure}

\begin{figure}[h!]
  \centering
  \includegraphics[width=0.48\textwidth]{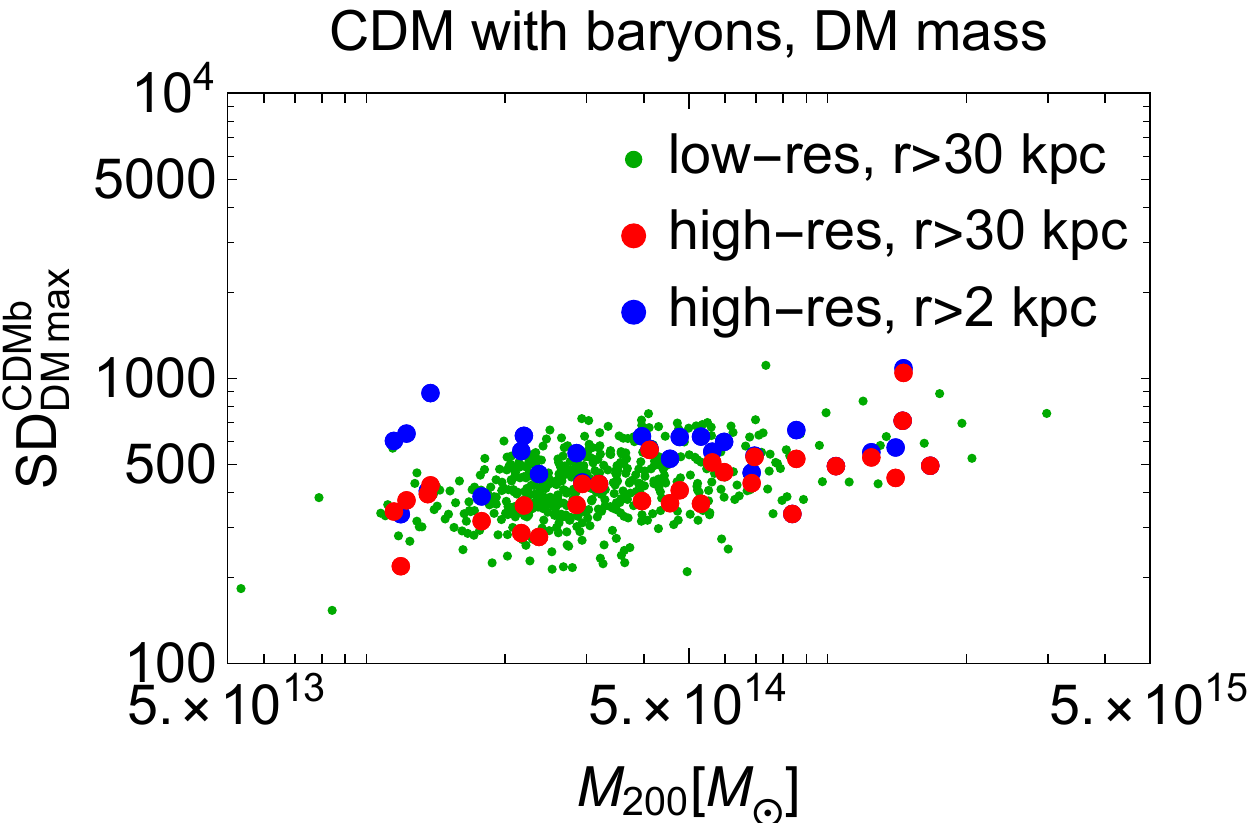}~\includegraphics[width=0.48\textwidth]{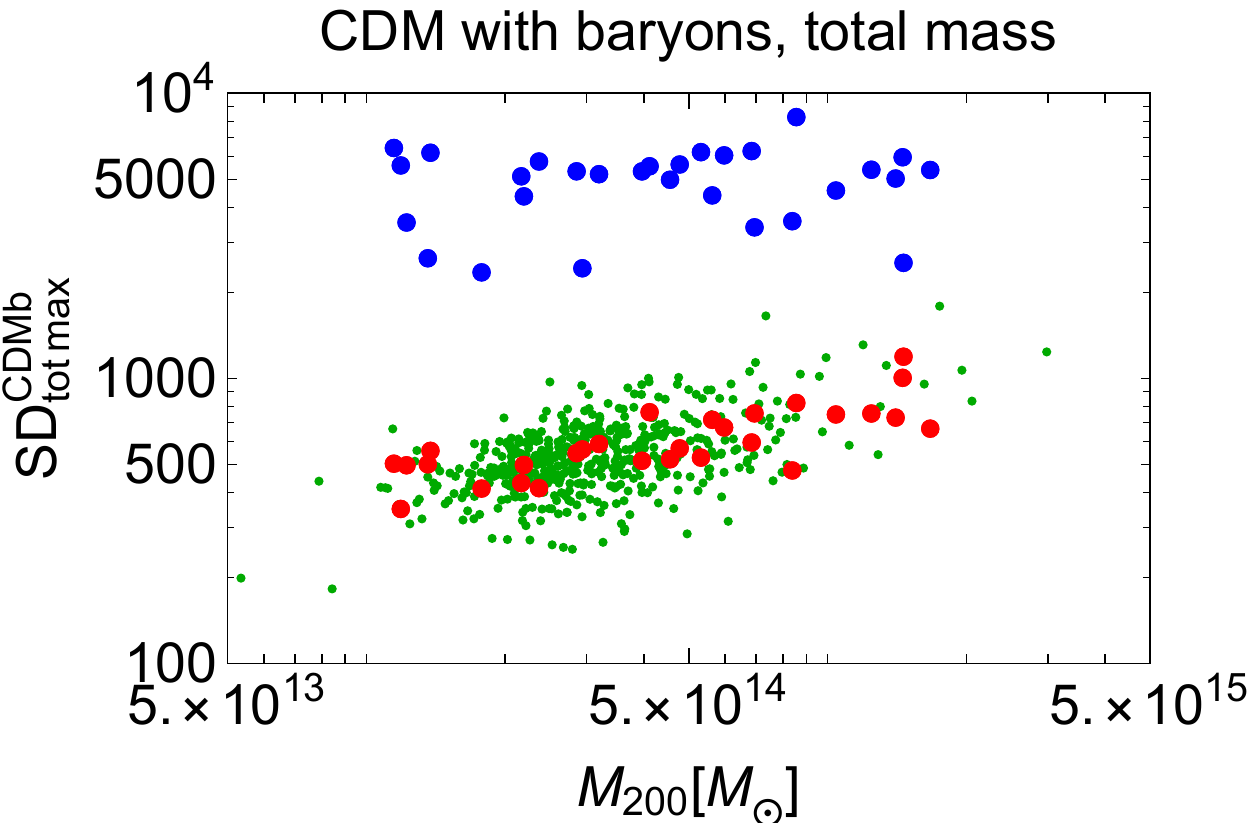}
    \caption{The maximum surface density for DM mass (left) and total mass (right) for CDMb simulations of clusters. Here we compare low resolution BAHAMAS simulations with $r>30$~kpc (green points), with high resolution C-EAGLE simulations for radii $r>30$~kpc (red points) and $r>2$~kpc (blue points).}
    \label{fig:SD_max_30kpc}
\end{figure}

\begin{figure}[h!]
  \centering
  \includegraphics[width=0.495\textwidth]{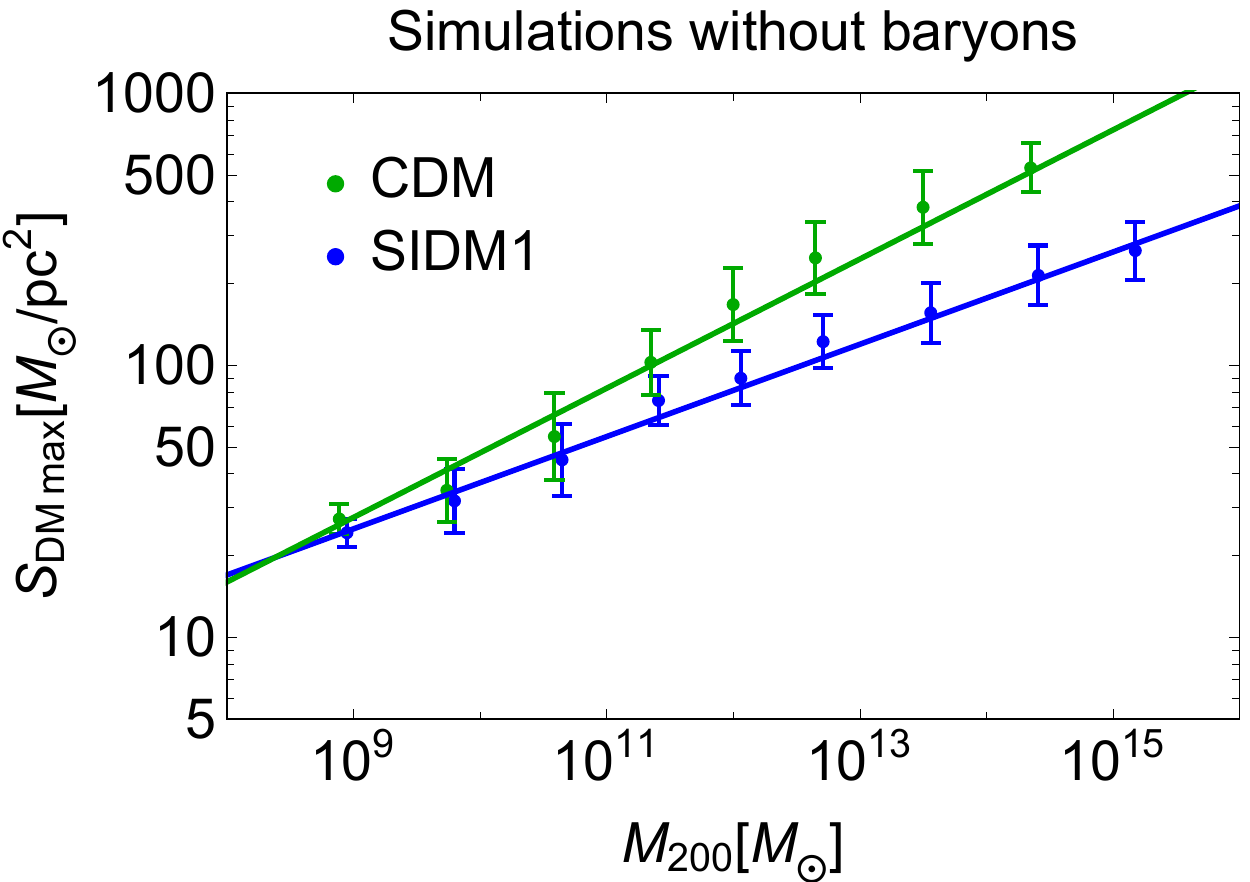}
  \includegraphics[width=0.495\textwidth]{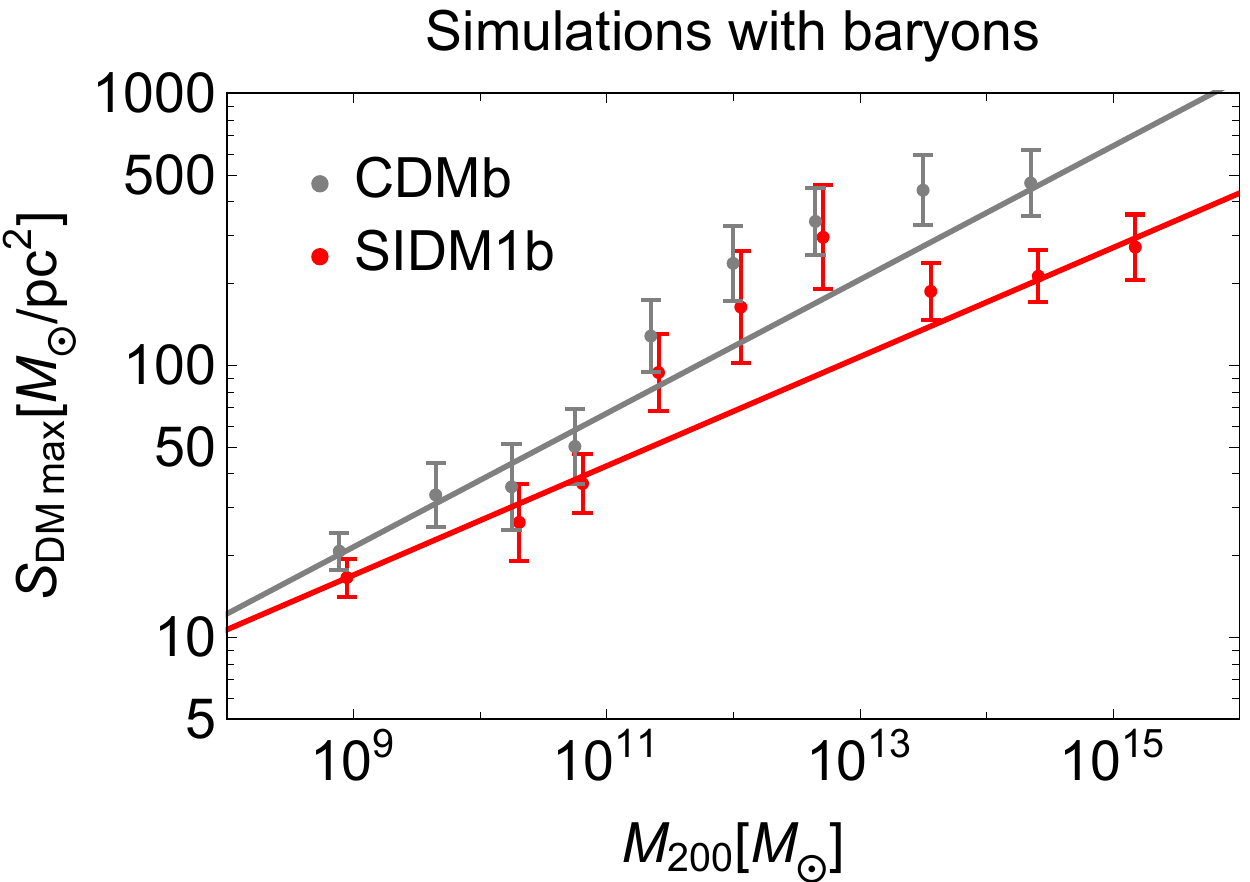}
  \caption{Mass binned representation of the maximal values of DM surface density in simulations without baryons (left) and with baryons (right) versus virial mass: green points for CDM, blue points for SIDM1, gray points for CDMb, red points for SIDM1b. The lines represent the best-fit power laws excluding objects from $10^{11}M_{\odot}$ to $10^{14}M_{\odot}$ (due to strong baryonic effects).}
  \label{fig:SDmax_baryons_bins}
\end{figure}

In Fig.~\ref{fig:contraction} we plot the ratio of the maximal surface density with and without baryons for our two different DM models.
We see that for CDM, baryonic effects result in slight increase in the average value
of the maximal surface density. For SIDM, baryonic contraction results in large (above a factor of 5 for some halos) increases to the maximum surface density, \textcolor{black}{ in agreement with analytic estimates done in~\cite{Kaplinghat:2013xca}}. 
This masks some of the difference (in DMO simulations) between the maximal DM surface densities in the two models, especially for galaxies with masses 
$10^{12}-10^{13}M_{\odot}$, see Fig.~\ref{fig:SDmax_baryons} and the right panel of Fig.~\ref{fig:SDmax_baryons_bins}. For the DM surface densities of clusters we use the BAHAMAS simulations with a trust radius of $30$~kpc. In the left panel of Fig.~\ref{fig:SD_max_30kpc} we compare the BAHAMAS simulations with high-resolution simulations of clusters from C-EAGLE and we see that there is not much difference between the maximum surface density outside of $30$~kpc and outside of $2$~kpc.

\begin{figure}[h!]
  \centering
  \includegraphics[width=0.48\textwidth]{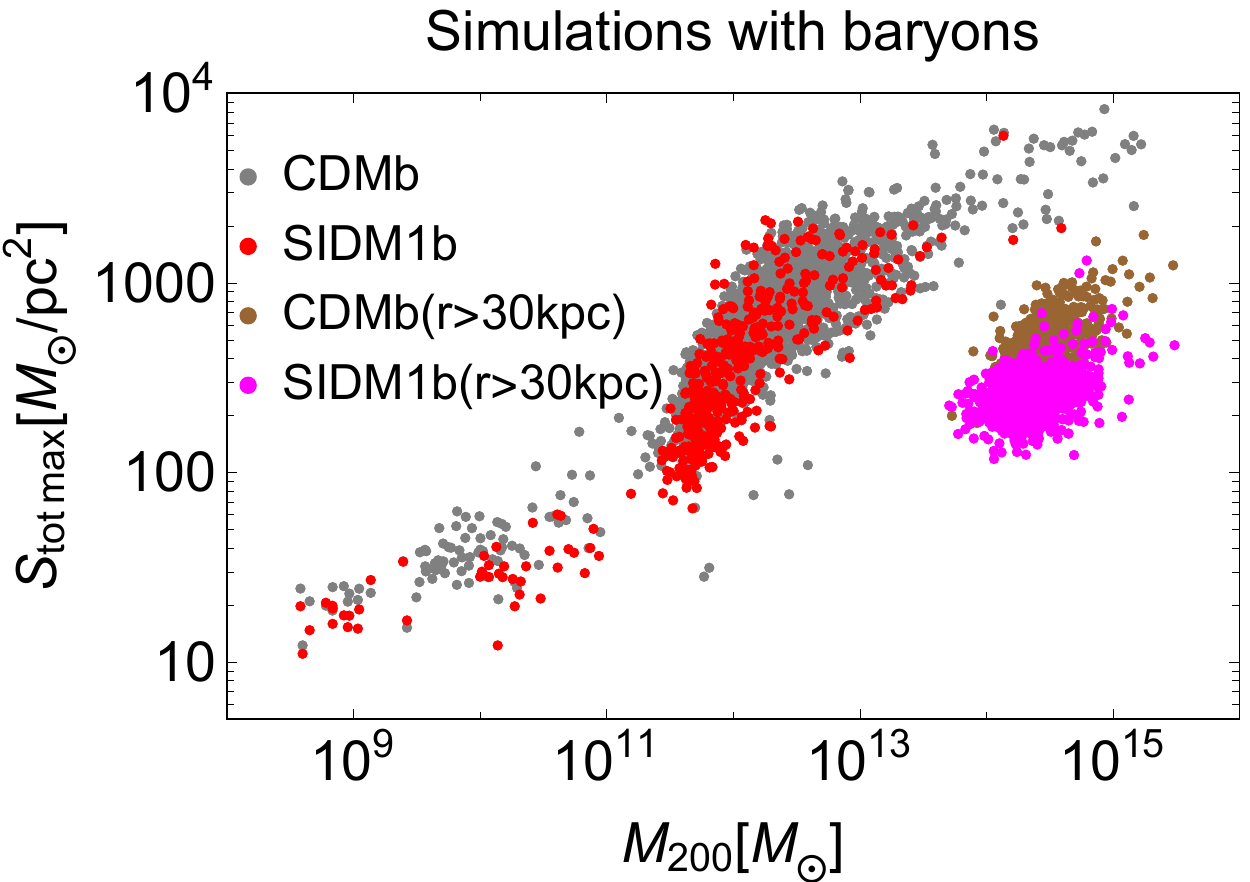}
  \includegraphics[width=0.51\textwidth]{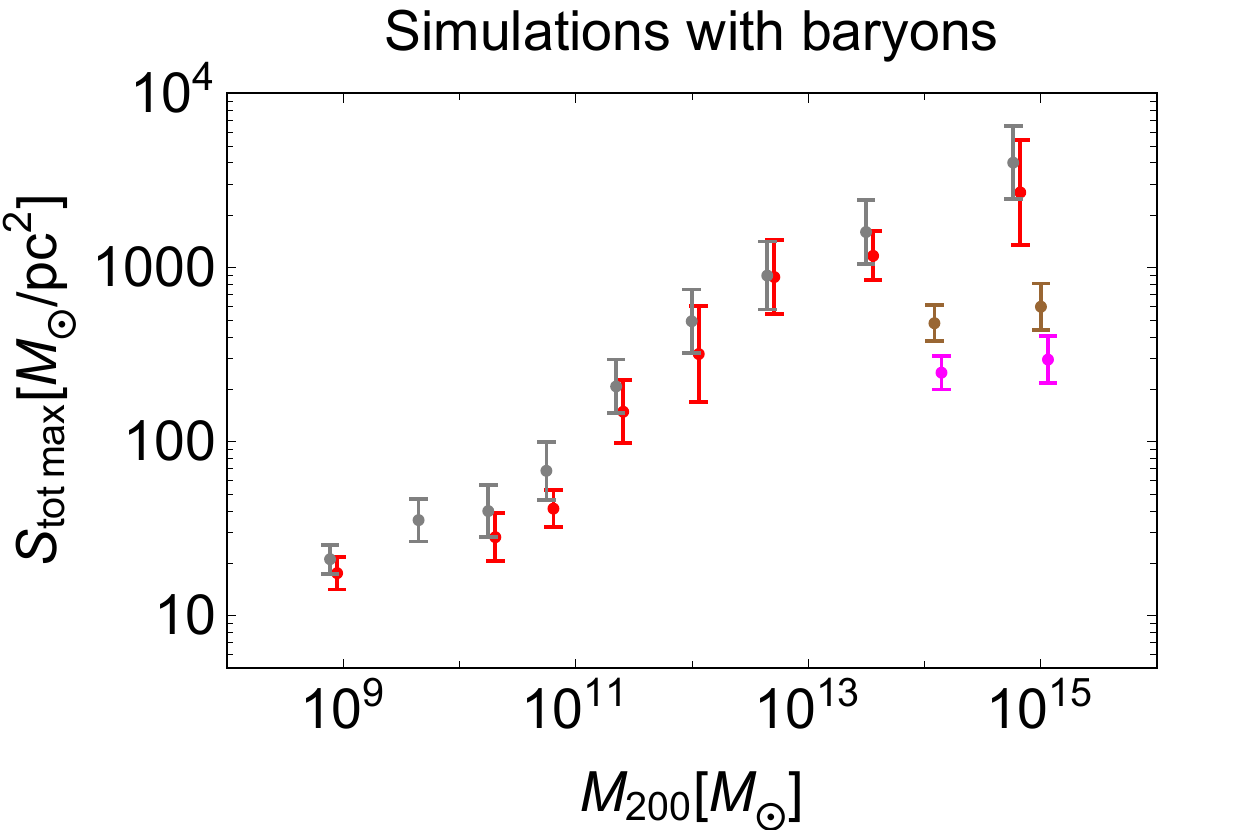}
  \caption{\textcolor{black}{\textit{Left panel:}} Maximum surface density for DM mass versus virial mass for simulations with baryons: gray points for CDMb, red points for SIDM1b. Magenta (SIDM1b) and black (CDMb) points are maximal values of the total mass surface density for
  BAHAMAS simulations calculated outside 
  $30$~kpc -- the trust radius of BAHAMAS, see
  Table~\ref{tab:simulations}. \textcolor{black}{\textit{Right panel:} The average and scatter of the maximum surface density from the left panel in different mass bins.}}
  \label{fig:SDtot_bins}
\end{figure}

\paragraph{Total mass surface density.} 
Another interesting quantity to study is the total mass surface density. SIDM1b BAHAMAS simulations demonstrate, that for  galaxy clusters 
the total mass surface density $S_{\text{tot}}(r)$ for $r>30$~kpc
has a maximum for most of the halos. High-resolution simulations
(available for 3 SIDM1b clusters only) show, however, that
even larger values of $S_{\text{tot}}(r)$ are achieved  in the inner part from of the halo, inside the bright central galaxy (BCG), growing toward the trust radius. 
We conclude that in most of the SIDM1b clusters, apart from the global maximum of $S_{tot}(r)$ near the center of BCG, there is 
also a local maximum, located at larger distances, between 30 kpc and the position of the maximum of DM surface density in the same halo (see Fig.~\ref{fig:SD_tot} for an example).  For most of the small galaxies ($<10^{11} M_{\odot}$) the maximum is also present and, in any case, the maximal value of the total mass surface density is systematically lower in SIDM1b than in CDMb (see Fig.~\ref{fig:SDtot_bins}).
This provides a possibility to use the total mass surface density to discriminate between DM models, not relying on decomposing the total mass into DM and baryons, which often requires complicated modelling, with the potential for systematic errors.
However, for most of the galaxies with virial masses in the range from $3\times 10^{11}M_{\odot}$ to $3\times 10^{13}M_{\odot}$
the total mass surface density $S_{\text{tot}}(r)$ grows monotonically towards the trust radius. Also, we have to be careful distinguishing the global and the local maximum in the total mass surface density for clusters.

\begin{figure}[h!]
  \centering
  \includegraphics[width=0.495\textwidth]{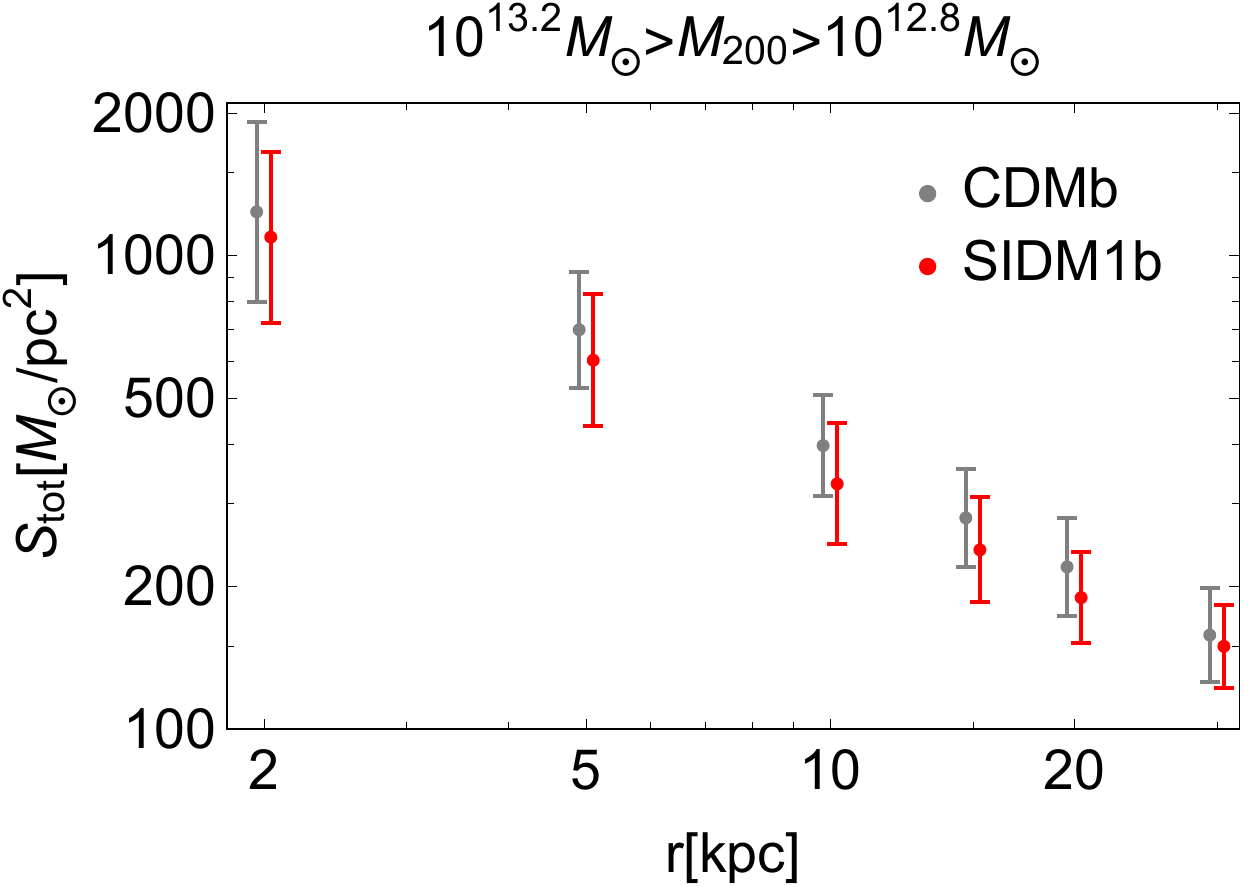}
  \caption{Comparison between total mass surface densities in CDMb (gray) and SIDM1b (red) simulations at different radius $r$ for objects with mass close to $10^{13}M_{\odot}$.}
  \label{fig:SDtot_13}
\end{figure}

The comparison between the values of the total mass surface density at the global maximum in CDMb and SIDM1b simulations is shown in Fig.~\ref{fig:SDtot_bins}. According to the right panel of Fig.~\ref{fig:SD_max_30kpc} we see large difference between high-resolution cluster simulation and BAHAMAS, so we indicate BAHAMAS simulation with different color.
Unfortunately, our ability to compare two models in simulations is limited. Indeed, the maximum of total mass surface density is located at small radii for both models. Such small distances are resolved only by high-res simulations, that are available for CDMb (where we can use Eagle 100 Mpc and C-Eagle simulations). For SIDM1b
we do not have Eagle 100 Mpc and only two high-resolution C-Eagle clusters plus one cluster from Eagle 50 Mpc are available. The total mass surface density of the simulated objects can be compared with observations (see Section~\ref{sec:compare}). However, if we want to see the difference between CDMb and SIDM1b we should not use the absolute maximum of the total mass surface density for the large haloes. Indeed, as we can see both in Fig.~\ref{fig:SD_tot} and in Fig.~\ref{fig:baryon_fraction} the maximal value for the heavy objects is mostly influenced by baryons. 
Instead, we should calculate the total mass surface density inside larger radii, closer to the maximum  of DM surface density of SIDM1b.
The magenta and brown points in Fig.\ref{fig:SDtot_bins} present the maximum of the total mass surface density for 
the galaxy clusters (BAHAMAS) in the range of radii outside 30 kpc.
We see that the total mass surface density for the two models is distinguishable for galaxy clusters. For massive elliptical galaxies the situation is different: the effects of SIDM are visible in the DM profiles on smaller scales, where the total mass is dominated by baryons. As a result, the total mass surface density is indistinguishable between the two models at all available radii in these objects, see Fig.~\ref{fig:SDtot_13}.

\paragraph{Summary of simulations.} 
In simulations, the maximal surface density of DM demonstrates a regular behavior (scaling law) over 6 order of magnitude in $M_{200}$, see Fig.~\ref{fig:SDmax_baryons_bins}. This is true for CDM as well as SIDM simulations, both DM-only and with baryons. In the mass range around $10^{12} M_{\odot}$ strong baryonic effects introduce a feature in this scaling law. We can use DM surface density to distinguish between SIDM and CDM, despite baryonic effects. For spiral galaxies the difference is not very visible, however for large halos the two models are clearly separated. 

The maximum of the total mass surface density also has a regular dependence on $M_{200}$.
While for small objects this maximum is a bit lower in SIDM1b than in CDMb (see Fig.~\ref{fig:SDtot_bins}), for larger objects this maximal value is dominated by baryons and seems to be indistinguishable between the two models. To see the difference between them we can use instead the total mass surface density calculated outside 30 kpc where the fraction of DM is more significant and we can see the difference between CDMb and SIDM1b.

We conclude that the available simulations predict that  $S_{\max}$ as a function of $M_{200}$ is expected to be different in CDM and SIDM, despite baryonic effects, and therefore we proceed to compare simulations with the available observations.

\section{Available observational data}
\label{sec:data}

In this section we will introduce the best estimates of the maximal surface density for observed systems, which we will compare with the simulated systems in Section~\ref{sec:compare}.

\begin{figure}[h!]
  \centering
  \includegraphics[width=0.51\textwidth]{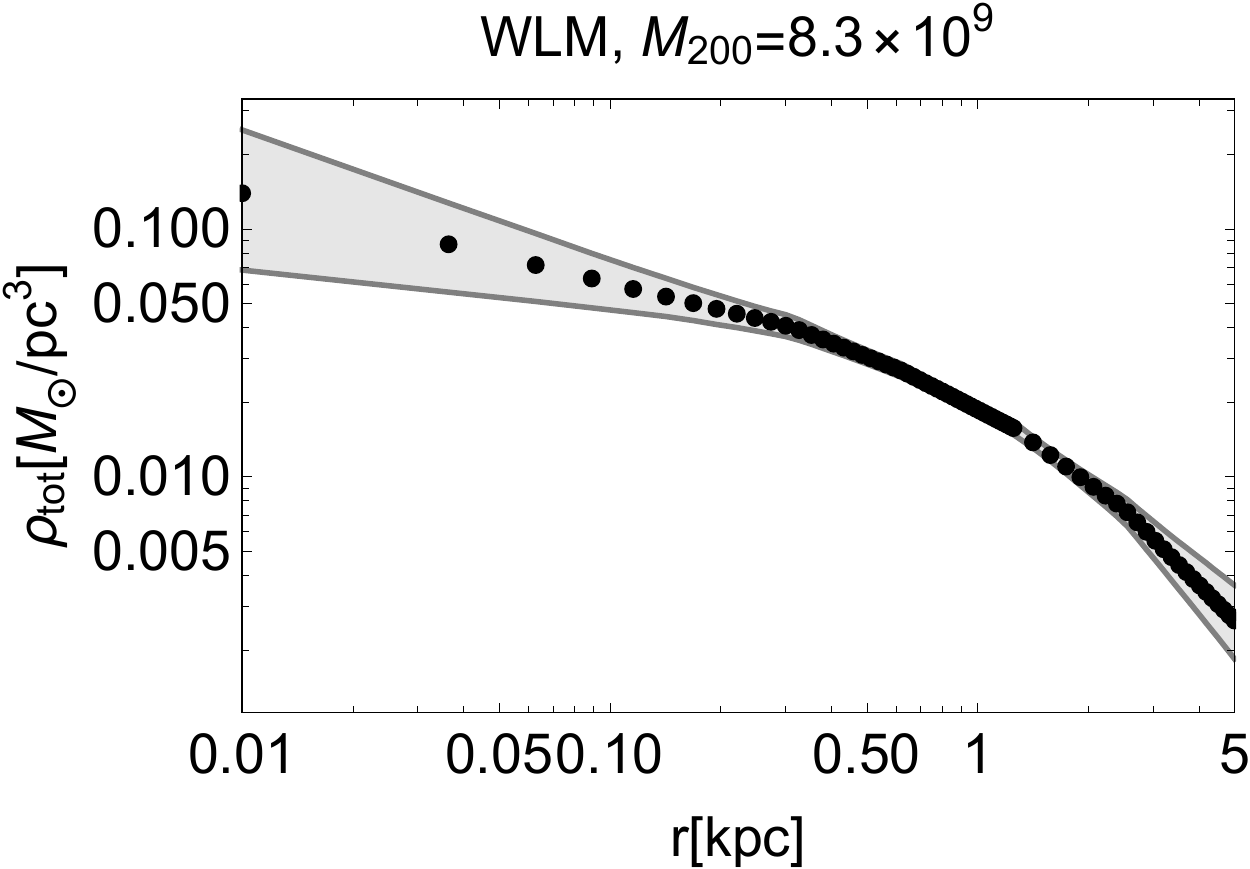}
  \includegraphics[width=0.48\textwidth]{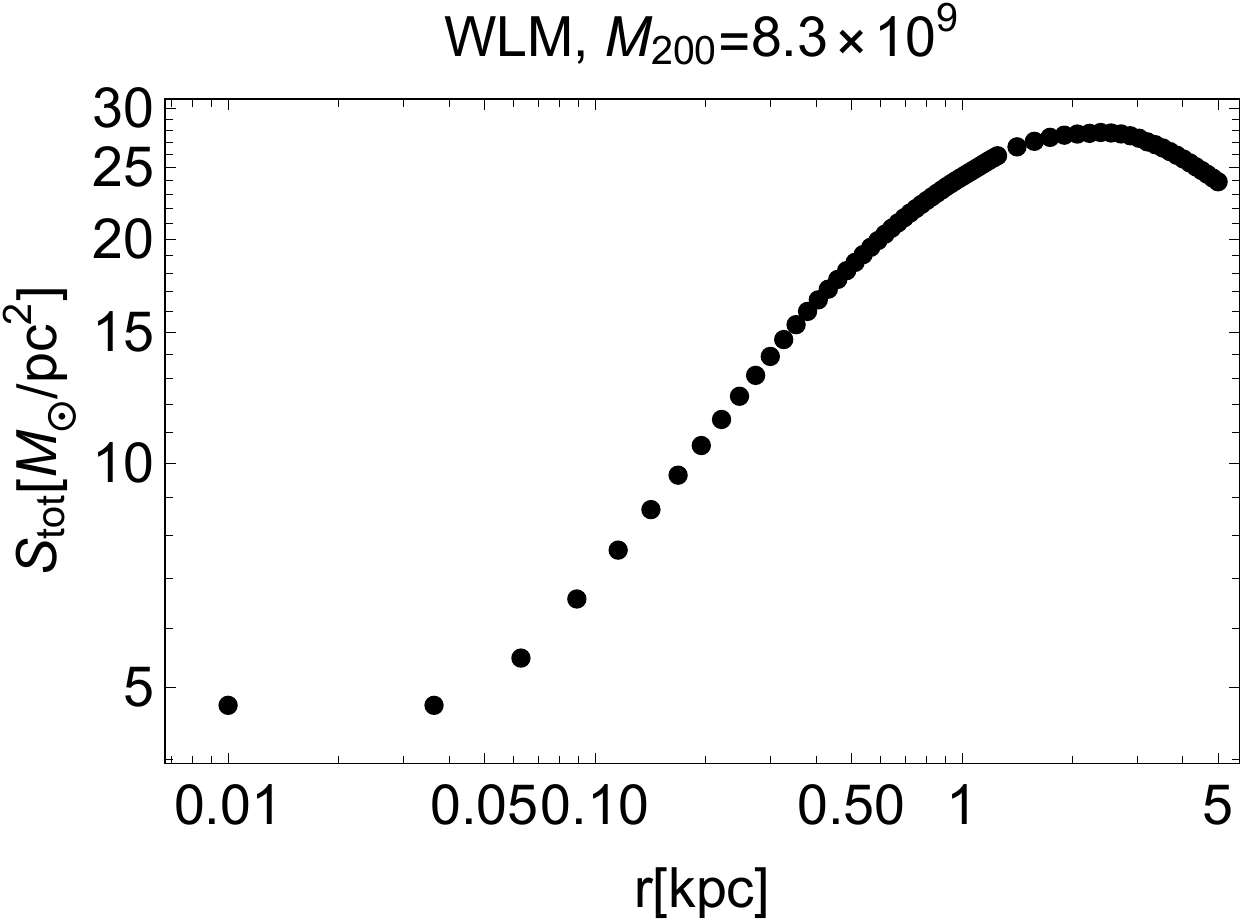}
  \caption{\textit{Left panel:} The best fit total mass density profile (black points) of WLM with 1 standard deviation (gray region). \textit{Right panel:} The total surface density calculated from the best fit values of the density profile.}
  \label{fig:Dwarfs_obs}
\end{figure}

\paragraph{Dwarf galaxies.} 
For the field dwarf galaxies we use the results from~\cite{Read:2018fxs} for 8 objects: CVnIdwA, DDO 52, DDO 87, DDO 154, DDO 168, DDO 210, NGC 2366, WLM. Here, the total enclosed mass profiles were calculated from either stellar kinematics or HI gas rotation curves. 
The main sources of uncertainty are: the distances to each halo, their inclinations, their ellipticities, and velocity anisotropy in the case of measurement of velocity dispersion. In Fig.~\ref{fig:Dwarfs_obs} we show an example of the density and surface density profiles, using WLM. We observe that for all these objects $S(r)$ has a clear maximum at 1--5 kpc from the center. All these objects except of CVnIdwA and DDO 168 are strongly DM dominated, so we do not distinguish between the total mass and DM mass~\cite{Oh:2015xoa}. For CVnIdwA and DDO 168 we found that the radii of maximal total mass surface density were $1.3$~kpc and $3.8$~kpc, respectively. The enclosed mass at these radii is dominated by DM, so we do not distinguish between the total mass and DM mass for them as well.

We also consider classical  dwarf spheroidal (dSphs) satellites of the Milky Way (MW). 
These objects have very high mass-to-light ratios and are DM dominated even in the central parts (see e.g.~\cite{Read:2018fxs} and references therein) so they are good objects in which to study DM properties. However,
they 
are sub-halos of the MW halo, and as such are affected by processes that do not act on isolated halos in `the field'. As such, when we compare them with simulations, we use simulated satellites of Milky Way-like galaxies.
 The main observable in dSphs is the line-of-sight velocities of stars. The dispersion of velocities is used to reconstruct the gravitational potential and the mass enclosed within a given radius. The main uncertainty of this method is the unknown stellar velocity anisotropy. This uncertainty can be minimized when the analysis is applied to the mass inside the half-light radius, $r_h$~\cite{Walker:2011zu,Wolf:2009tu,Amorisco:2011hb}. So we use the surface density calculated at the half-light radius as a lower bound on the maximal surface density in dShps, see Appendix~\ref{app:obs_data} for details.

\begin{figure}[h!]
  \centering
  \includegraphics[width=0.51\textwidth]{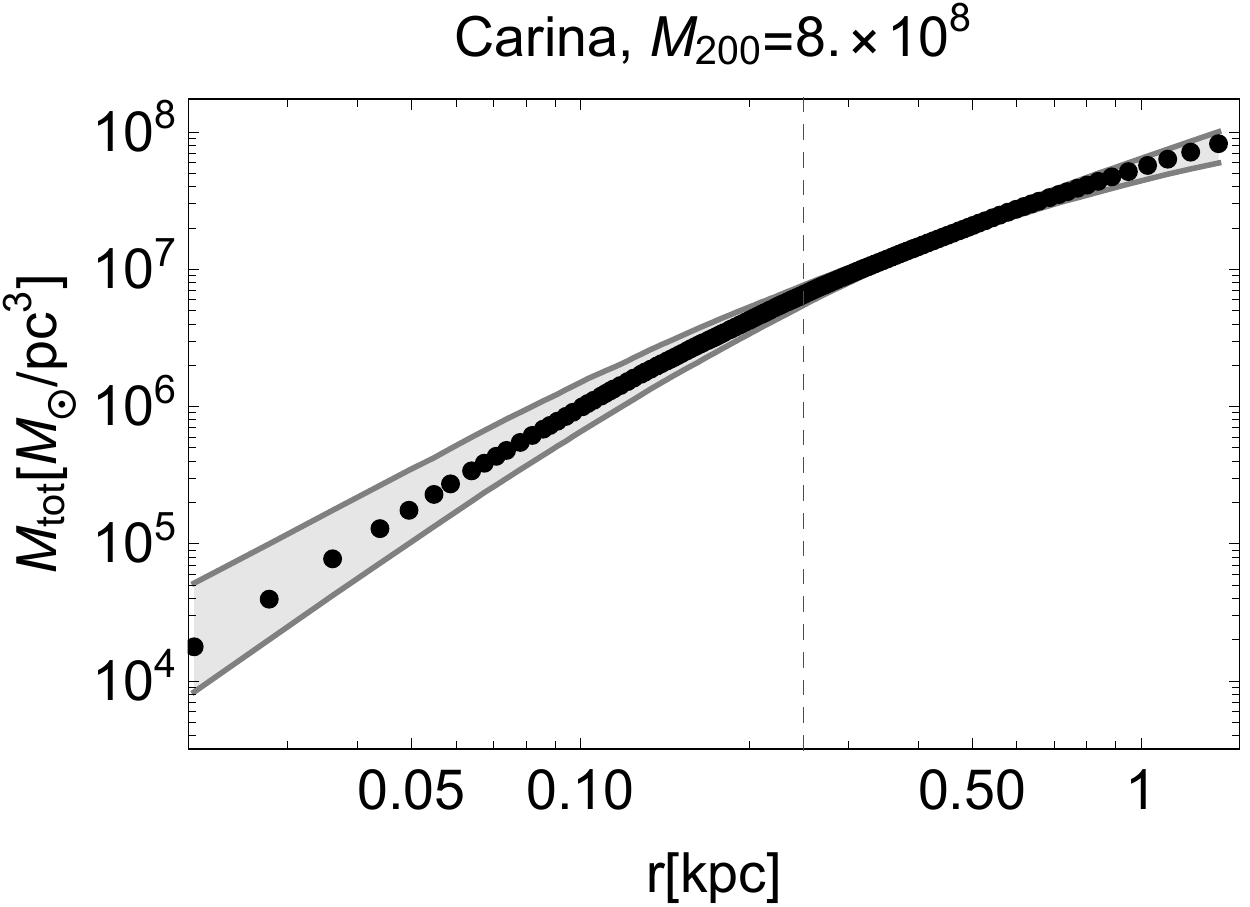}
  \includegraphics[width=0.48\textwidth]{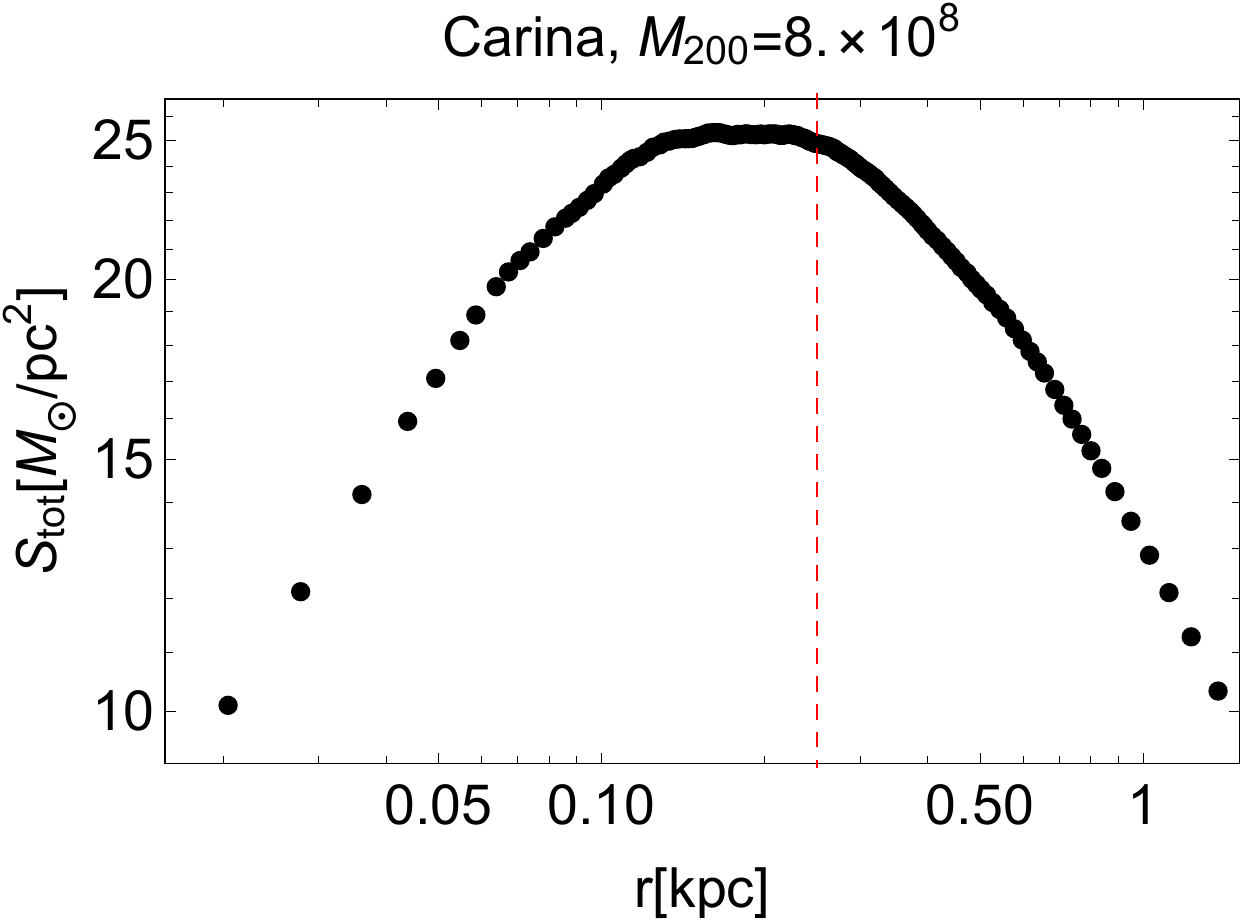}
  \caption{\textit{Left panel:} The best fit total mass density profile (black points) of Carina with 1 standard deviation (gray region). \textit{Right panel:} The total surface density calculated from the best fit values of the density profile. The red dashed line shows the half-light radius.}
  \label{fig:Dwarfs_class_obs}
\end{figure}

In~\cite{Read:2018fxs} a method using not only the velocity dispersion, but also higher moments of the velocity distribution function was applied to dwarf galaxies. This approach was tested using CDM and SIDM simulations in~\cite{Genina:2019job} and showed good accuracy. Here we use the results of~\cite{Read:2018fxs} for the MW dSphs, see Table~\ref{tab:dSphs-data-old} for the data we used in this work. Recently, new dSphs were discovered by the SDSS and DES surveys~\cite{York:2000gk,Abbott:2017wau}. However, many of these objects lack high-resolution spectroscopic observations so we do not include them in our analysis.

\paragraph{Spiral galaxies.}
The main observables in spiral galaxies are the line-of-sight velocities of stars and neutral hydrogen, from which rotation velocities can be inferred if the inclination of the disk with respect to the line-of-sight is known. This is challenging for disks that are close to face on (where the velocities are perpendicular to the line of sight and do not produce a Doppler shift) or edge on (where the velocities of stars at different locations in the disk appear blended together).

Because the central regions of spiral galaxies are dominated by stars, a large source uncertainty in measuring the DM distribution comes from the uncertain modelling of the baryonic contribution to the total mass. As a result, the DM profile can often be fitted (almost) equally well with NFW and cored profiles, see e.g. the rotation curves displayed in Ref.~\cite{deBlok:2008wp}. Another source of uncertainty is the distance to each galaxy: if a galaxy contains `standard candles' in it we can measure the distance to that galaxy with high precision. However, for most of the galaxies the method used to measure the distance is the Hubble law, which can be highly uncertain for nearby galaxies.

\begin{figure}[h!]
    \centering
     \includegraphics[width=0.48\textwidth]{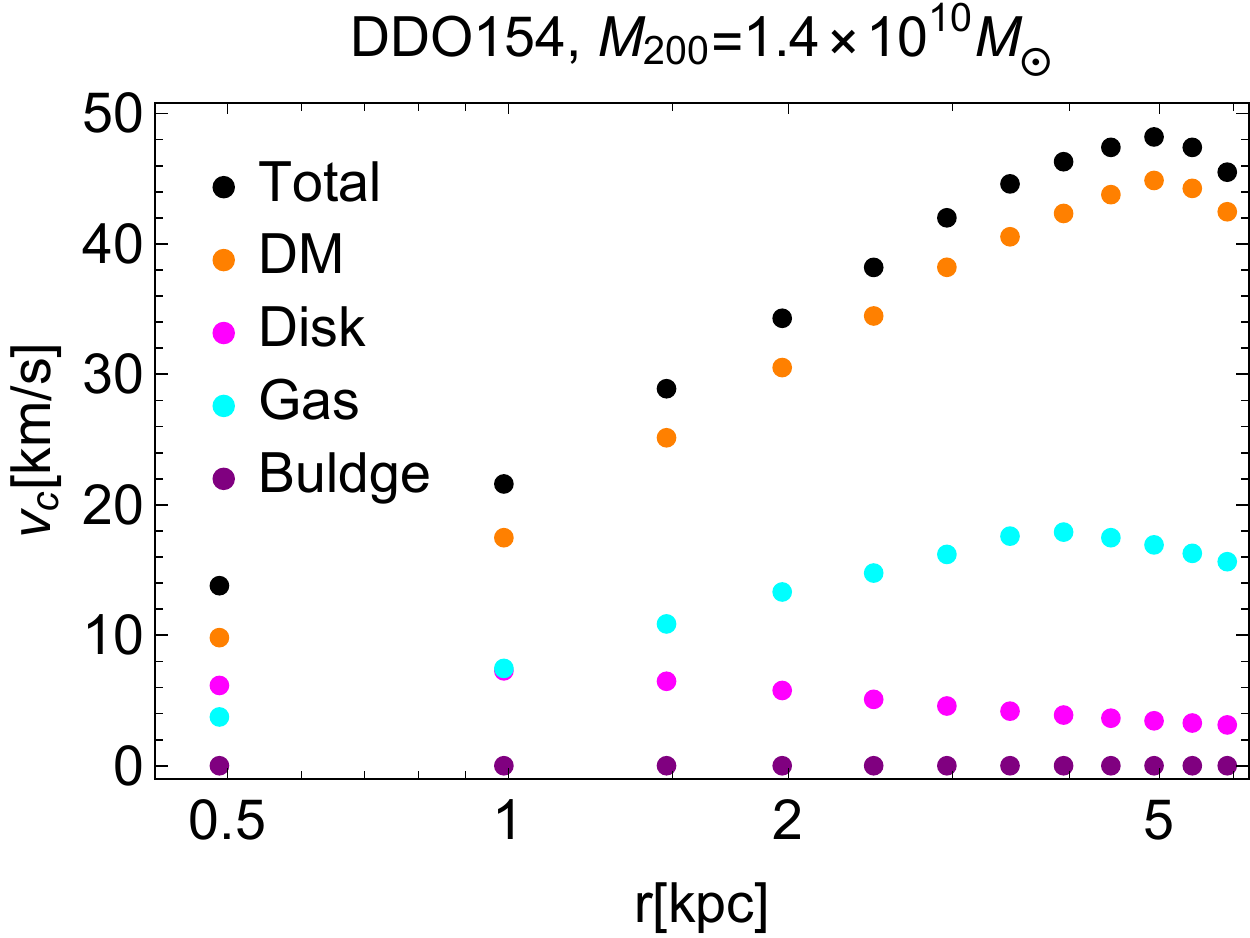}~\includegraphics[width=0.48\textwidth]{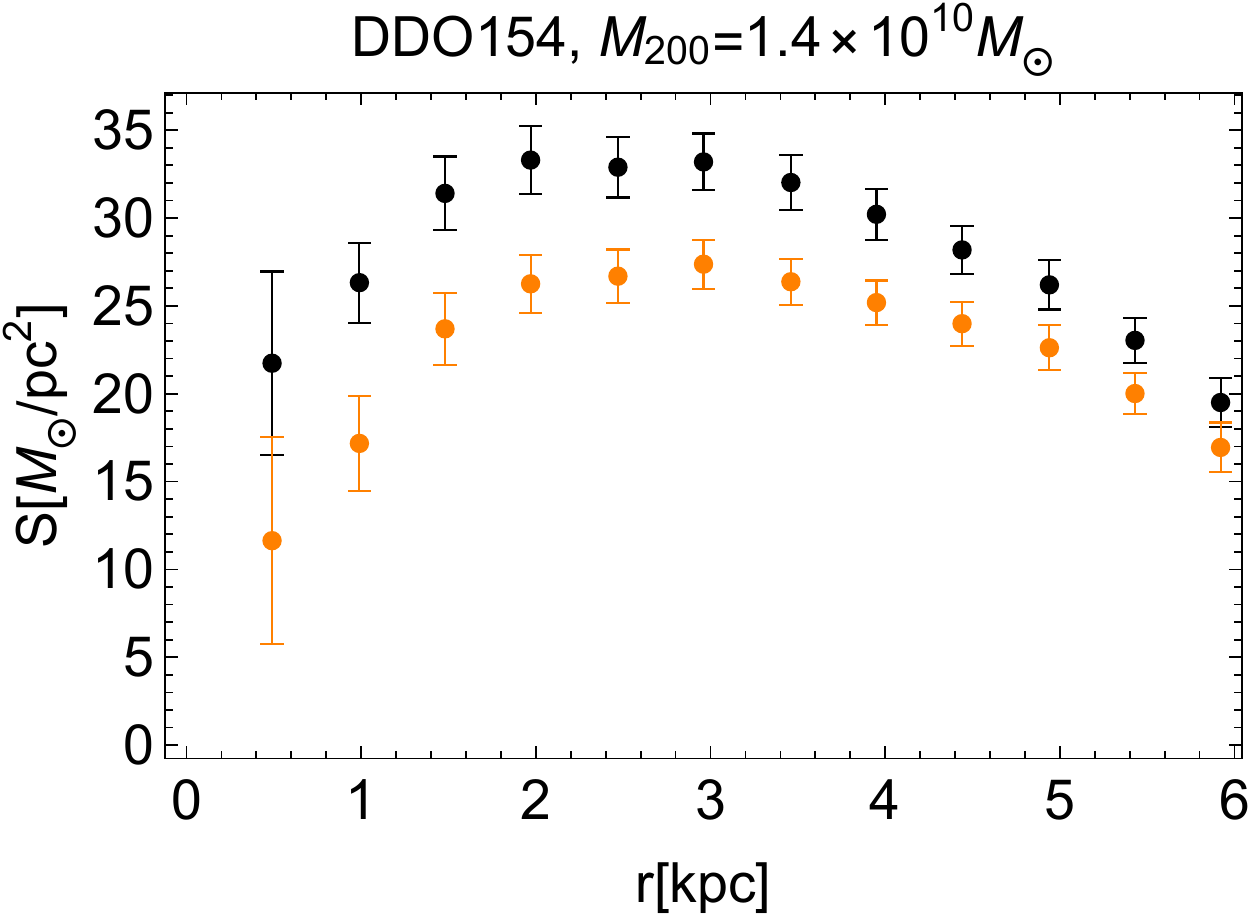}
     \\
     \includegraphics[width=0.48\textwidth]{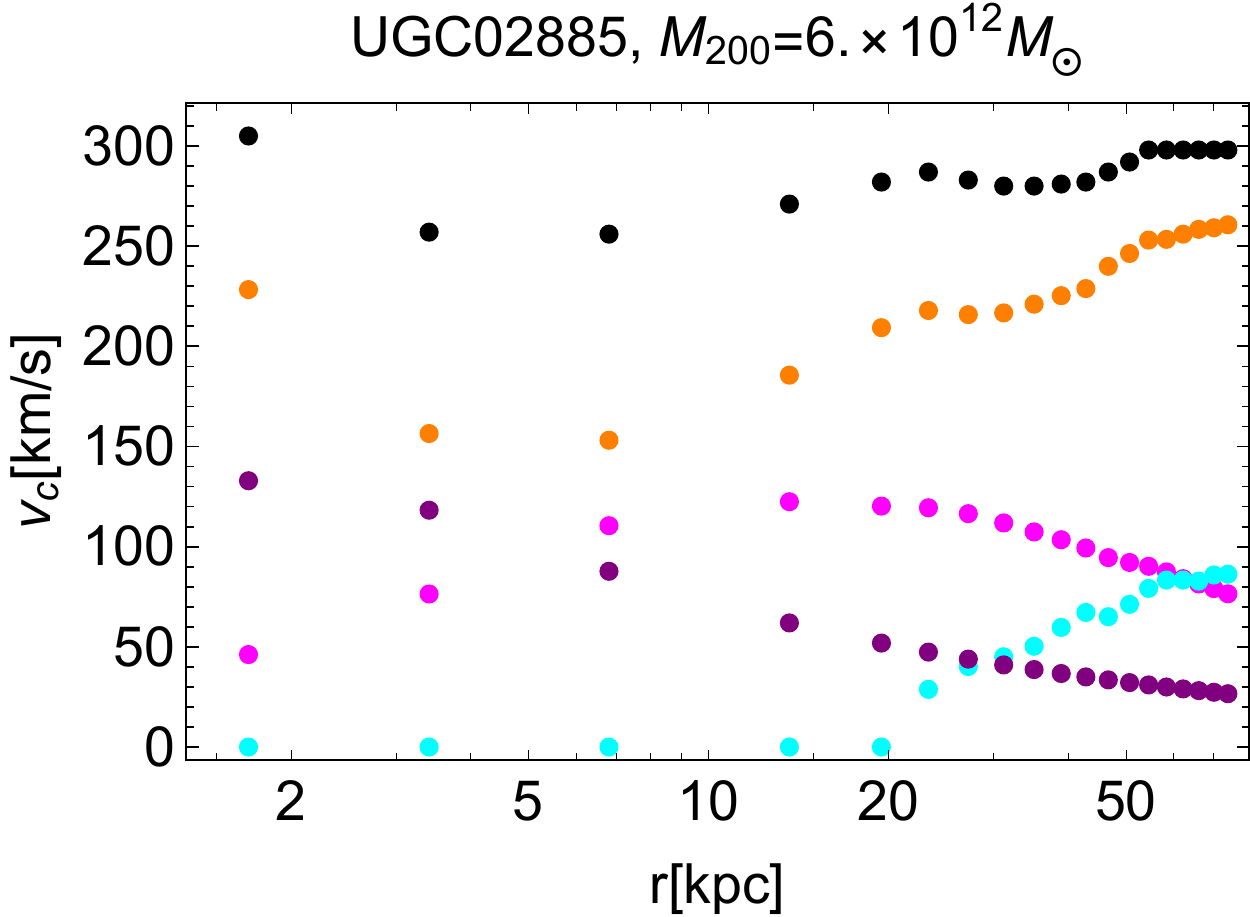}~\includegraphics[width=0.48\textwidth]{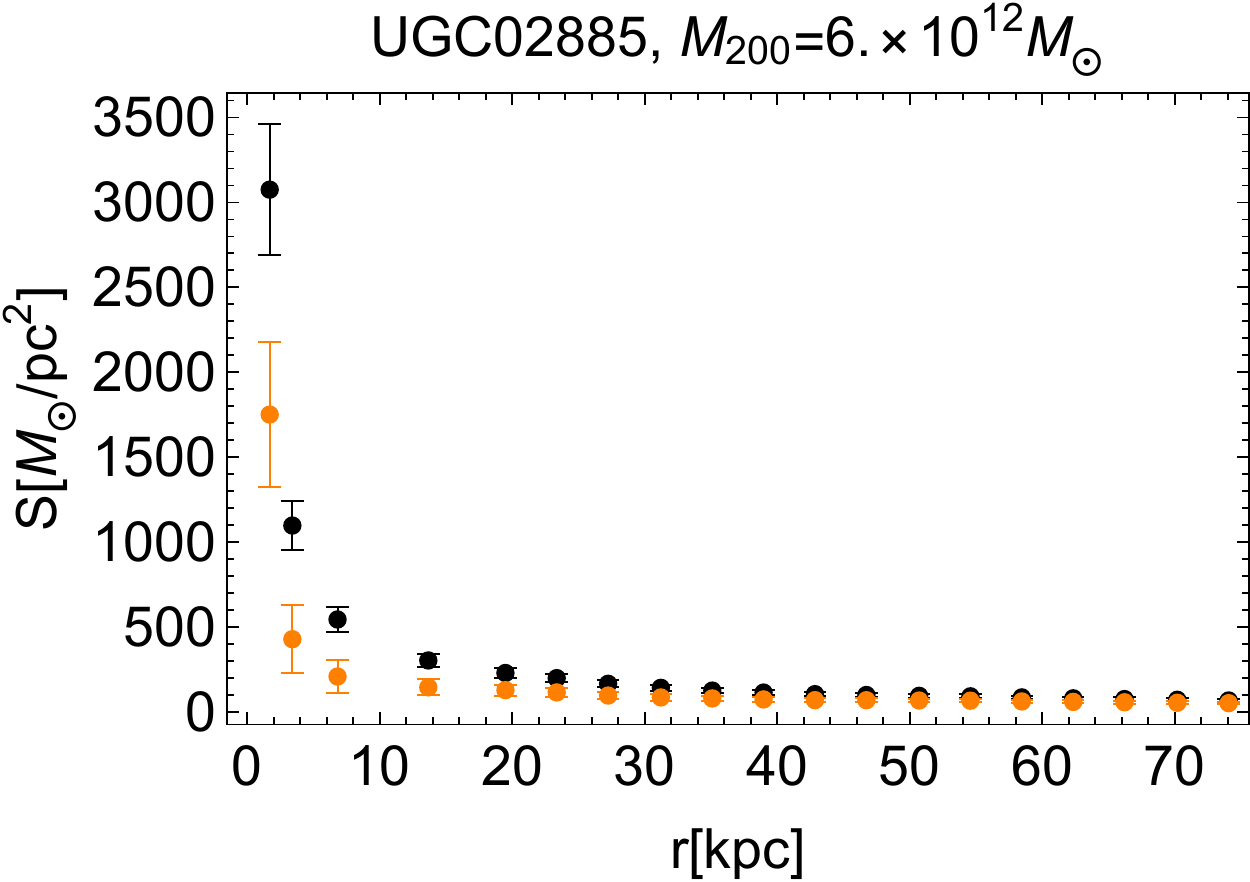}
    \caption{Examples of SPARC objects from the third group (the group of good objects, see the main text for an explanation). \textit{Left panel}: the circular velocity due to different mass components. \textit{Right panel}: the total surface density (black points) and DM surface density (orange points).}
    \label{fig:SPARCthirdgroup}
\end{figure}

We took 175 spiral galaxies from the SPARC catalogue~\cite{Lelli:2016zqa} which provides us with models of the baryons based on the method described in~\cite{2019MNRAS.483.1496S}. Our selection criteria are: an uncertainty on the distance measurement less than $15\%$, a galaxy disk inclination more than $30^{\circ}$ and a quality flag that is equal to 1 or 2 (which means the best objects, see details in~\cite{Lelli:2016zqa}). After applying these cuts we are left with 83 objects. These objects can be divided into 3 different groups:
\begin{enumerate}
    \item Objects with anomalies, e.g. the baryonic mass is larger than the total mass inferred from the rotation curve, or there is no flat part in the rotation curve (12 haloes);
    \item Objects with too little data, which we define as less than 10 radial data points (22 haloes);
    \item Good objects (49 haloes).
\end{enumerate}
The list of selected objects is given in Appendix~\ref{app:obs_data}. An example of an object from the third group is given in Fig.~\ref{fig:SPARCthirdgroup}.

\begin{figure}[h!]
    \centering
    \includegraphics[width=0.5\textwidth]{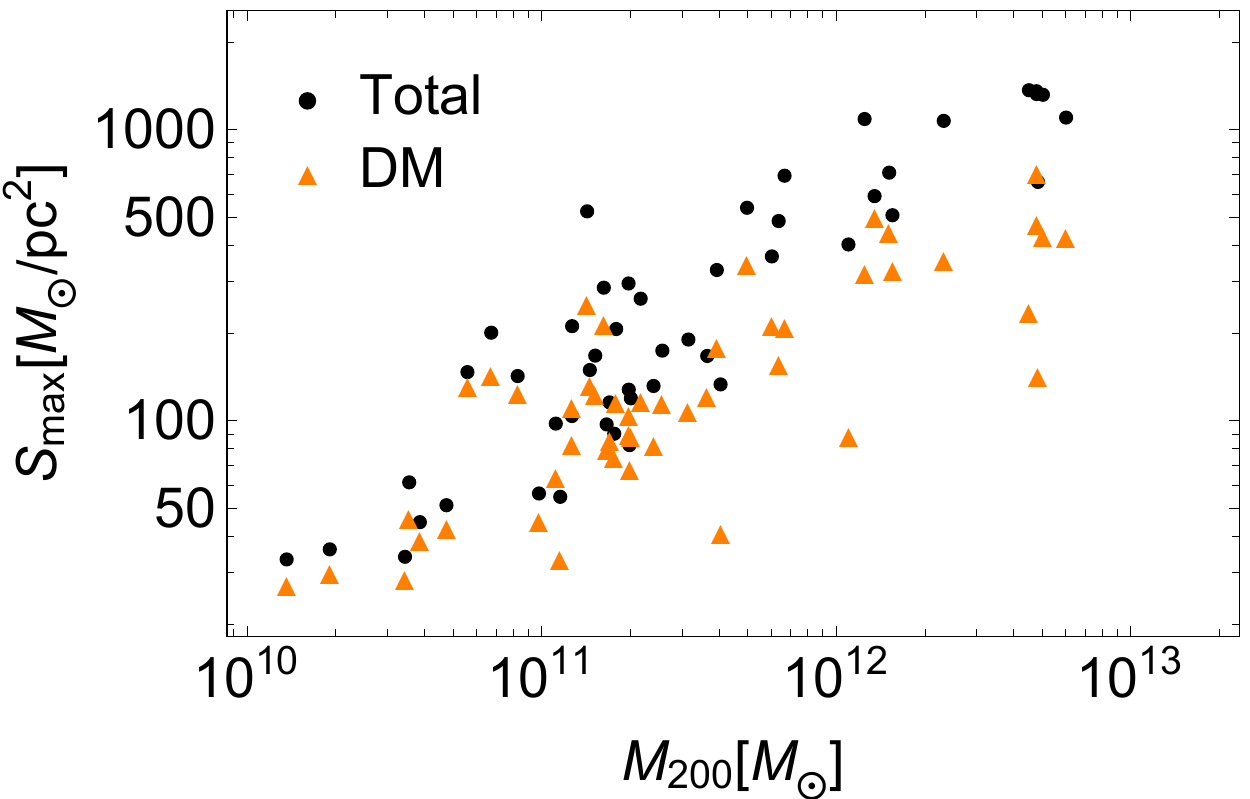}~\includegraphics[width=0.47\textwidth]{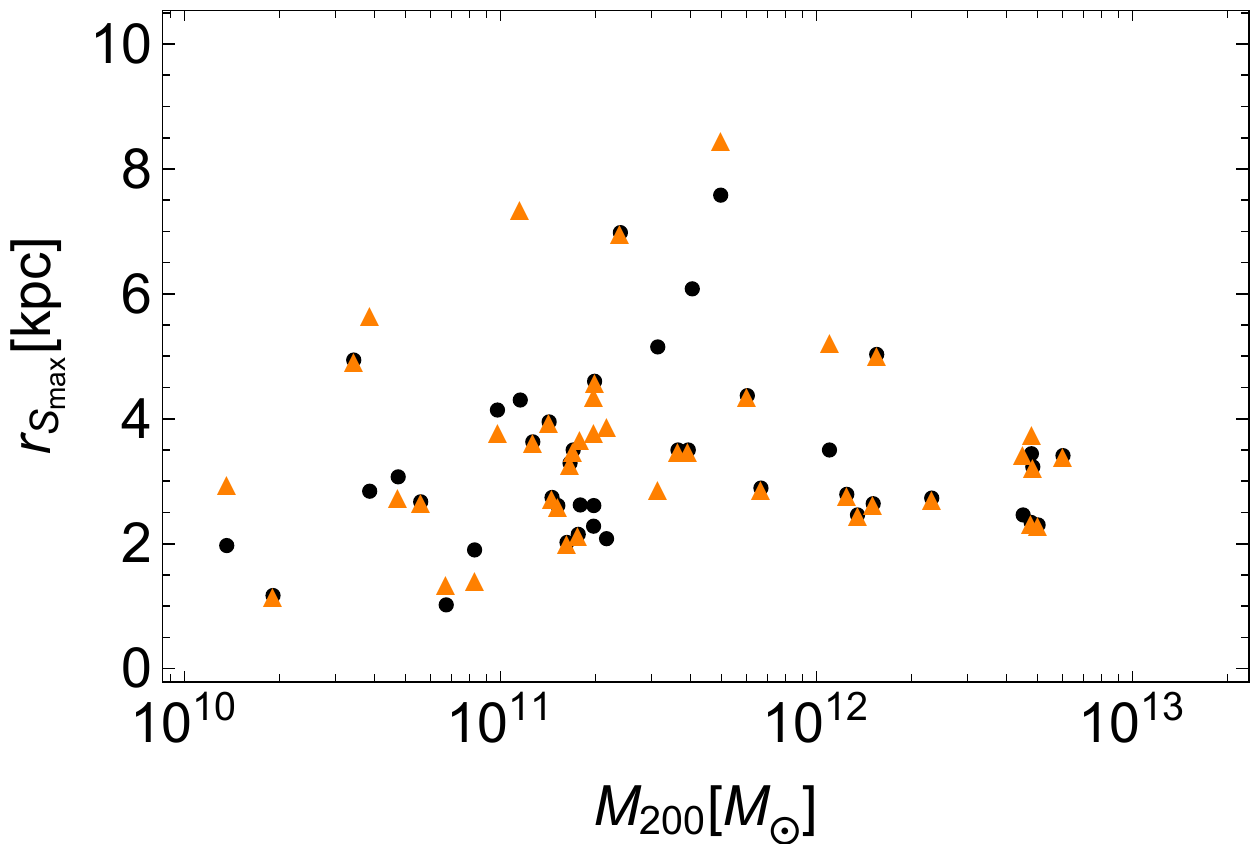}
    \caption{\textit{Left panel}: Maximal total surface density (black dots) and maximal DM surface density (orange triangles) versus $M_{200}$ for objects from SPARC. \textit{Right panel}: Radii of the maximal surface density for the total mass (black dots) and DM surface density (orange triangles) versus $M_{200}$ for objects from SPARC.}
    \label{fig:SPARCmaxSD}
\end{figure}

Using objects from the third group we show the maximum of the surface density and the radius of the maximum of the surface density as a function of the virial mass, see Fig.~\ref{fig:SPARCmaxSD}. In this figure we find the maximum of the surface density only at radii that are larger than the trust radius of the simulations that cover the relevant mass range: $r>1$~kpc for $M< 10^{11}M_{\odot}$ and $r>2$~kpc for $M> 10^{11} M_{\odot}$, see Table~\ref{tab:simulations}. The estimation of the virial mass is taken from the best fit Einasto profiles from paper~\cite{Li:2018rnd}. We see that both the total and DM maximal surface densities have a regular dependence on halo mass, while the radii at which these maxima are achieved have large scatter and no clear trend with mass.

\paragraph{Elliptical galaxies and groups of galaxies.} The mass profiles of large elliptical galaxies can be reconstructed using various observational data: X-ray measured density and temperature profiles, kinematical data of stars and globular clusters, spectroscopic measurements of the neutral hydrogen dynamics, and strong and weak gravitational lensing (we provide references to the papers that we used below). These objects are baryon-dominated in the central part making the separation of the DM contribution very uncertain. Therefore we use only total mass profiles for them.\footnote{This of course requires to assume for the comparison that simulations have realistic baryonic profiles. We will see in the next Section that the agreement between data and simulations is in general quite good.}

\begin{figure}[h!]
  \centering
  \includegraphics[width=0.51\textwidth]{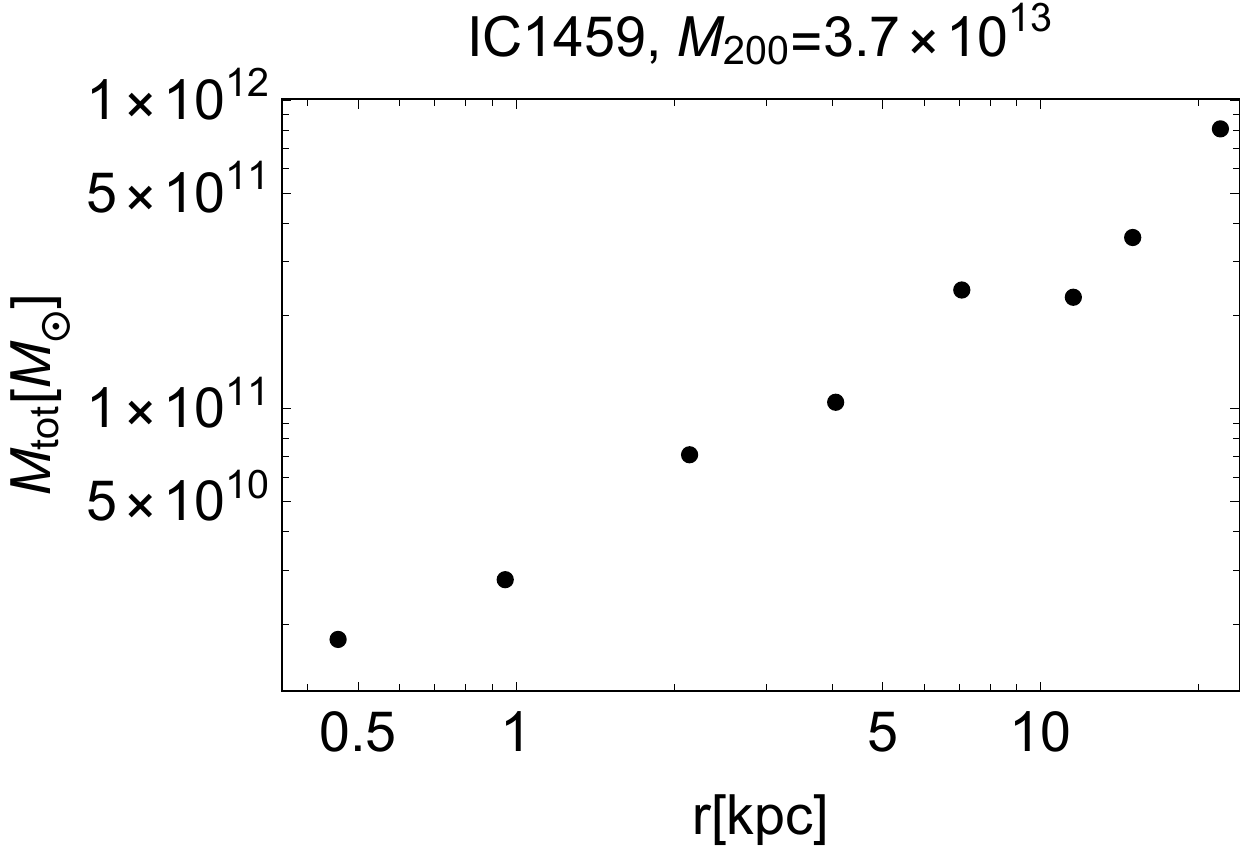}
  \includegraphics[width=0.48\textwidth]{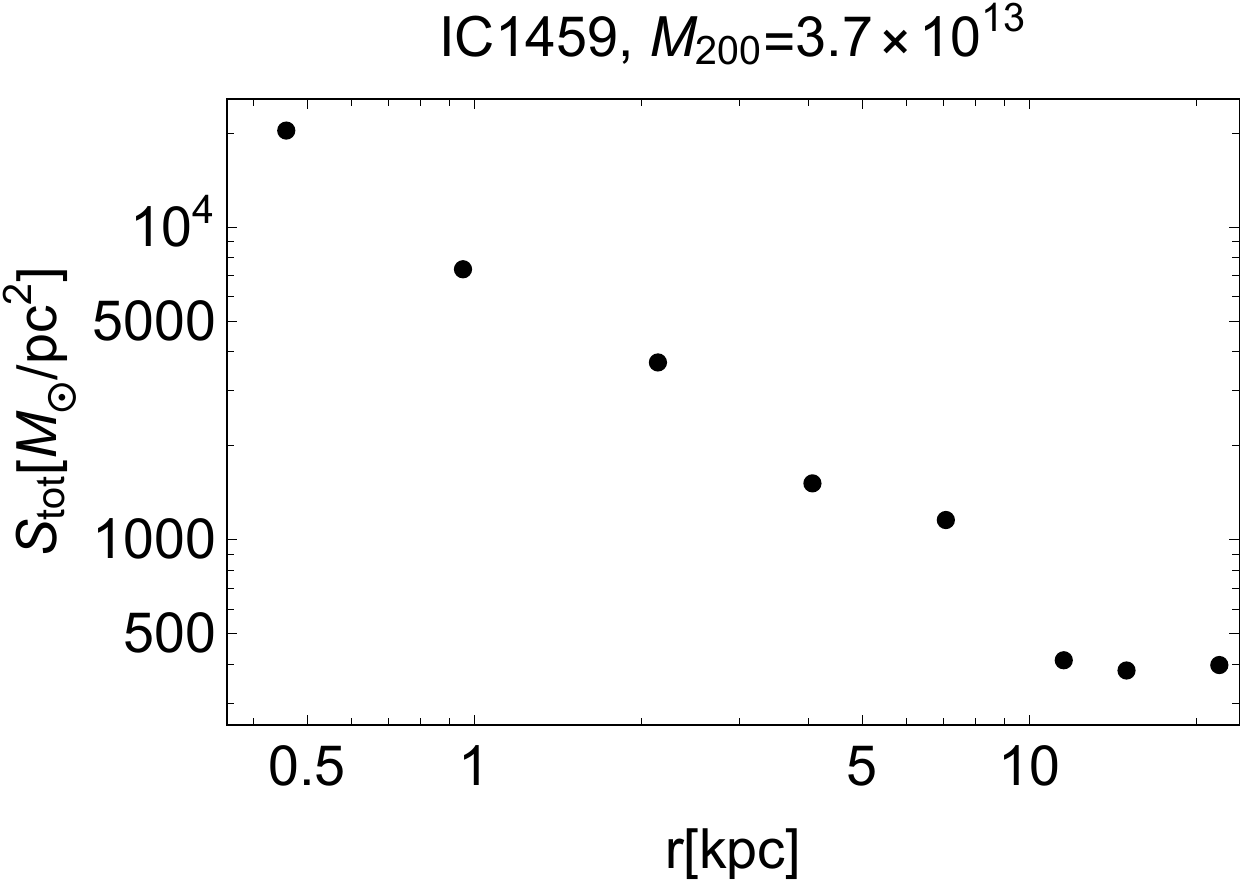}\\
  \includegraphics[width=0.51\textwidth]{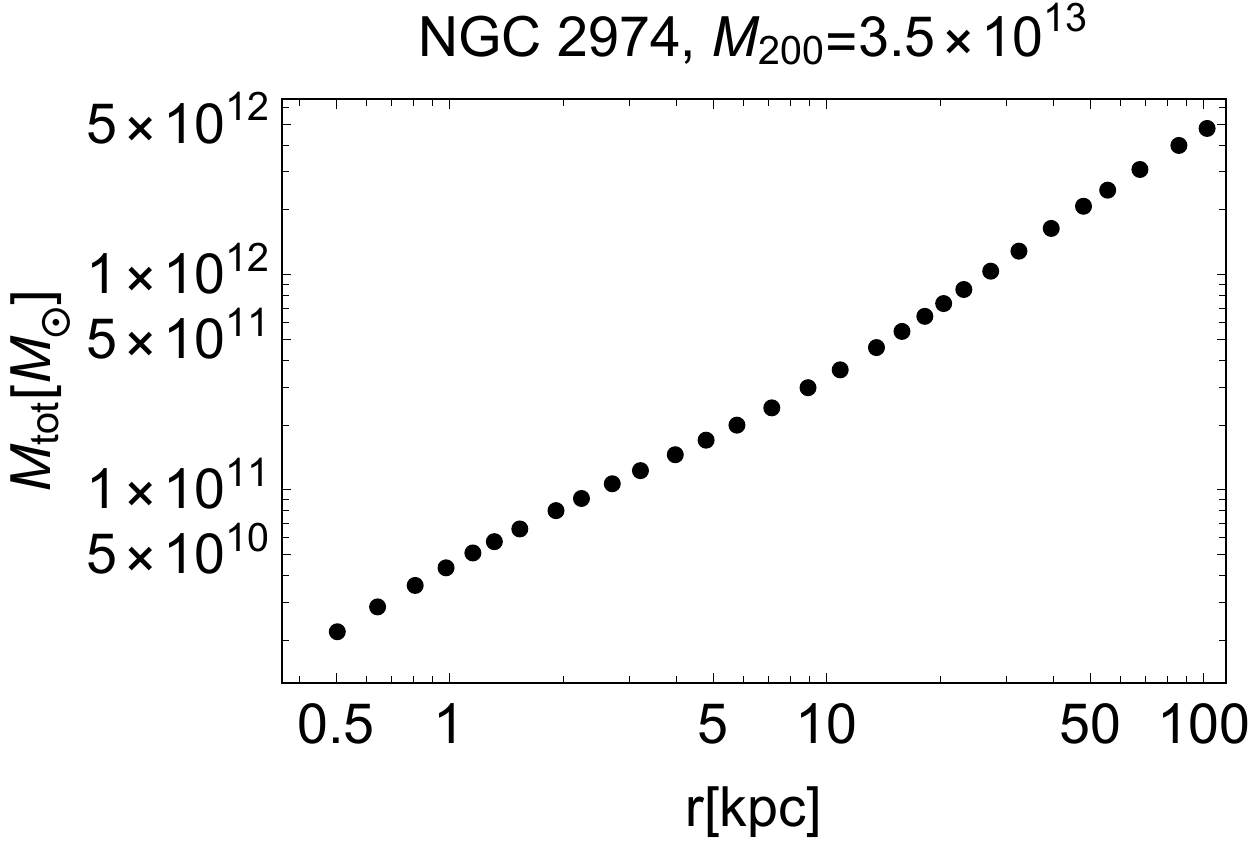}
  \includegraphics[width=0.48\textwidth]{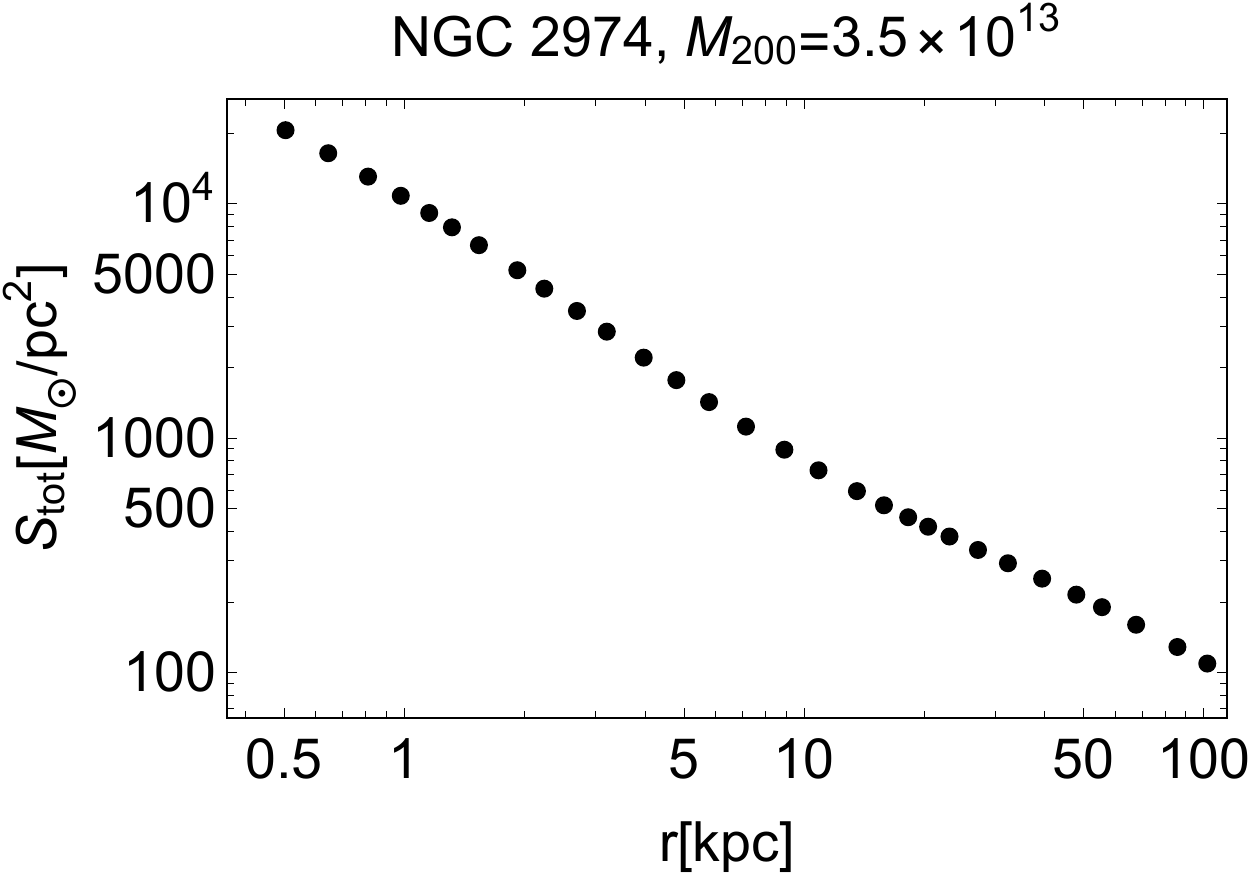}\\
  \includegraphics[width=0.51\textwidth]{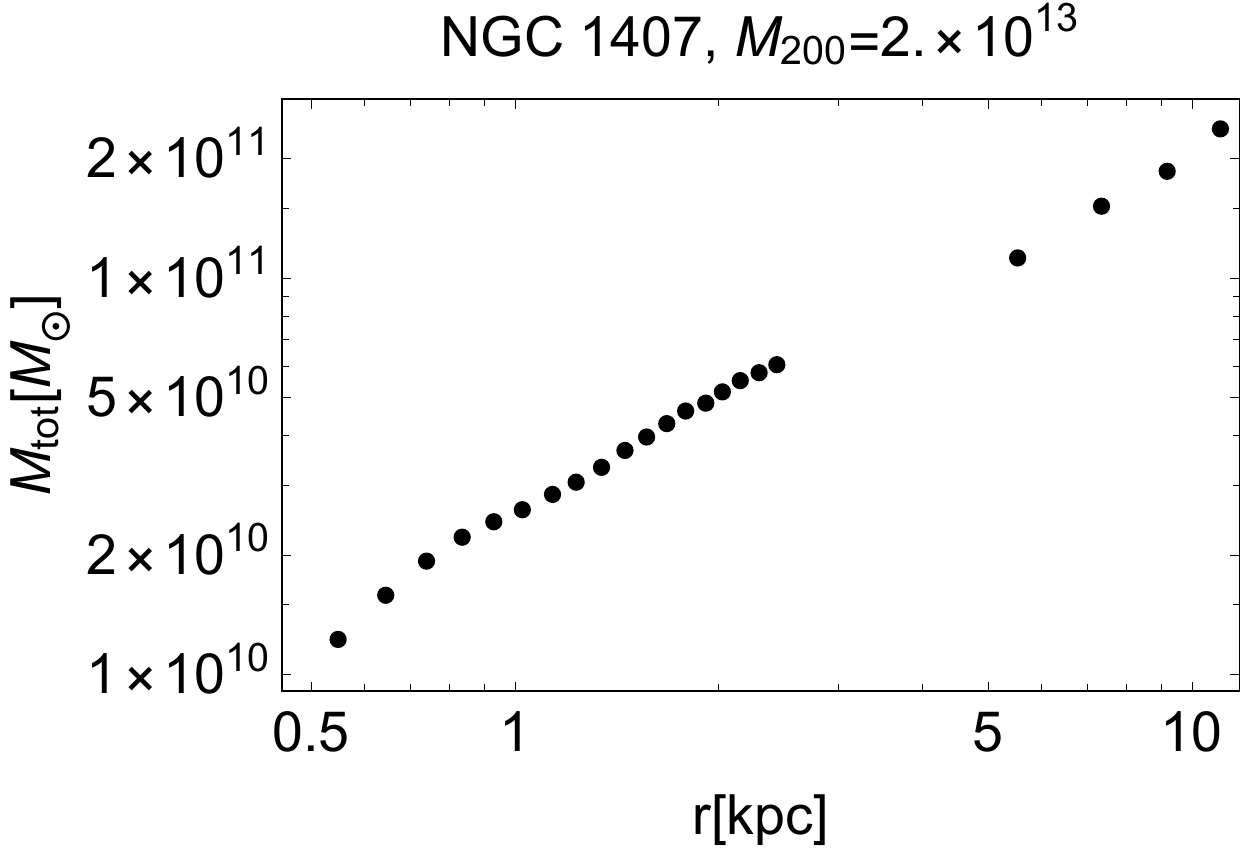}
  \includegraphics[width=0.48\textwidth]{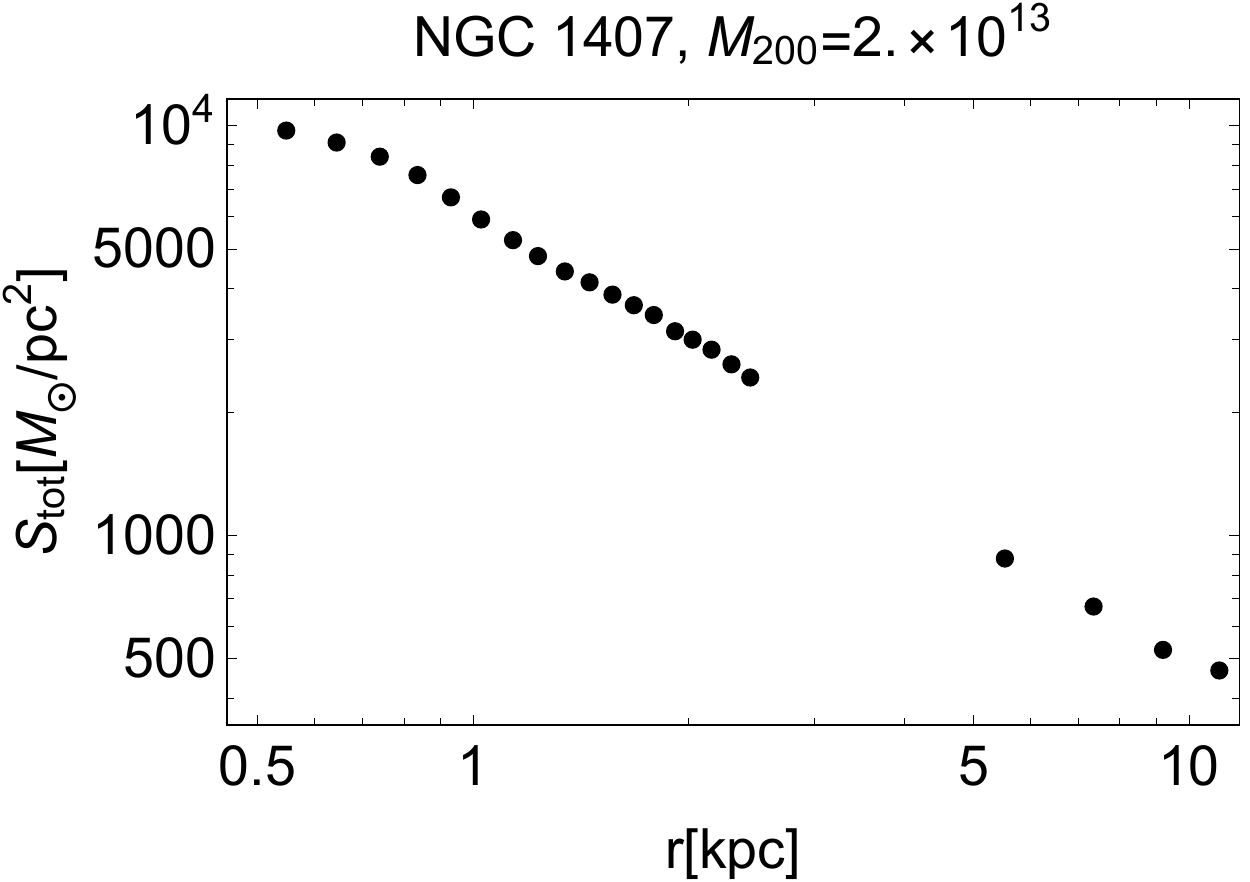}
  \caption{\textit{Left panel:} The best fit total mass density profile (black points). \textit{Right panel:} The total surface density calculated from the best fit values of the density profile.}
  \label{fig:Elliptical_obs}
\end{figure}

\emph{X-ray data.} In our analysis we use the results derived from   X-ray observations of 18 elliptical galaxies~\cite{thesisNagino,Nagino:2009em} to calculate the maximum total surface density for each object. For the halo masses we used the data from~\cite{Forbes:2016rpi,Humphrey:2009mh,2020A&A...635A...3I}. An example of the mass and surface density profiles (for IC 1459) is shown in the top panels of Fig.~\ref{fig:Elliptical_obs}. The main factors of uncertainty are the possible ellipticity of the X-ray gas as well as the assumed distances to the objects. The list of selected objects is in Appendix~\ref{app:obs_data}. 

\emph{Strong lensing.} Another dataset that we consider contains enclosed masses in the central regions of 12 objects reconstructed from strong lensing~\cite{Oldham:2018yte}. The shape of the mass profiles may depend in this case on the parametric model that is used, therefore we take the surface density inside the Einstein radius as a lower  bound on the maximum of total mass surface density.\footnote{Of course, the directly measured quantity is a 2D mass. In fact we use the critical lensing density reported by the observers to calculate the mass inside $r_{\text{Ein}}$}
The values of the Einstein radii $r_{\text{Ein}}$ for these objects are between $1.4$~kpc and $4.9$~kpc, see Appendix~\ref{app:obs_data} for details. We estimated the halo masses using the so-called Moster stellar mass -- halo mass relation~\cite{Moster:2009fk}.

Also we use 2 individual objects from~\cite{Weijmans:2007qx,Wasserman:2017vnt}. In the paper~\cite{Weijmans:2007qx} the rotation curve of NGC 2974 was obtained from spectroscopic measurements of the neutral hydrogen dynamics, see the middle panel of Fig.~\ref{fig:Elliptical_obs}. In the paper~\cite{Wasserman:2017vnt} the velocity dispersions of stars and globular clusters were used to infer the mass profile of NGC 1407, see the lower panel of Fig.~\ref{fig:Elliptical_obs}. Important caveats with such an analysis are the assumption of the unknown stellar velocity anisotropy $\beta(r)$ as well as the specific functional form of the parametric profiles used for DM and stars. For these two objects we use the maximal values of the total mass surface density outside 2~kpc (that is actually achieved at 2~kpc). We do not consider smaller radii as they are not resolved in the simulations with which we compare the observed systems. 

The main conclusion about elliptical galaxies is that we do not see clear maxima of the total mass surface densities. This is consistent with the simulations, where even with SIDM1 the cores formed are only evident at small radii, where the total mass is dominated by baryons. Therefore massive elliptical galaxies are not the best objects in which to see core like effects, see Fig.~\ref{fig:SDtot_13}. We will see, nevertheless, that the total mass surface density in these objects is very well reproduced in the simulations (both with CDM and SIDM).

\paragraph{Galaxy clusters.}
To calculate the surface densities of galaxy clusters we use strong and weak lensing data, in some cases supplemented with stellar kinematics and/or X-ray data.

\emph{7 objects from}~\cite{Newman:2012nv,Newman:2012nw}. In these papers the data on strong and weak gravitational lensing were combined with stellar kinematics within the BCG. These data were fitted using parametric models for the stellar and DM components, using two different ansatzes for the functional form of the DM density profile. This results in large uncertainties on the DM mass in the central parts of the halos where stars dominate. Therefore we use only data with $r>30$~kpc and compare them to the BAHAMAS simulation that have a $30$~kpc trust radius. In this region the resulting dark matter and total masses profiles agree well with each other.  

\emph{CLASH.} We also use 8 objects from the CLASH strong lensing survey~\cite{Merten:2014wna}  selected by the requirement that
the so called mean critical line distance\footnote{For CLASH objects there are  several strongly lensed sources  at different redshifts, with correspondingly different Eistein radii. The mean critical line distance is roughly the average value of these Eistein radii.} for them is between 25 and 50 kpc. The reason for this requirement is that we are going
to compare these objects with the BAHAMAS simulations and therefore we are interested in the maximum of total mass (or surface density) outside apertures of 30 kpc. For the objects with mean critical line distances in the above-mentioned range, the Einstein radii for some of the lensed sources are close to 30 kpc and therefore the total mass within 30 kpc of the cluster centre should be well determined. The values used in this work are given in Appendix~\ref{app:obs_data}, the $M_{200}$ values are from~\cite{Merten:2014wna}.

\emph{Abell S1063}. We use the data for the cluster Abell S1063 from~\cite{Sartoris:2020sdh} where the data on the velocity dispersion from the stellar kinematics of the BCG was combined with X-ray data to reconstruct mass profiles for the stellar mass of member galaxies, the hot gas component, the BCG stellar mass, and the DM. As with the other observed galaxy clusters, we use only data with $r>30$~kpc in our analysis, and as this cluster was part of the CLASH sample, we also compare the kinematics plus X-ray mass-profile measurement with that derived from strong and weak gravitational lensing.

\begin{figure}[h!]
  \centering
  \includegraphics[width=0.51\textwidth]{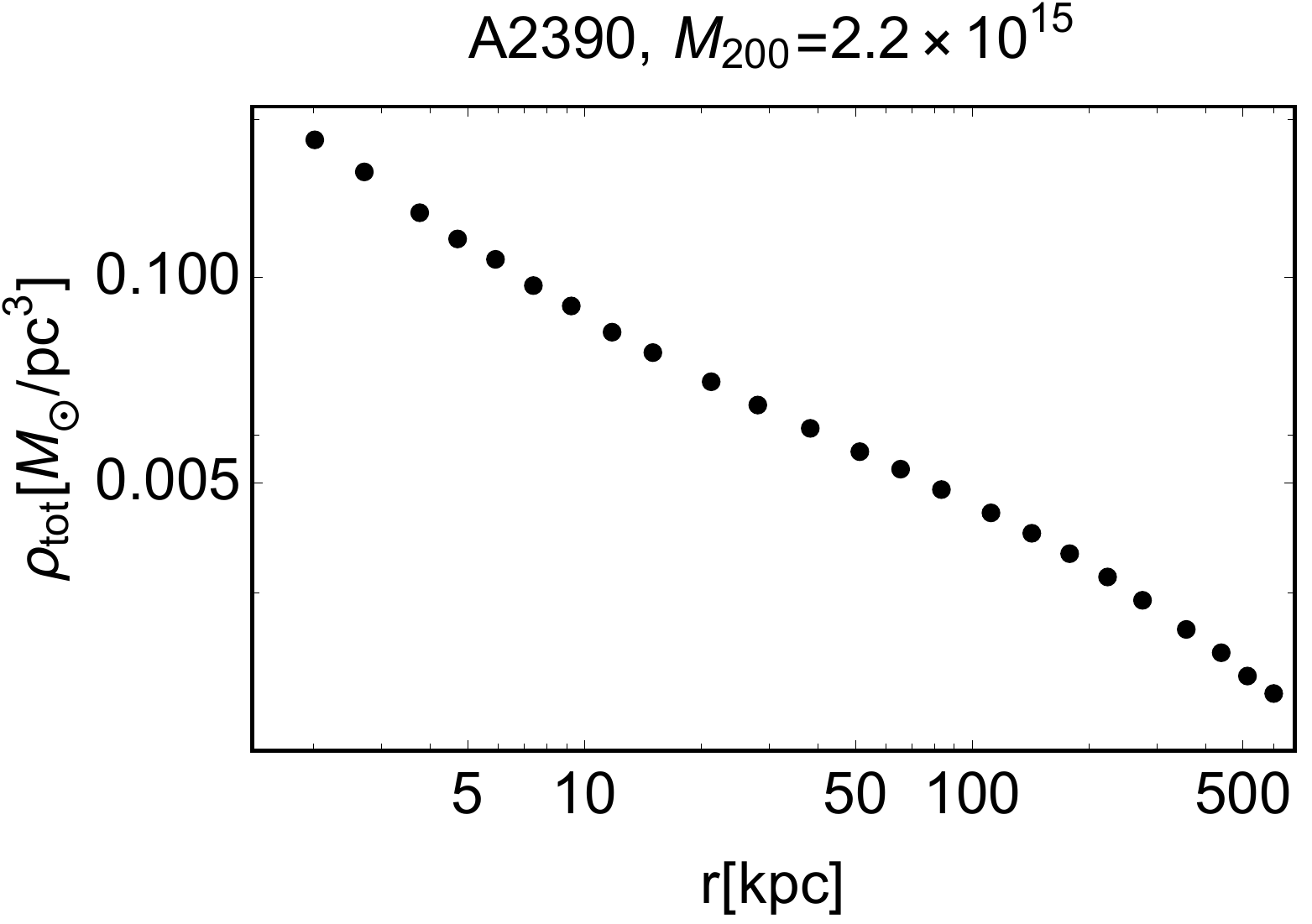}
  \includegraphics[width=0.48\textwidth]{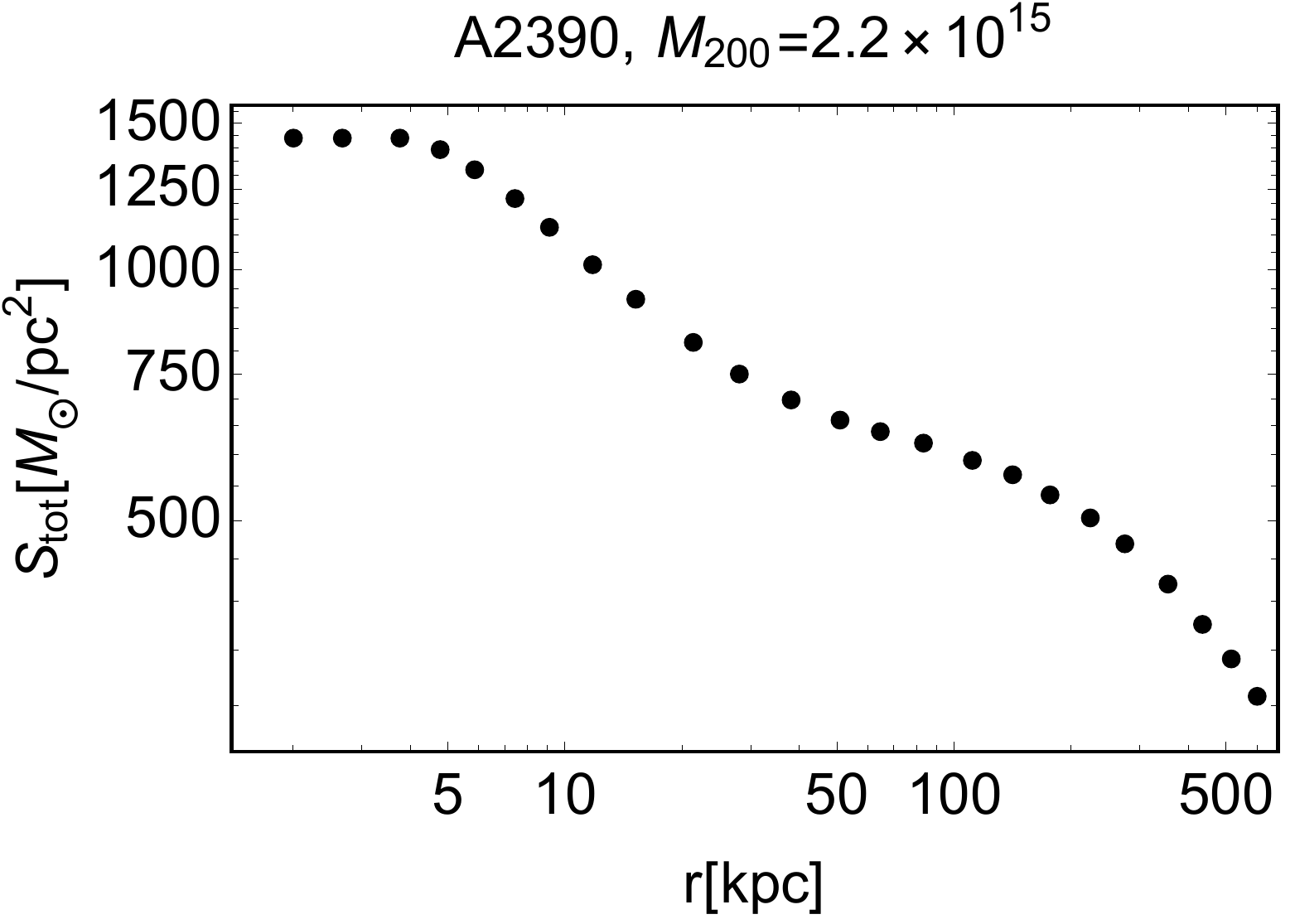}\\
  \includegraphics[width=0.51\textwidth]{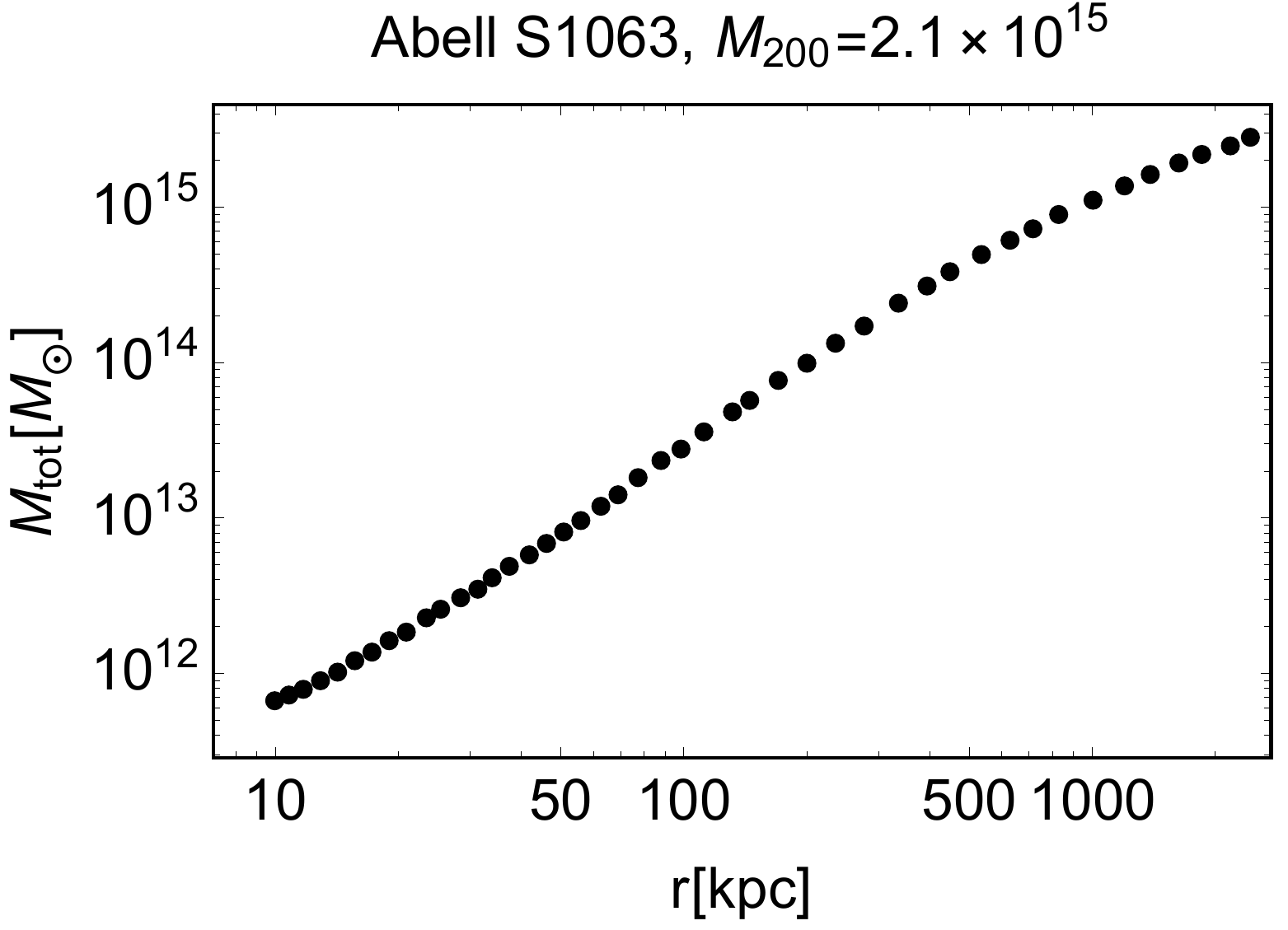}
  \includegraphics[width=0.48\textwidth]{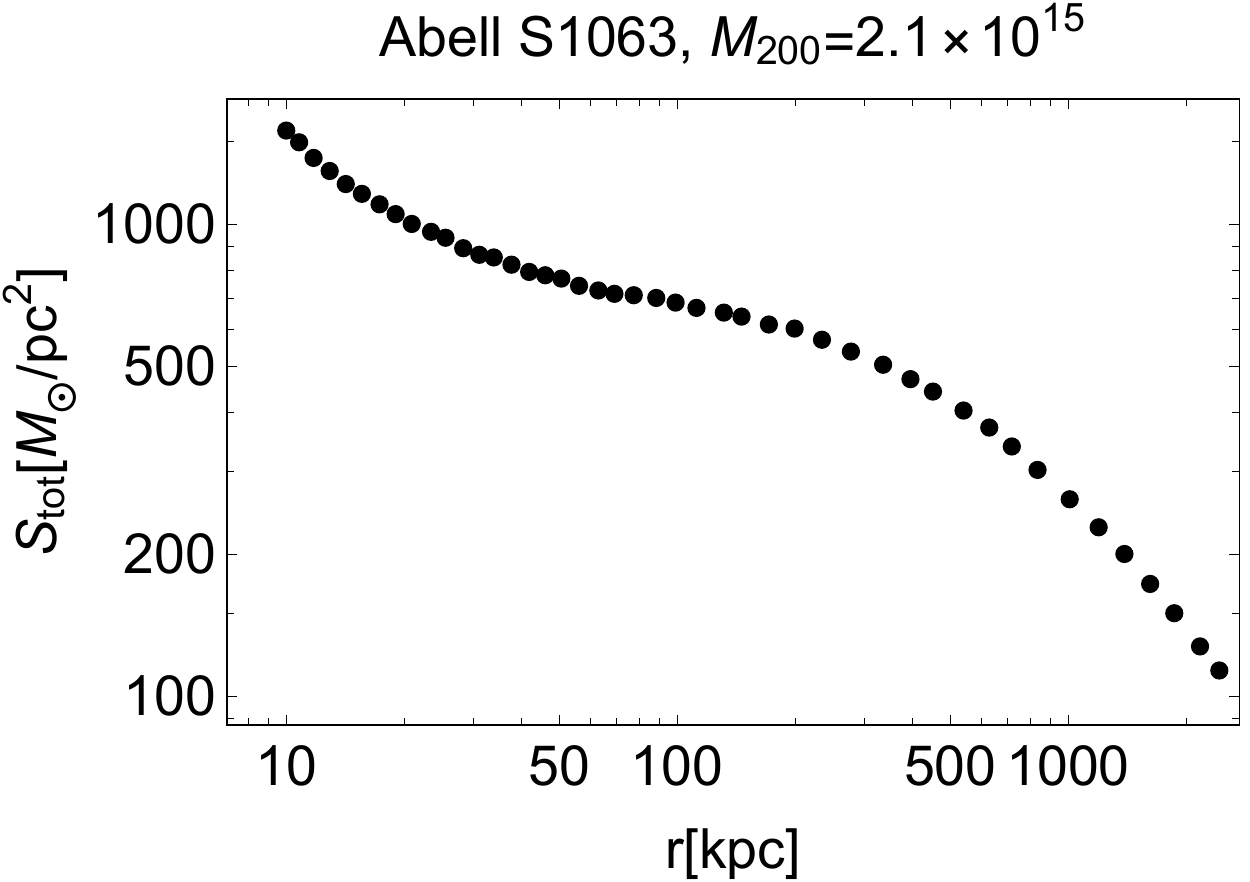}
  \caption{\textit{Left panel:} The best fit total mass density profile (black points). \textit{Right panel:} The total surface density calculated from the best fit values of the density profile.}
  \label{fig:Cluster_obs}
\end{figure}

\emph{Weak lensing data.} We take weak lensing data from a sample of 52 massive clusters from the Canadian Cluster Comparison Project (CCCP)~\cite{Hoekstra:2012ks} with redshifts $0.15 < z < 0.55$. The CCCP is an X-ray selected sample of massive galaxy clusters for which there is deep imaging from  the Canada-France-Hawaii Telescope (CFHT). From this sample we selected the most massive clusters by adopting a constraint on the gas temperature,  $T_{X} > 5$ keV. We also rejected 3 clusters (Abell 115, Abell 223 and Abell 1758) due to the fact that these objects are experiencing mergers~\cite{Hoekstra:2012ks}.

\begin{figure}[h!]
    \centering
    \includegraphics[width=0.5\textwidth]{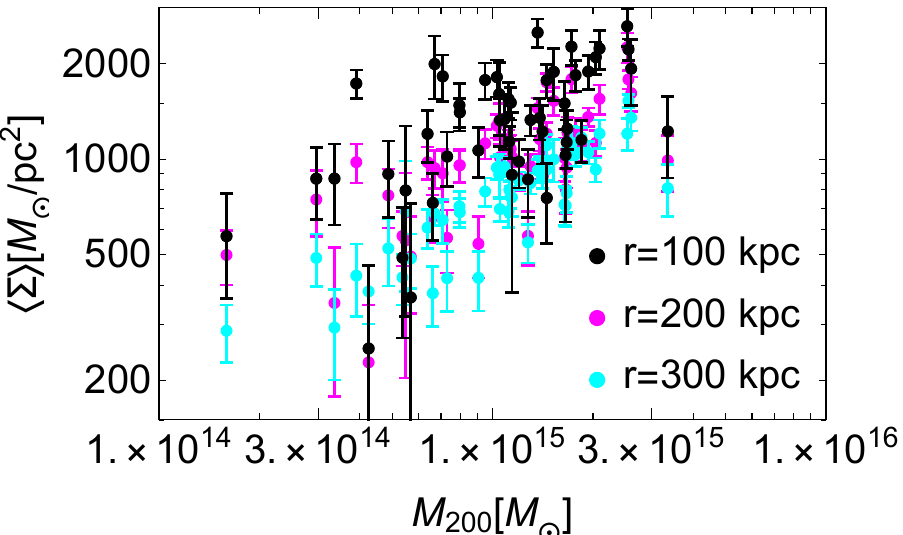}
    \caption{The 2-dimensional surface density (see Eq.~\eqref{eq:2dSD}) measured using weak lensing for clusters from CCCP sample at 3 different radii.}
    \label{fig:Clusters_2DSD}
\end{figure}

Weak leansing measures the 2-dimensional surface density (projected mass) that is given by
\begin{equation}
    \langle\Sigma\rangle(R) = \frac{M_{2D}(R)}{\pi R^2},
    \label{eq:2dSD}
\end{equation}
where $M_{2D}(R)$ is the mass inside a cylinder with  radius $R$.
For each cluster we take from the observational data three data points:
$M_{2D}(100 \text{kpc})$, $M_{2D}(200 \text{kpc})$ and $M_{2D}(300 \text{kpc})$. We illustrate this quantity for CCCP clusters in Fig.~\ref{fig:Clusters_2DSD}. For the final analysis we use the mass inside the smallest available radius - 100 kpc.

\section{Comparison of surface density between observations and simulations}
\label{sec:compare}

\begin{figure}[t!]
  \centering
  \includegraphics[width=0.48\textwidth]{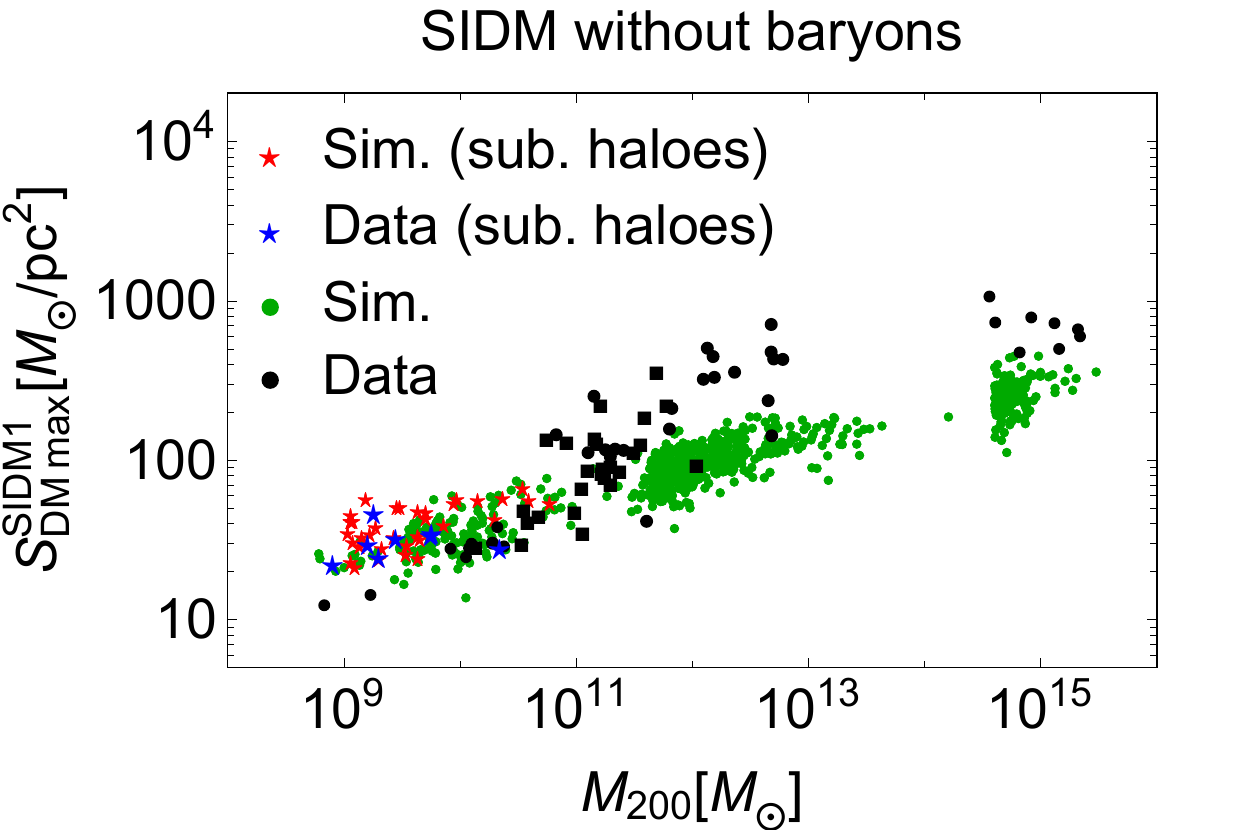}~\includegraphics[width=0.48\textwidth]{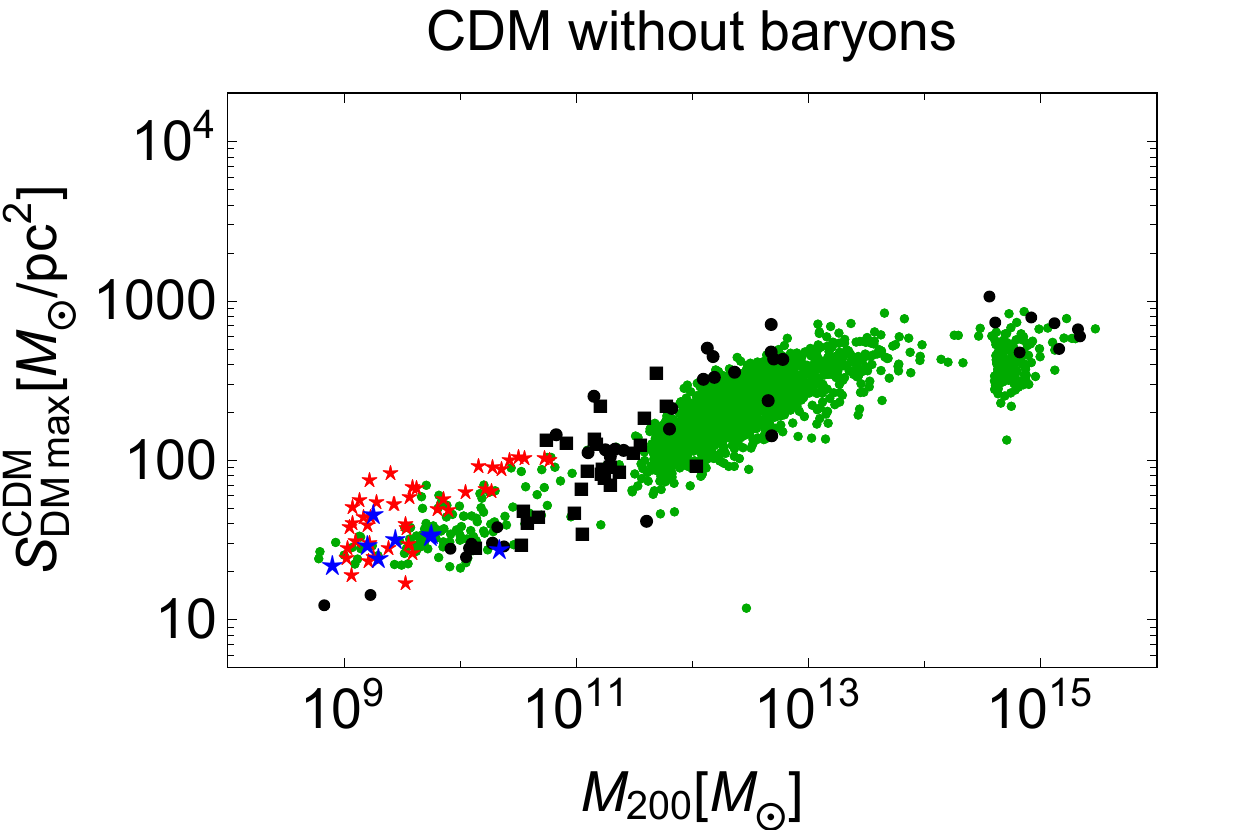}
  \\
  \includegraphics[width=0.48\textwidth]{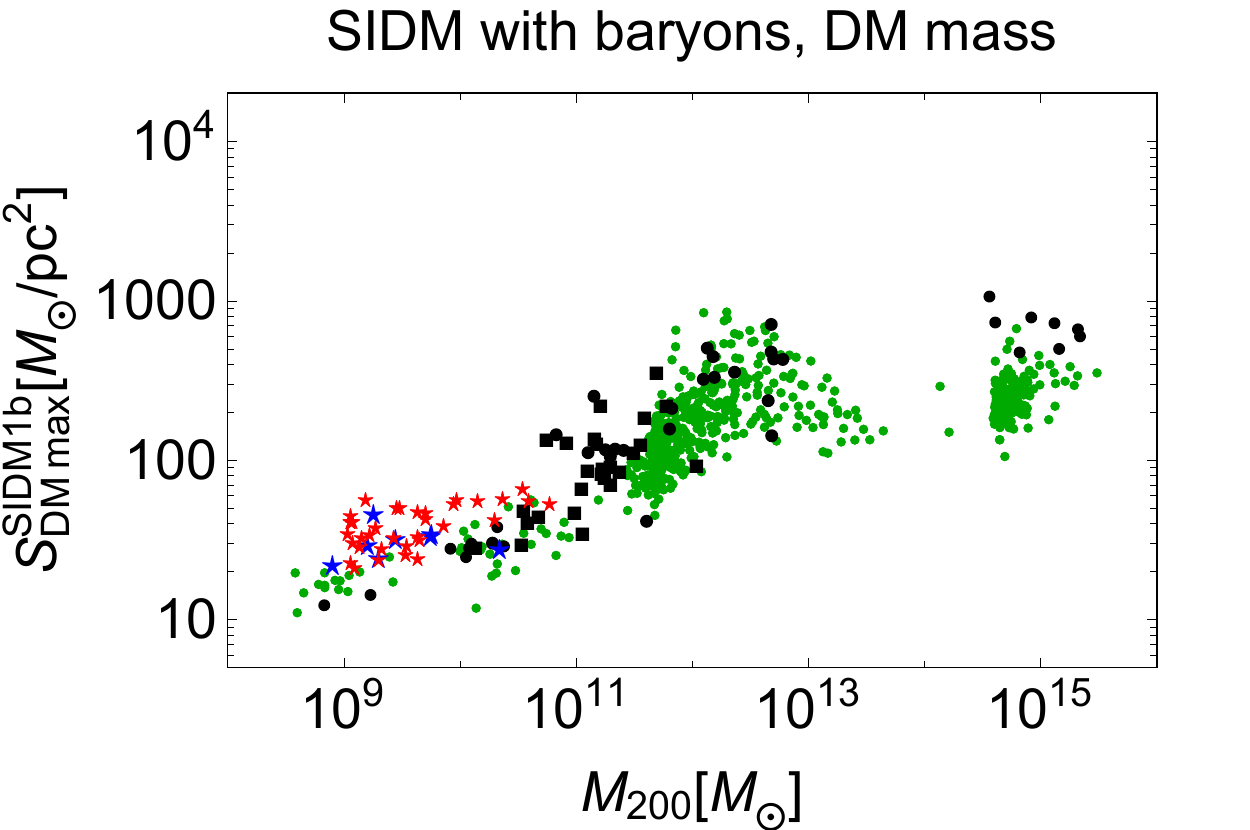}~\includegraphics[width=0.48\textwidth]{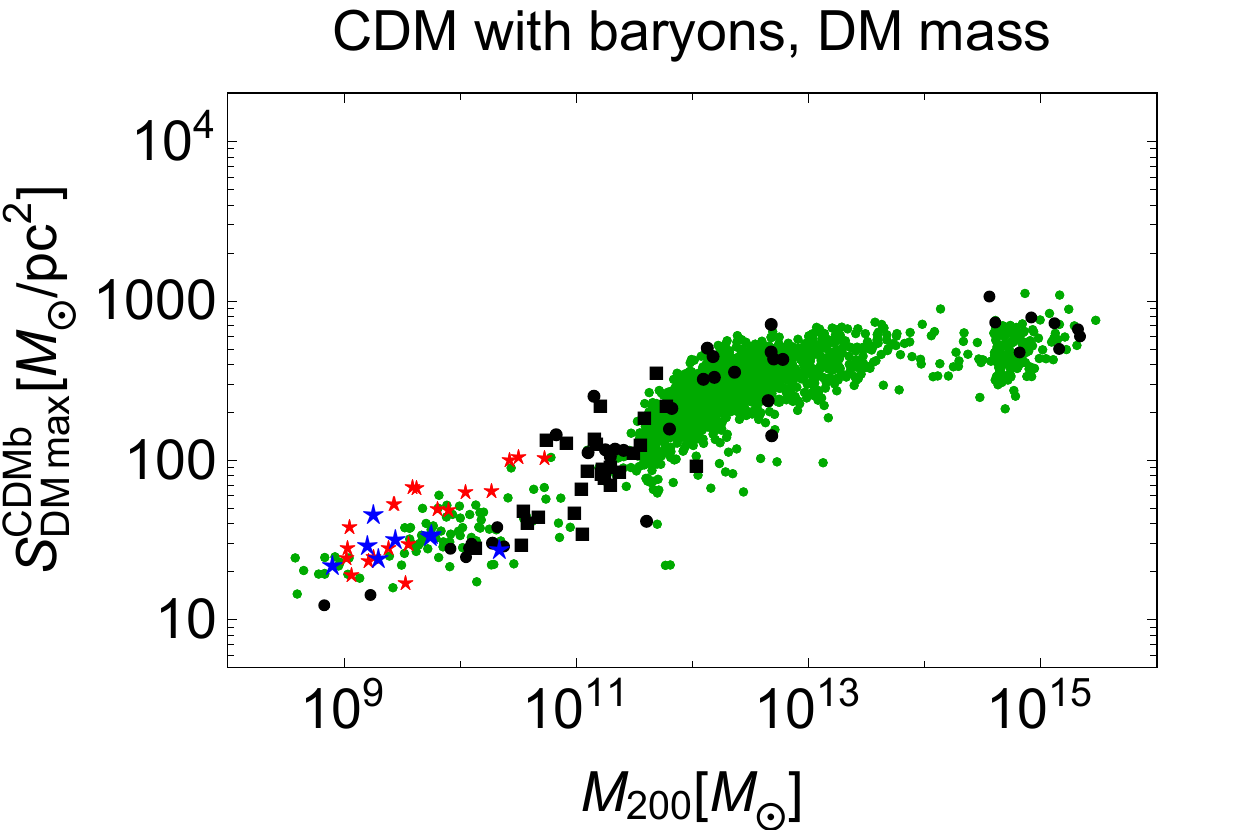}
  \\
  \includegraphics[width=0.48\textwidth]{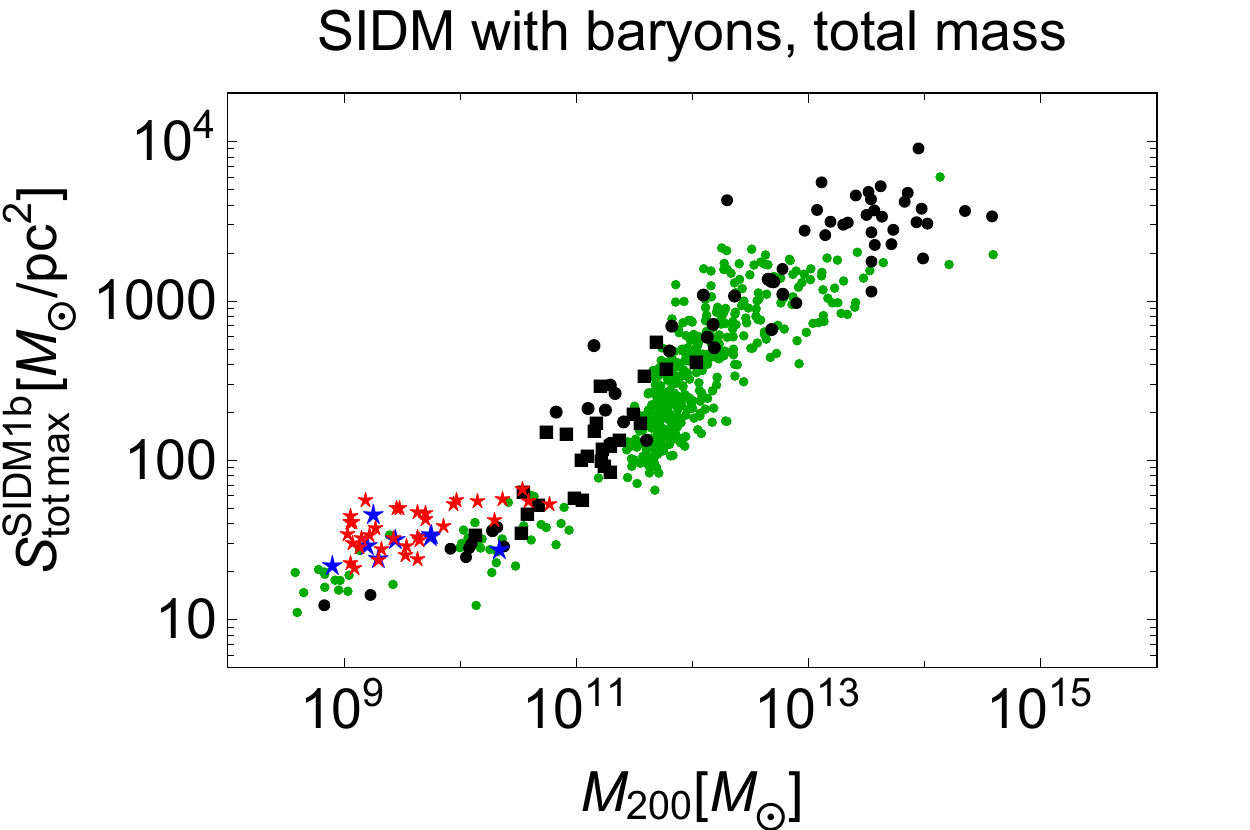}~\includegraphics[width=0.48\textwidth]{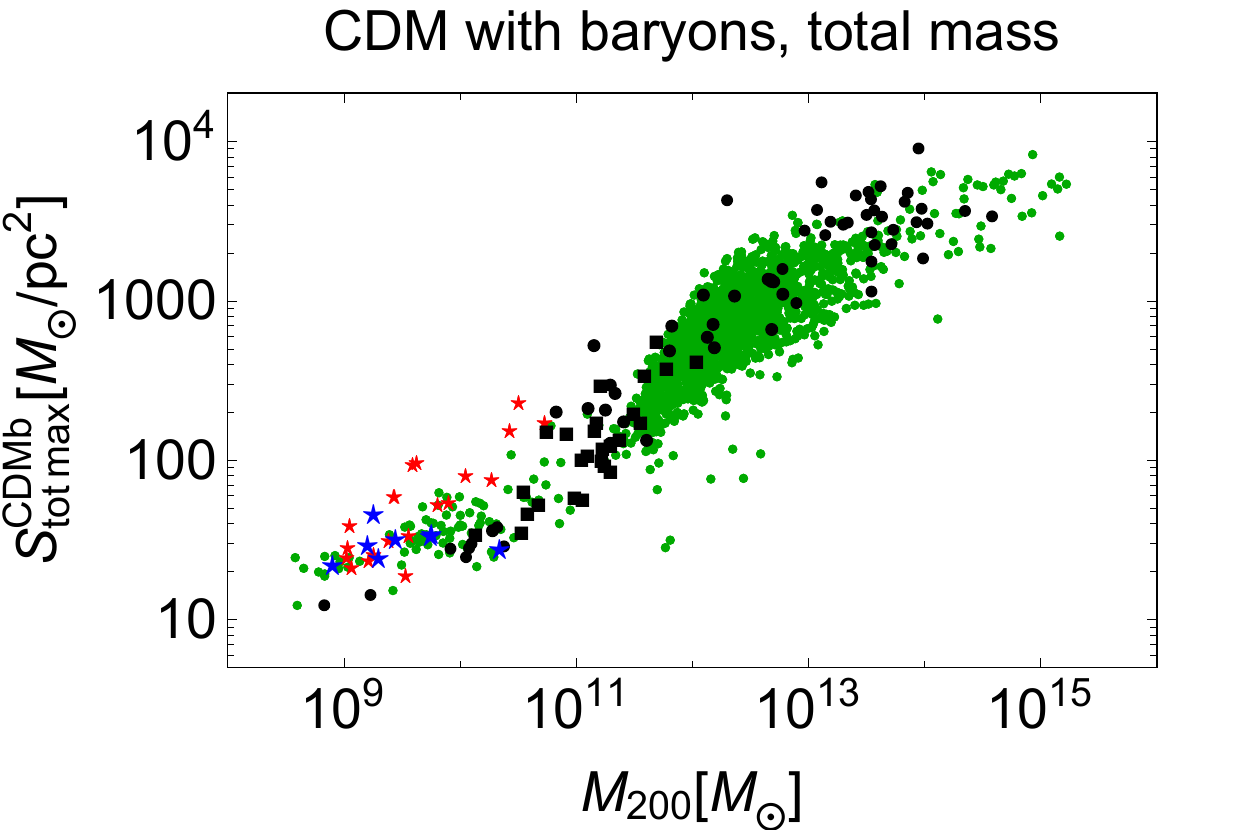}
    \caption{The maximal surface density of DM mass or total mass versus $M_{200}$ for observed isolated haloes (black points), observed subhalos (blue stars) and simulated isolated haloes (green points) and subhalos (red stars). Three left panels represent SIDM, three right panels represent CDM. Top two panel are for dark matter only simulations, two middle are for  DM surface density in the simulations with baryons, two lower panels are for the total mass surface density. 
    Observed objects include galaxy clusters, elliptical galaxies, spiral galaxies, isolated dwarf galaxies and classical dwarf satellites of the Milky Way. 
    Observed objects where the real local peak in $S(r)$ was detected are represented by black squares rather than black circles (see Section~\ref{sec:data} for details). 
    Simulation data is described in Table~\ref{tab:simulations}. For the BAHAMAS simulations we choose only massive haloes, $M_{200}>4 \times 10^{14} M_{\odot}$ that correspond to the masses of the observed halos that we use.
    For SIDM1b we do not have simulated subhalos and use SIDM1 DMO subhalos.
    }
    \label{fig:SDmax_new}
\end{figure}

In this section we finalize our analysis and compare the observational data with simulations of CDM and SIDM with a cross section of $1$ cm$^2/$g.
As stated earlier, our goal is to preform this comparison in a maximally model-independent way.
The quantity that we compare between haloes from our ensembles of simulated and observed systems is the surface density. We use both the \emph{total mass} surface density and (when available) the DM surface density.

We find that the maximal total surface density in the central parts of large halos, where the profile is dominated by baryons, is in good agreement with the observations (both for CDMb and SIDM1b). 
We also compare  the maximum surface density outside 30~kpc from the centre, where DM contribution is more significant and the difference between two DM  models is more visible.

\begin{figure}[h!]
  \centering
  \includegraphics[width=0.55\textwidth]{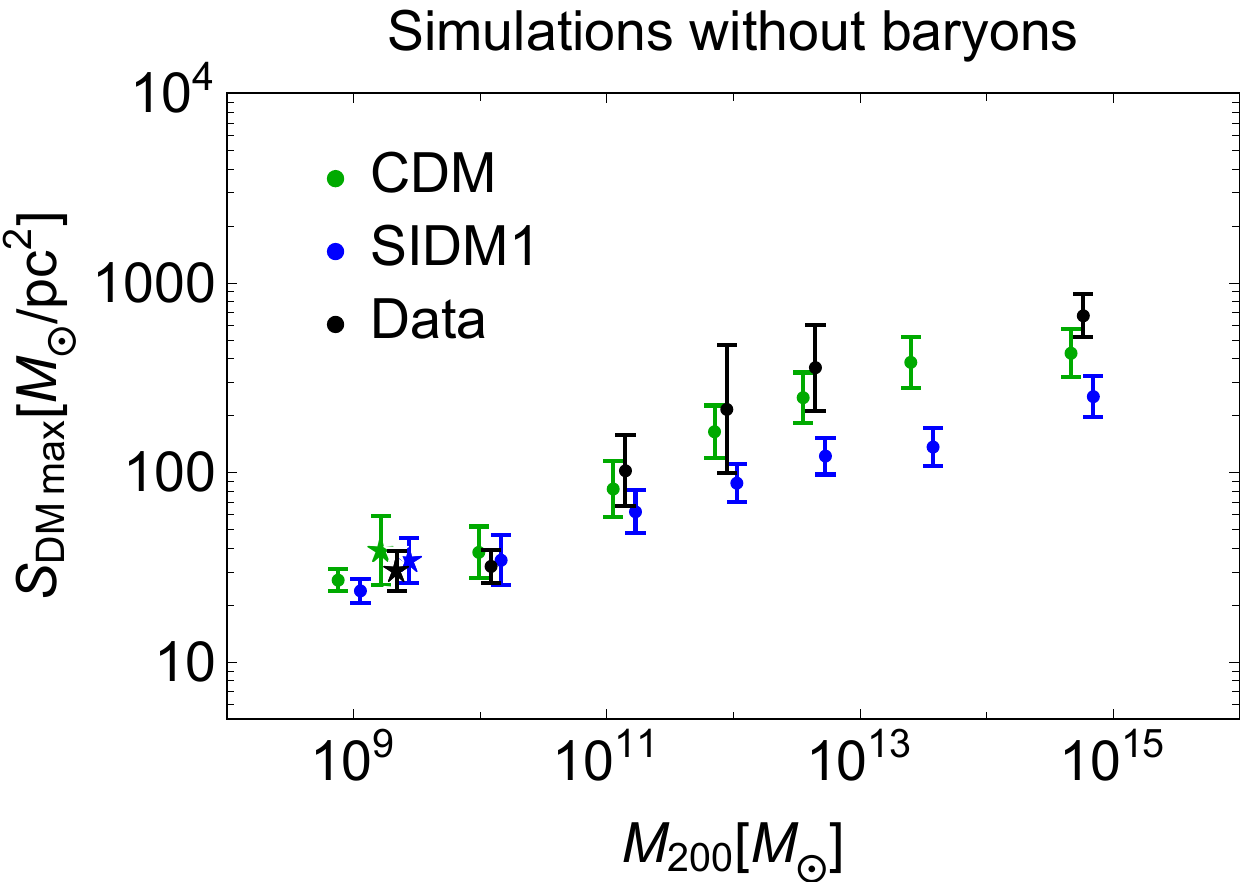}\\
  \includegraphics[width=0.55\textwidth]{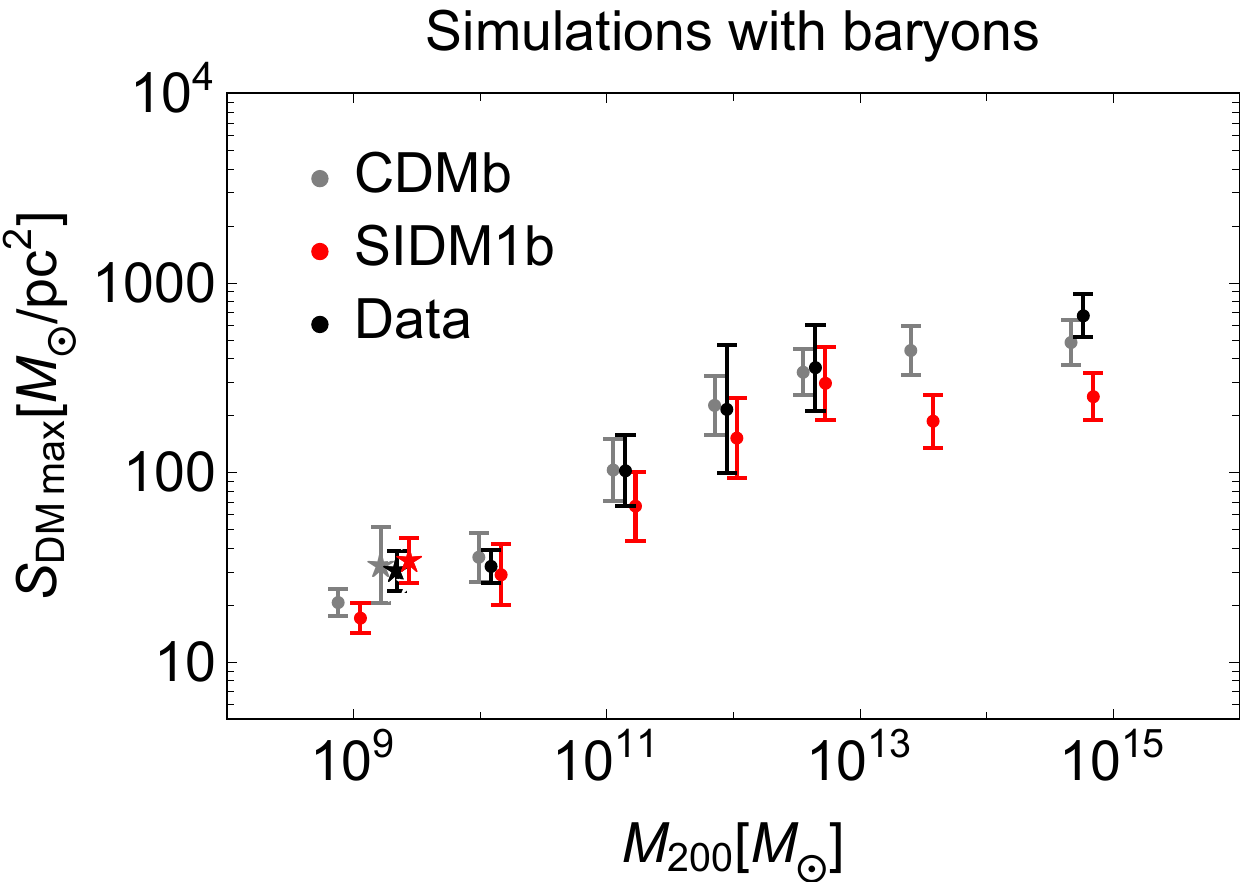}
  \\
  \includegraphics[width=0.55\textwidth]{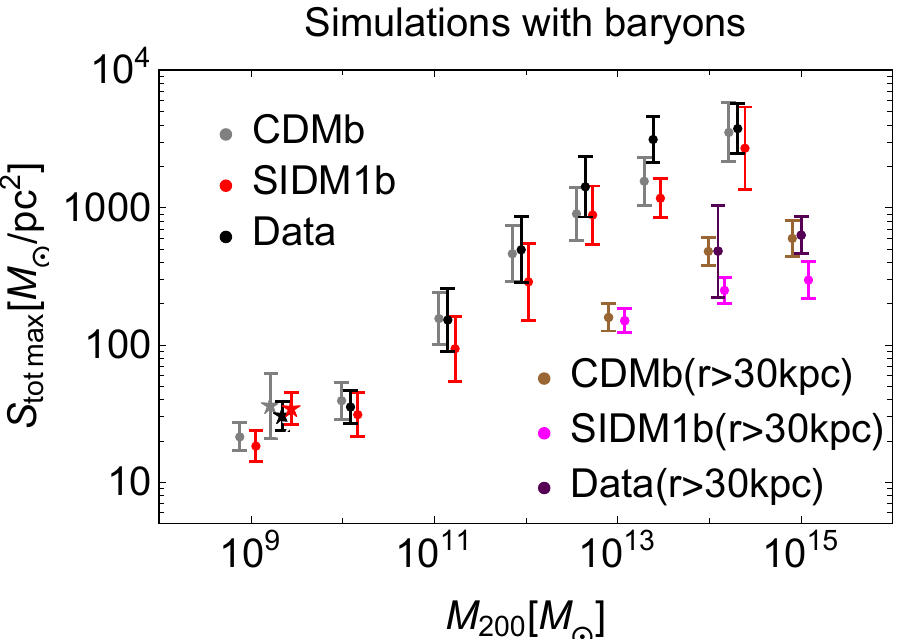}
  \caption{The same as in Fig.~\ref{fig:SDmax_new} for mass bins. Error bars correspond to the standard deviation. SIDM1 is plotted in blue, CDM in green, SIDM1b in red, CDMb in gray and observational data in black. Isolated halos are marked by points, subhalos are marked by stars. By brown, magenta and purple color in the bottom panel we present CDMb, SIDM1b and observational data for large halos looking for total mass surface density maximum outside $30$~kpc.}
    \label{fig:SDmax_new_v2}
\end{figure}

On the observational side we observe a real maximum in the total mass surface density profiles -- a clear and model-independent signature of a cored density profile -- only in spiral galaxies  with $M_{200} \lesssim 3 \cdot 10^{11} M_{\odot}$.
For all other objects we use a lower bound on the maximum of the surface density -- the value at the smallest available radius. This procedure is discussed in details in Section~\ref{sec:data}.
Our main results are presented in Fig.~\ref{fig:SDmax_new} and  Fig.~\ref{fig:SDmax_new_v2}.

From the upper row of Fig.~\ref{fig:SDmax_new}
(\textbf{DMO simulations} for SIDM1 (left) and CDM(right)) we conclude that, even if the overall trend in $S_{\text{DM max}}(M_{200})$ is roughly correct in both models, SIDM1 without baryons is not consistent with the data. For CDM the situation is slightly better, however, the simulated maximum surface densities still lie below the observed ones.

\begin{figure}[h!]
  \centering
  \includegraphics[width=0.48\textwidth]{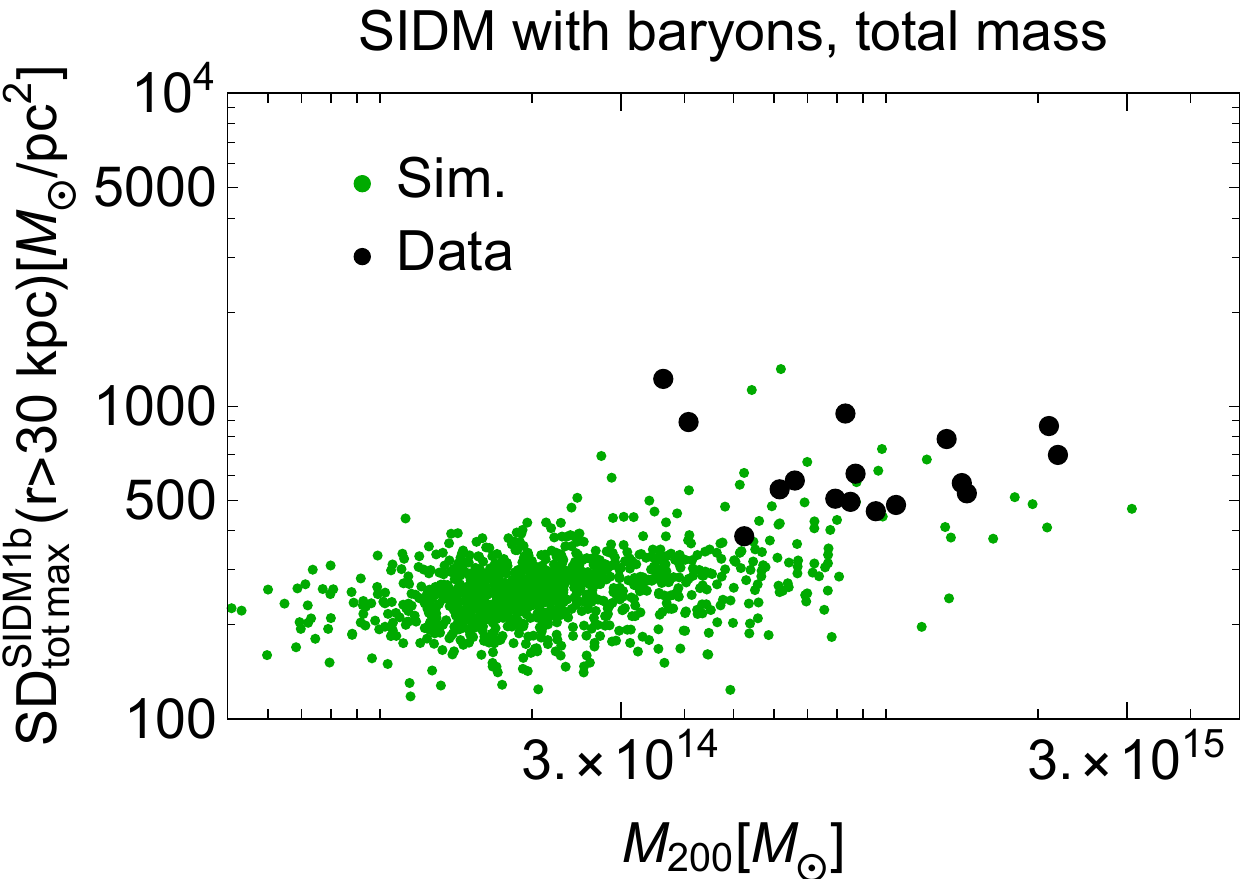}~\includegraphics[width=0.48\textwidth]{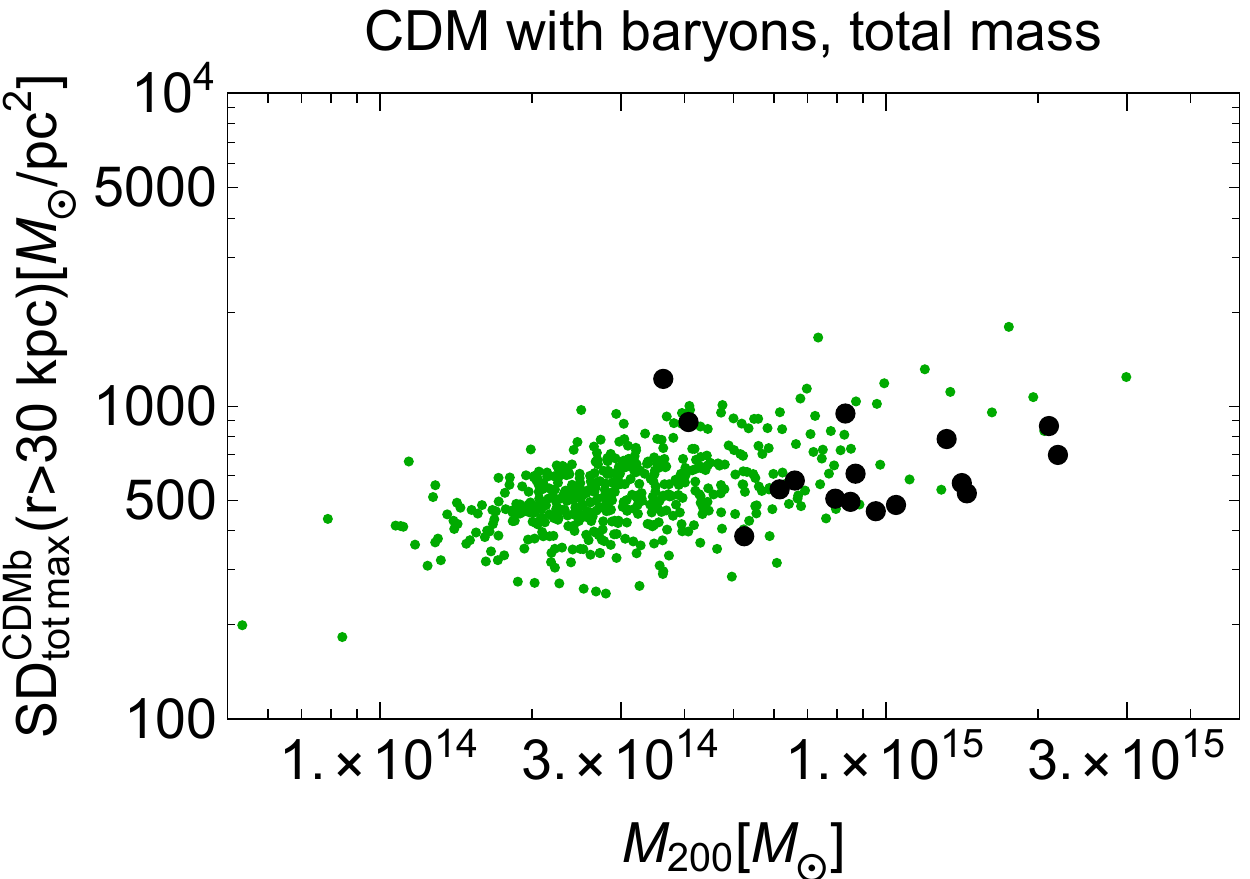}
    \caption{The maximum surface density for the total mass for SIDM1b (left) and CDMb (right) BAHAMAS simulations. Black points shows the observational data from Newman's clusters~\cite{Newman:2012nw} and one cluster from CLASH-VLT~\cite{Sartoris:2020sdh} for radii $r>30$~kpc that correspond to the trust radius in simulations.}
    \label{fig:SD_30}
\end{figure}

\begin{figure}[h!]
  \centering
  \includegraphics[width=0.48\textwidth]{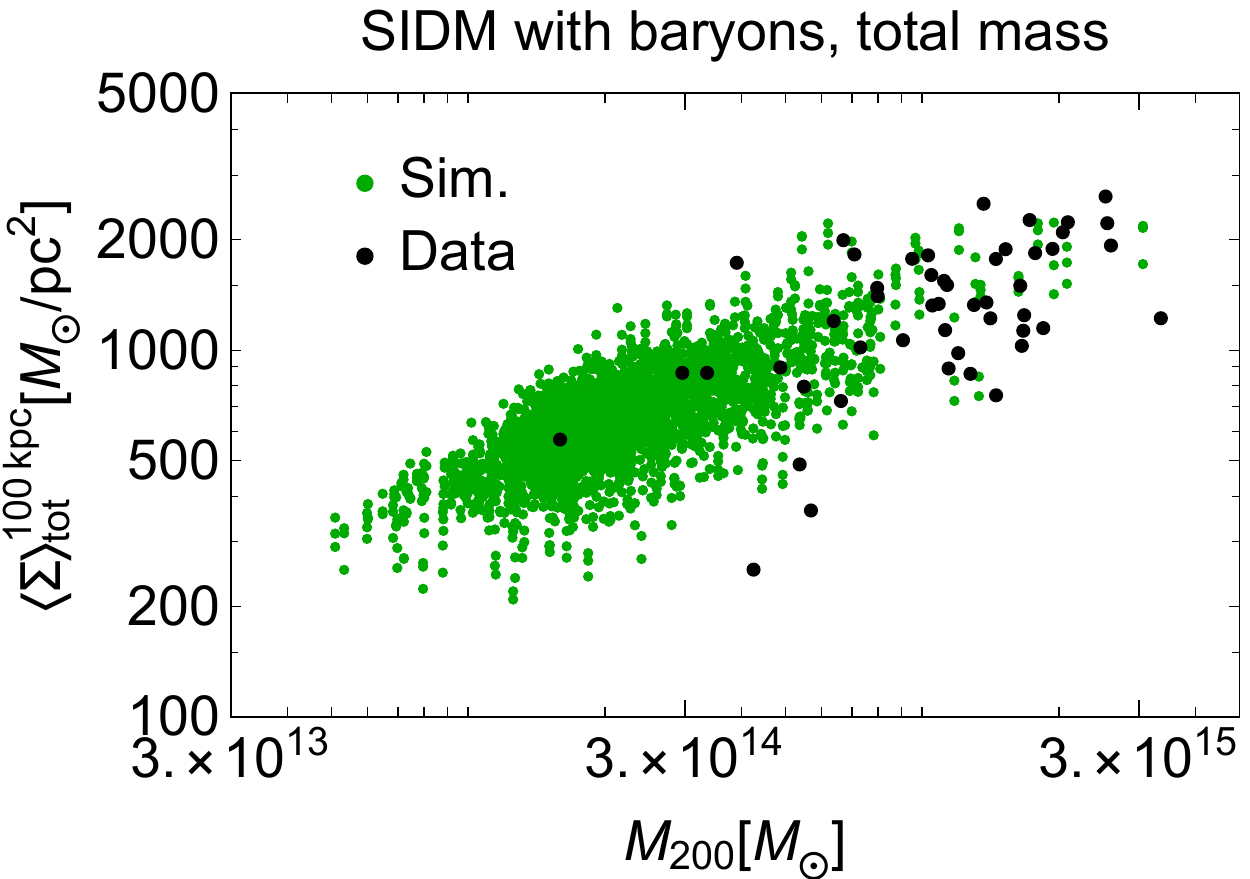}~\includegraphics[width=0.48\textwidth]{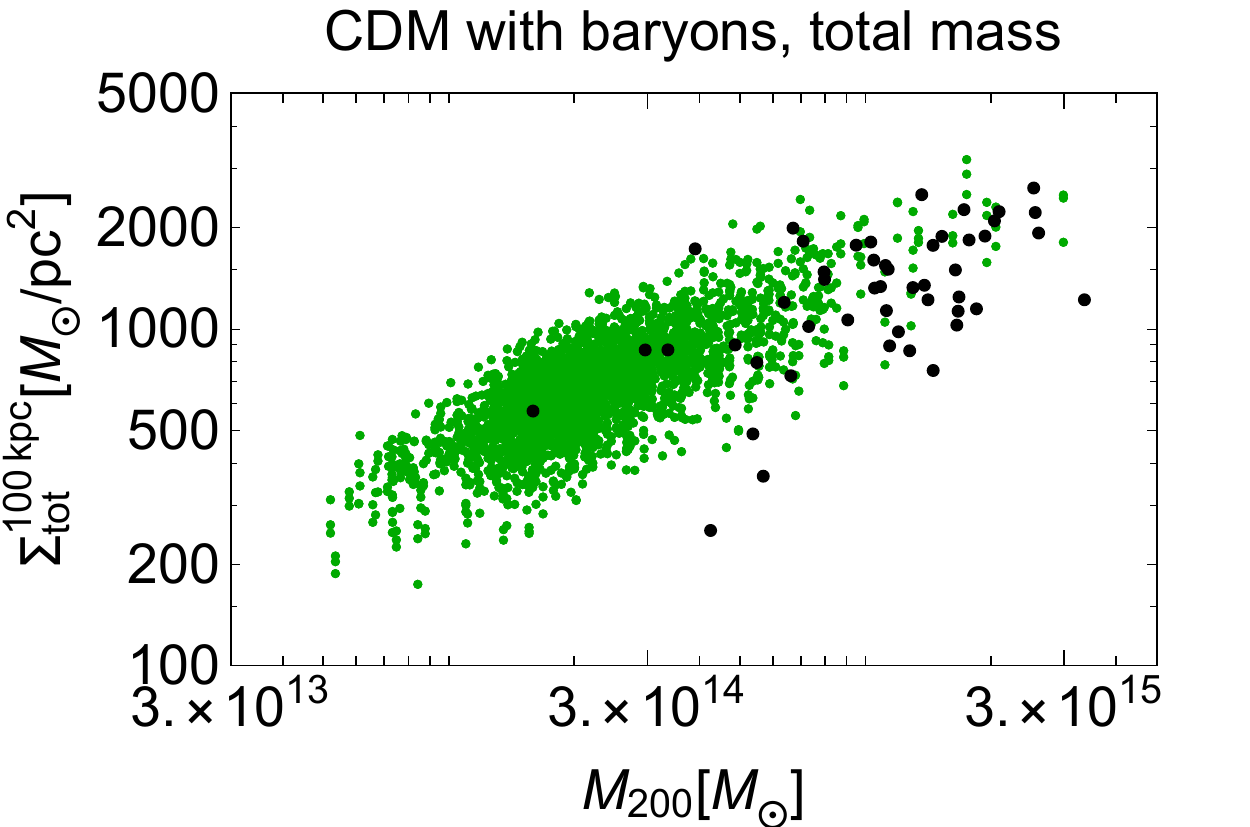}
    \caption{Comparison of the 2-dimensional surface densities (see Eq.~\eqref{eq:2dSD}) between simulations (green points) and observations (black points) at the smallest measured radius $100$~kpc. Left panel shows SIDM1b simulations, right panel shows CDMb simulations, observational points are taken from CCCP sample, see Fig.~\ref{fig:Clusters_2DSD}.}
    \label{fig:SD2D_clusters}
\end{figure}

The agreement between simulated and observed systems is improved substantially
for the \textbf{dark matter surface density in simulations with baryons}. Looking at the middle row of Fig.~\ref{fig:SDmax_new}, the EAGLE-based simulations provide a much better match to the observations than do the DMO simulations, both for CDM and SIDM1. In both models, there are strong baryonic effects on the DM distributions within spiral galaxies with masses around $10^{12} M_{\odot}$. With these effects taken into account, the simulated and observed distributions have significant overlap, for both models.  We see that the difference between CDM and SIDM is most pronounced in galaxy clusters, and we find that CDM with baryons provides a visibly better match to the observed systems than does SIDM (see a more quantitative discussion below).

We also observe in Fig.~\ref{fig:SDmax_new} that for sub-halos (red stars for simulations and blue stars for observations) the DM surface density at fixed $M_{200}$ is higher than in isolated halos (see e.g.~\cite{Boyarsky:2009rb,Boyarsky:2009af} for discussion).

The maximal values of the \textbf{total mass surface density} are virtually indistinguishable between the two DM models, and in both cases are in good agreement with the observations. As discussed before, for the most massive haloes these maximal surface density values occur at radii within which the mass distribution is dominated by baryons.
Therefore, we compare the maximal surface density at large radii (> 30~kpc) in Fig.~\ref{fig:SD_30}, where the difference between the two models is still visible in the total mass.
For this case we again see that  CDM is in better agreement with simulations than
SIDM. Finally, in Fig.~\ref{fig:SD2D_clusters} we compare the 2D mass inside 100~kpc directly derived from the weak lensing data with simulations. We see that both simulations agree with the data pretty well, and it does not seem to be possible to distinguish between the models at these scales.

In Fig.~\ref{fig:SDmax_new_v2} we re-plot the same data as in Fig.~\ref{fig:SDmax_new} introducing mass bins and calculating the average values and the standard deviations in each bin. We see that DMO haloes with SIDM1 produce maximal surface densities at odds with the observations. 
For the simulations with baryons the difference between the DM surface density in the two models is really visible only in the two most-massive bins. In the bin with the heaviest masses, the observations seem to be inconsistent with the predictions of SIDM1b. In the next largest bin the difference between the two models is still clear, but unfortunately we do not have good quality observational data here. For the total mass the situation is similar -- a significant difference between the data
and SIDM1b is present only for galaxy clusters with the mass around $10^{15} M_{\odot}$ and for the data outside $30$~kpc (magenta and brown points). In the next largest mass bin ($10^{14} M_{\odot}$) there is still a difference between the two models, but the available data do not allow us to distinguish between them. In the mass bins $M_{200} \sim 10^{13} M_{\odot}$  the difference between the models is not visible anymore, see the discussion above.

\section{Conclusions}
\label{sec:conc}

Our primary conclusion is that a self-interacting dark matter (SIDM) cross-section of 1 cm$^2/$g is marginally excluded, based on a comparison between the maximal surface density of simulated halos and that inferred for observed halos. We used the data for the observed and simulated (both CDM and SIDM1) objects of different sizes (over 7 orders of magnitude in total mass, from dwarf galaxies to galaxy clusters).
The constraints come from galaxy clusters, with less massive systems being relatively less affected by a velocity-independent SIDM cross-section.

\textcolor{black}{In this paper we choose to directly compare the haloes in realistic  SIDM simulations with those that are observed, rather than using a semi-analytic model (see ~\cite{Bondarenko:2017rfu} and \cite{Sokolenko:2018noz} as well as recent paper~\cite{Robertson:2020pxj} for the comparison between simulations and semi-analytic models).}
The current state of the art in simulations, including models for the baryonic physics relevant for galaxy formation, allows this to be done. In this work we collected ensembles of haloes across a wide range of halo masses, simulated with velocity-independent SIDM with a cross section 1 cm$^2/$g. These simulations were performed both with and without baryonic physics. Baryonic material can affect the distribution of DM within haloes, so it is important to include it in simulations that are to be compared directly with observations.
Analyzing these simulations we were able to make robust predictions that can be directly compared with observations. 

On the observational side, we used stellar velocity dispersion data for dwarf spheroidal galaxies, HI rotation curve data for spiral galaxies and strong and weak leaning data for galaxy clusters (some clusters had additional stellar kinematics or X-ray data). We demonstrated that it is possible to compare the mass distributions inferred from these data with SIDM simulations without fitting parametric models to the DM density distributions in either case. Specifically, we used a non-parametric quantity, the maximum value of the surface density ($S(r) = \left\langle \rho(<r)\right\rangle r$), to compare between observed and simulated systems. We believe that the analysis presented here is a successful proof of concept for obtaining robust constraints on SIDM from the inner properties of DM haloes.

Our results demonstrate that the velocity-independent cross section $\sigma=1$ cm$^2/$g is marginally disfavoured by our data.  These results are mainly based on the data from galaxy clusters, as at the relevant distances from the center these objects are more or less DM dominated.
For spiral galaxies the situation is different. Although the observational data implies a clear signature of constant-density DM cores (the total mass surface density demonstrates a  maximum at 2-5 kpc
from the center), there is a very significant contribution of baryonic matter inside these regions.
In simulations, baryons have a significant effect on the DM distribution, and tend to erase the differences between CDM and SIDM.
This means that realistic modelling of baryonic effects is crucial if we want to use the data from spiral galaxies to constrain SIDM (or maybe other DM models). In our work, we used a number of different baryonic physics models, reflecting the fact that a sample of simulated halos covering a wide range of halo masses requires a wide range of simulation resolutions, which in turn requires different methods for modelling baryons.
However, at fixed mass we did not investigate the differences between alternative models for baryonic physics, which will be important going forward. 

\acknowledgments

We gratefully acknowledge Torsten Bringmann for collaboration and useful discussions, and also the University of Oslo for hospitality. 
We thank Mike Boylan-Kolchin, Riccardo Herbonnet, Henk Hoekstra, Federico Lelli, Andrii Magalich, Justin Read, Matthieu Schaller, Joop Schaye, Matthew Walker for help, discussions and comments. We also thank Ian McCarthy for proving us with data from the BAHAMAS simulations, and Till Sawala, Azadeh Fattahi, Julio Navarro, Carlos Frenk and Kyle Oman for letting us use the APOSTLE simulations.

AS is supported by the New Frontiers program of the Austrian Academy of Sciences. KB and AB are supported by the European Research Council (ERC) Advanced Grant ``NuBSM'' (694896). AR is supported by the European Research Council's Horizon2020 project `EWC' (award AMD-776247-6). DH acknowledges support by the ITP Delta foundation.

\appendix

\section{Description of the simulations}
\label{sec:simulations}

\paragraph{GEAR.}
For our simulated low-mass dwarf galaxies we used the GEAR simulations \citep{revaz:2018,Revaz:2011yt,2016A&A...588A..21R}. GEAR includes models for gas cooling, chemical evolution, star formation, hydrogen self-shielding and Type Ia and II supernova yields and thermal blastwave-like feedback. The parameters of these \emph{subgrid} models were calibrated to reproduce not only the observed kinematics of stars in low-mass dwarves, but also the metallicty gradients, abundance ratios (Mg/Fe), and the observed kinetically distinct stellar populations. The SIDM version of the GEAR code was introduced in \cite{harvey:2018dwarf}, and used the SIDM scattering algorithm described in \cite{2017MNRAS.465..569R, 2017MNRAS.467.4719R}, with a velocity-independent cross-section of $10\,$cm$^2$/g. For this work, these simulations were re-run with a cross-section of $1\,$cm$^2$/g. 

Our GEAR sample includes 15 low-mass halos, each of which has a star-formation history that quenches before z=0. The DM and gas particle masses are $22,462 \, M_\odot/h$ and $4,096 \, M_\odot/h$ respectively, while the star particle mass is $1024 \, M_\odot/h$. The Plummer-equivalent gravitational softening lengths of DM and gas particles are 50 and 10 pc/h respectively at z = 0.

\paragraph{EAGLE}
The EAGLE simulations \cite{2015MNRAS.446..521S, 2015MNRAS.450.1937C}  are cosmological hydrodynamical simulations of galaxy formation, with sub-grid physics models for gas cooling, star formation, and feedback from both stars and active galactic nuclei. The simulations use a \cite{2014A&A...571A..16P} cosmology, with $\Omega_{m} = 0.307$, $\Omega_{b} = 0.04825$, $\Omega_{\Lambda} = 0.693$, $\sigma_8 = 0.8288$, $n_s$ = 0.9611 and $h$ = 0.6777. The DM and initial baryon particle masses are $1.8 \times 10^6 M_{\odot}$ and $9.7 \times 10^6 M_{\odot}$ respectively, and the Plummer-equivalent gravitational softening length is 0.7 kpc. The SIDM implementation within EAGLE was introduced in \cite{2018MNRAS.476L..20R}, based on the SIDM simulation method described in \cite{2017MNRAS.465..569R, 2017MNRAS.467.4719R}. 

\textit{EAGLE 100 Mpc.} The 100 Mpc EAGLE box was the flagship EAGLE simulation presented in \cite{2015MNRAS.446..521S}. As the simulation requires a large amount of computational resources, there is not an SIDM equivalent of this simulation, but it does exist both as a CDM-only simulation and with CDM plus EAGLE galaxy formation physics. We took the $1500$ most massive friend-of-friends groups from each of the CDM and CDMb simulations as our sample of haloes. These objects span a range of $M_{200}$ values, from $2.2\times 10^{11} M_{\odot}$ to $3.8\times 10^{14} M_{\odot}$.

\textit{EAGLE 50 Mpc.} As well as the large 100 Mpc box, a smaller 50 Mpc box exists. As well as exisiting with CDM and CDMb, this box has now been simulated with SIDM1 and SIDM1b [maybe REF Robertson et al. 2020 in prep?]. Within these simulations we choose the $450$ most massive friend-of-friends groups for CDM, SIDM1 and CDMb runs while for SIDM1b we take the $400$ most massive friend-of-friends groups. The halo mass range spanned by these haloes is from approximately $1.6\times 10^{11}M_{\odot}$ to $1.6\times 10^{14}M_{\odot}$.

\textit{C-EAGLE.} The relatively modest volume of the EAGLE simulation boxes leads to only a few massive galaxy clusters. We therefore supplement our haloes from the EAGLE 100 and 50 Mpc boxes with galaxy clusters from C-EAGLE \cite{2017MNRAS.470.4186B,2017MNRAS.471.1088B}. This project involved re-simulating massive clusters from a large DM-only box, using a zoom technique where the zoom region had the same mass and spatial resolution as the EAGLE simulations and used a very similar model of galaxy formation. Our sample of haloes from C-EAGLE includes 29 cluster for CDMb in the mass range from $1.1 \times 10^{14}M_{\odot}$ to $1.7 \times 10^{15}M_{\odot}$, and two clusters for SIDM1b with $M_{200}=1.4$ and $3.9 \times 10^{14} \, M_\odot$.

\paragraph{APOSTLE.}
APOSTLE (A Project Of Simulating The Local Environment) is a suite of cosmological hydrodynamic simulations of 12 zoom-in volumes that each contain a pair of haloes that approximately match the Local Group (the Milky Way and Andromeda)~\cite{2016MNRAS.457..844F,Sawala:2015cdf}. The simulations use the EAGLE model of galaxy formation, although they adopt a slightly different (WMAP-7 \cite{2011ApJS..192...18K}) cosmology. APOSTLE simulations exist at different resolutions, and we use both Level 1 (L1) and Level 2 (L2) simulations in this work, which we call `high res' and `low res' respectively.

\textit{APOSTLE low res.} One of the low-res APOSTLE volumes has been resimulated with $1$cm$^2$/g SIDM both with and without baryons, and so it is this volume that we use for all models in this work (CDM, CDMb, SIDM1, SIDM1b). For each model, we selected all friends-of-friends halos with $M_{200} > 10^{10}M_{\odot}$. We excluded halos that were contaminated by any low-resolution particles (these are the particles that trace the evolution of the volume outside of the zoom-in region). After applying these criteria we obtained $39$ haloes for CDM, $39$ for SIDM1, $33$ for CDMb, and $31$ for SIDM1b with masses up to $2.4 \times 10^{12} M_{\odot}$ (corresponding to the Milky Way or Andromeda-like galaxy in the simulation volume). Low-res APOSTLE has gas (DM) particle masses of approximately $1.2(5.9) \times 10^5 \, M_{\odot}$, and a Plummer-equivalent gravitational softening length of $300\,$pc. We adopt a trust radius of $1\,$kpc.

\textit{APOSTLE high res.} There are not any high-res APOSTLE volumes simulated with $1$cm$^2$/g SIDM and baryons, but there is a DM-only volume with $1$cm$^2$/g SIDM. As such, we only include the three models: CDM, CDMb and SIDM1 for APOSTLE high-res. We used a lower minimum halo-mass than with low-res, of $3 \times 10^{9} M_{\odot}$, reflecting the improved mass resolution. Again, we excluded halos that were contaminated by low-resolution particles. This leads to $63$ haloes for CDM, $86$ for SIDM1, and $44$ for CDMb. L1 APOSTLE has DM (gas) particle masses of approximately $1.0(5.0) \times 10^4\, M_{\odot}$, and a Plummer-equivalent gravitational softening length of $130\,$pc. The trust radius for these halos we took as $0.4$~kpc~\cite{Genina:2019job}.

\textit{APOSTLE subhaloes.} Given that the APOSTLE volumes are Local Group-like they contain satellite galaxies of the Milky Way and Andromeda, which are analogues of the observed Milky Way dwarf spheroidal galaxies. Using the high-res APOSTLE volumes discussed above, we extracted data for the most massive subhalos around either of the two main haloes, including all subhaloes with masses (as identified by the SUBFIND algorithm \cite{2001MNRAS.328..726S}) larger than $10^{9} M_{\odot}$. After applying this criteria we obtained $22$ subhaloes for CDM, $18$ for SIDM1, and $11$ for CDMb with masses up to $6 \times 10^{10} M_{\odot}$.

\paragraph{BAHAMAS.}
BAHAMAS simulations (BAryons and HAloes of MAssive Systems)~\cite{McCarthy:2016mry} are hydrodynamical simulations with a boxsize of $400$~Mpc$/$h. There are $1024^3$ DM particles with masses $5.5\times 10^9 \, M_{\odot}$, and the same number of gas particles with initial masses of $1.1\times 10^9 \, M_{\odot}$. The simulations use a WMAP-9 cosmology \cite{2013ApJS..208...19H}. As well as the CDM plus baryons simulation, this volume has been simulated with SIDM1 plus baryons \cite{Robertson:2018anx}; there are also DM-only versions of both the CDM and SIDM1 simulations.

To compare the properties of CDM and SIDM haloes we selected the $1000$ most massive friends-of-friends groups for CDM and SIDM1, $500$ for CDMb, and $999$ for SIDM1b.

\section{Observational data}
\label{app:obs_data}

\paragraph{Dwarf irregular galaxies.} For the field dwarf galaxies we use the results from~\cite{Read:2018fxs} for 8 objects: CVnIdwA, DDO 52, DDO 87, DDO 154, DDO 168, DDO 210, NGC 2366, WLM.

\begin{table}[!t]
    \centering
    \begin{tabular}{|l|c|c|c|c|c|}
        \hline
         Object   & $r_h$, pc & $M(r_h)$, $M_\odot$ & $S(r_h)$, $M_\odot/\text{pc}^2$ & $M_{200}$, $10^{9}M_{\odot}$ \\
         \hline
         Carina & 250 & $6.5\times 10^6$ & 24.9 & 0.8 \\
         Draco & 221 & $1.0\times 10^7$ & 50.6 & 1.8 \\
         Fornax & 710 & $5.7\times 10^7$ & 27.1 & 21.9 \\
         Leo I & 251 & $9.7\times 10^6$ & 36.7 & 5.6 \\
         Leo II & 176 &  $6.0\times 10^6$ & 46.2 & 1.6 \\
         Sculptor & 283 & $1.1\times 10^7$ & 33.6 & 5.7 \\
         Sextans & 695 & $3.6\times 10^7$ & 17.9 & 2.0 \\
         Ursa Minor & 181 & $4.2\times 10^6$ & 30.8 & 2.8 \\
         \hline
    \end{tabular}
    \caption{Parameters of classical dSphs from  paper~\protect\cite{Read:2018fxs}. The columns are: 1) Object name, 2) half-light radius 3) mass at half-light radius 4) surface density at half-light radius.}
    \label{tab:dSphs-data-old}
\end{table}

\paragraph{Classical dwarfs.} In Table~\ref{tab:dSphs-data-old} we present data for dShps from~\cite{Read:2018fxs} used in this work.

\paragraph{Spiral galaxies:} After selection of objects in SPARC catalogue~\cite{Lelli:2016zqa} discussed in Section~\ref{sec:data} we have selected the following objects (good objects list): 
IC 4202, D631-7, DDO 154, DDO 168, ESO 563-G021, F574-1, F568-3, F568-V1, NGC 0024, NGC 0055, NGC 0247, NGC 0300, NGC 0801, NGC 2403, NGC 2683, NGC 2841, NGC 2915, NGC 2955, NGC 2976, NGC 2998, NGC 3109, NGC 3198, NGC 3741, NGC 3877, NGC 3893, NGC 3917, NGC 3972, NGC 4010, NGC 4100, NGC 4157, NGC 4183, NGC 5005, NGC 5907, NGC 6195, NGC 6503, NGC 7793, NGC 7814,  UGC 01281, UGC 02885, UGC 06614, UGC 06917, UGC 06930, UGC 06983, UGC 07151, UGC 07524, UGC 08286, UGC 08490, UGC 09037, UGC 12506.

\paragraph{Elliptical galaxies and groups of galaxies.} List of the selected X-ray galaxies from~\cite{thesisNagino}: IC 1459, NGC 720, NGC 1316, NGC 1332, NGC 1395, NGC 1399, NGC 3607, NGC 3665, NGC 3923, NGC 4365, NGC 4472, NGC 4526, NGC 4552, NGC 4636, NGC 4649, NGC 5044, NGC 5322, NGC 5846.

Additionally, we use data for 12 strong lensed early-type galaxies from~\cite{Oldham:2018yte}, see data for them in Table~\ref{tab:EllipticalDavid}. Also, we use 2 individual objects NGC 2974 from~\cite{Weijmans:2007qx} and NGC 1407 from~\cite{Wasserman:2017vnt}.

\begin{table}[]
    \centering
    \begin{tabular}{|l|c|c|c|c|}
    \hline
    Object & $r_{\text{Ein}}$, kpc &$M(r_{\text{Ein}})$, $10^{11}M_{\odot}$  & $S(r_{\text{Ein}})$, $M_{\odot}/\text{pc}^2$ & $M_{200}$, $10^{13}M_{\odot}$ \\ \hline
    J0837& 3.1 & 1.5 & 3656 & 22.4\\
    J0901& 3.1 & 1.2 & 3090 & 2.2\\
    J0913& 2.3 & 1.0 & 4751 & 7.2\\
    J1125& 4.9 & 3.4 & 3382 & 38.2\\
    J1144& 3.5 & 1.4 & 2784 & 5.4\\
    J1218& 3.2 & 1.3 & 3046 & 10.6 \\
    J1323& 1.4 & 0.40 & 4574 & 2.6 \\
    J1347& 2.3 & 0.77 & 3453 & 3.2 \\
    J1446& 1.9 & 0.47 & 3129 & 1.6 \\
    J1605& 2.9 & 1.2 & 3366 & 4.3 \\
    J1606& 2.7 & 0.96 & 3108 & 8.6 \\
    J2228& 2.3 & 0.81 & 3777 & 9.5 \\ \hline
    \end{tabular}
    \caption{Parameters of 12 strong lensed early-type  galaxies from~\cite{Oldham:2018yte}. The columns are: 1) the object name; 2) Einstein radius; 3) mass at the Einstein radius; 4) the total mass surface density at the Einstein radius; 5) the virial mass. The virial mass was estimated from the stellar mass using the Moster relation~\cite{Moster:2009fk}.}
    \label{tab:EllipticalDavid}
\end{table}

\begin{table}[]
    \centering
    \begin{tabular}{|l|c|c|c|}
         \hline
         Object & $M(30\text{ kpc})$, $10^{12}M_{\odot}$  & $S(30\text{ kpc})$, $M_{\odot}/\text{pc}^2$ & $M_{200}$, $10^{14}M_{\odot}$ \\ \hline
        A383 & 2.3  & 608 & 8.7 \\ 
        A209 & 1.7 & 461 & 9.5  \\ 
        A2261 & 2.1  & 567 & 14.1 \\ 
        RXJ2129 & 2.0  & 541 & 6.2 \\ 
        A611 & 1.9  & 494 & 8.5 \\ 
        MS2137 & 1.8  & 483 & 10.5 \\ 
        RXJ1532 & 1.4  & 383 & 5.2 \\ 
        MACSJ0429 & 1.9  & 505 & 7.9 \\ \hline
    \end{tabular}
    \caption{Parameters of CLASH clusters from~\cite{Merten:2014wna}. The columns are: 1) the object name; 2) mass at 30 kpc; 3) the total mass surface density at 30 kpc; 4) the virial mass.}
    \label{tab:CLASH}
\end{table}

\paragraph{Clusters of galaxies.} 
From CLASH catalogue~\cite{Merten:2014wna} we selected 8 objects with mean critical line distance between 25 and 50 kpc, see data used in this work in Table~\ref{tab:CLASH}. 
From~\cite{Newman:2012nv,Newman:2012nw} we use best-fit data for the total mass and DM mass gNFW profile for 7 galaxy cluster.
Also, we use an individual object Abell S1063 from~\cite{Sartoris:2020sdh}.

\paragraph{Weak lensing data.} We use weak lensing data for 52 massive clusters from the Canadian Cluster Comparison Project (CCCP)~\cite{Hoekstra:2012ks} with redshifts $0.15 < z < 0.55$. We rejected 3 clusters (A115, A223 and A1758) due to the fact that these objects experience a merger. The full list of objects that we use is: 3C295, A68, A209, A222, A267, A370, A383, A520, A521, A586, A611, A697, A851, A959, A963, A1234, A1246, A1689, A1763, A1835, A1914, A1942, A2104, A2111, A2163, A2204, A2218, A2219, A2259, A2261, A2390, A2537, CL0024, CL0910, CL1938, MACS0717, MS0016, MS0440, MS0451, MS0906, MS1008, MS1224, MS1231, MS1358, MS1455, MS1512, MS1621, RXJ1347, RXJ1524.

\bibliographystyle{JHEP} %
\bibliography{SIDM.bib}

\end{document}